\documentclass[12pt]{article}
\usepackage[bf]{caption}
\usepackage{cmap}
\usepackage{lscape,isoent,cyrillic,amssymb,revsymb}
\newcommand{\cyrrm}{\fontencoding{OT2}\selectfont\textcyrup}
\newcommand{\cyrit}{\fontencoding{OT2}\selectfont\textcyrit}

\usepackage[square,comma,numbers,compress]{natbib}

\usepackage[linktocpage=true,dvips,bookmarksnumbered=true,breaklinks=true,pagebackref=true,colorlinks=true,urlcolor=blue,citecolor=red,linkcolor=magenta]{hyperref}
\usepackage{hypernat}
\usepackage{url}
\usepackage{breakurl}

\begin{document}
\bibliographystyle{plainnat}
\def\refname{\Large References

\noindent
{\rm\small 
\begin{minipage}[t]{15cm}
[For references in Cyrillic letters, 
we apply the (new) {\it Mathematical Reviews} transliteration
(transcription) scheme
(to be found at the end of index issues of {\it Mathematical Reviews}).]
\end{minipage}}\\[-0.3cm]}

\begin{flushright}
NIKHEF/2010-005
\end{flushright}

\ \\[-0.2cm]

\renewcommand{\thefootnote}{\fnsymbol{footnote}}
\begin{center}
{\LARGE\baselineskip0.9cm
Nonlinear Bogolyubov-Valatin transformations: 

2 modes\\[1.5cm]}

{\large 
K. Scharnhorst$^{1\;}$\footnote[2]{E-mail: {\tt k.scharnhorst@vu.nl}}, 
J.-W.\ van Holten$^{2,1\;}$\footnote[3]{E-mail: 
{\tt t32@nikhef.nl}}
}\\[0.3cm]

{\small 
$^1$ Vrije Universiteit Amsterdam,
Faculty of Sciences, Department of Physics and Astronomy,
De Boelelaan 1081, 1081 HV Amsterdam, The Netherlands

$^2$ NIKHEF, P.O.\ Box 41882, 1009 DB Amsterdam,
The Netherlands }\\[1.5cm]

\begin{abstract}
Extending our earlier study of nonlinear Bogolyubov-Valatin 
transformations (canonical transformations for fermions) for
one fermionic mode, in the present paper we perform a thorough
study of general (nonlinear) canonical transformations for two 
fermionic modes. We find that the Bogolyubov-Valatin group for
$n=2$ fermionic modes which can be implemented 
by means of unitary $SU(2^n = 4)$ 
transformations is isomorphic to $SO(6;\mathbb{R})/\mathbb{Z}_2$.
The investigation touches on a number of 
subjects. As a novelty from a mathematical point of view, we 
study the structure of nonlinear basis transformations in a Clifford 
algebra [specifically, in the Clifford algebra $C(0,4)$] entailing
(supersymmetric) transformations among multivectors 
of different grades. A prominent
algebraic role in this context is being played by bipa\-ra\-vectors
(linear combinations of products of Dirac matrices, 
quadriquaternions, sedenions) and 
spin bivectors (antisymmetric complex matrices). 
The studied biparavectors are equivalent to Eddington's
$E$-numbers and can be understood in terms of the tensor product
of two commuting copies of the division algebra of 
quaternions $\mathbb{H}$. From a 
physical point of view, we present a method to diagonalize any 
arbitrary two-fermion Hamiltonian. Relying on Jordan-Wigner
transformations for two-spin-$\frac{1}{2}$ and single-spin-$\frac{3}{2}$
systems, we also study nonlinear spin transformations and the 
related problem of diagonalizing arbitrary two-spin-$\frac{1}{2}$ 
and single-spin-$\frac{3}{2}$ Hamiltonians. Finally, from a 
calculational point of view, we pay due attention to explicit 
parametrizations of $SU(4)$ and $SO(6;\mathbb{R})$ matrices 
(of respective sizes $4\times 4$ and $6\times 6$) and their mutual relation.
\end{abstract}

\end{center}

\renewcommand{\thefootnote}{\arabic{footnote}}
\thispagestyle{empty}

\newpage

\tableofcontents

\newpage

\section{Introduction}

Like  canonical transformations in classical mechanics, unitary
transformations of quantum dynamical degrees of freedom often
simplify the dynamical equations, or allow to introduce sensible
approximation schemes. 
Such methods have wide-ranging applications,
from the study of simple systems to many-body problems in
solid-state or nuclear physics and quantum chemistry, up to
the infinite-dimensional systems of quantum field theory
\citep{blai1,wagn1,ring1,bish1}.
Linear (unitary) canonical transformations
(i.e., transformations preserving
the canonical anticommutation relations (CAR))
for fermions have been introduced by Bogolyubov and Valatin
(for two fermionic modes)
in connection with the study of the
mechanism of superconductivity
\citep{bogo1,bogo2,vala1,pine1,bogo9}.
These (linear) Bogolyubov-Valatin
transformations have been extended, initially by Bogolyubov and
his collaborators \citep{bogo3,bogo5}, \citep{bogo6a},
Appendix II, p.\ 123 (\citep{bogo6b}, p.\ 116, \citep{bogo6c}, p.\ 679),
to involve $n$
fermionic modes [so-called generalized linear Bogolyubov-Valatin
transformations, see, e.g., \citep{bogo11}, 
{\cyrrm \protect{ch}ast\cprime} III [chast' III]/[part III], p.\ 247
(\citep{bogo11b}, p.\ 341, \citep{bogo11c}, p.\ 259,
\citep{bogo14}, p.\ 235)]. Such linear canonical transformations
are important from a physical as well as from a mathematical
point of view. Mathematically, they allow to relate quite
arbitrary Hamiltonians quadratic in the fermion creation and
annihilation operators to collections of Fermi oscillators whose
mathematics is very well understood. From a physical point of view,
canonical transformations implement the concept of quasiparticles
in terms of which the physical processes taking place can
be described and understood in an effective and transparent manner.
To apply the powerful tool of canonical transformations
to the physically interesting class of non-quadratic Hamiltonians,
however, requires to go beyond linear Bogolyubov-Valatin transformations.
Certain aspects of nonlinear
Bogolyubov-Valatin transformations
have received some attention over time
\citep{solo1,solo2,haar1,brem1,ichi1}, \citep{schr1}, Sec.\ 2-6, p.\ 52,
\citep{varg1,kuze1,fuku2,fuku3,fuku4,
colp1,fuku1,nish1,nish2,zasl1,fuku5,bulg1,gunn1,suzu2,ostl1,holt1,nish3,
abe1,katr1,caba1,vanh1,caba2,ilie1,caba3,ostl4,nish5}
(We disregard here
work done within the framework of the coupled-cluster method (CCM)
\citep{bish1} which is nonunitary.).\footnote{Up to the referencing, 
this paragraph 
is identical to the introductory paragraph of our 
earlier article \citep{vanh1} on the one-mode case of nonlinear 
Bogolyubov-Valatin transformations.}
However, a systematic analytic study of general (nonlinear)
Bogolyubov-Valatin transformations had not been undertaken until the 
publication of our article \citep{vanh1}.\\

In \citep{vanh1} we have 
initiated a systematic analytic study of general (nonlinear) 
Bo\-go\-lyu\-bov-Valatin transformations by studying in detail 
the prototypical case of {\it one} fermion mode. The mathematics
of one fermion mode is closely relelated to the 
mathematics of the division algebra of quaternions $\mathbb{H}$.
In the present paper, we now investigate the most general (nonlinear) 
Bogolyubov-Valatin transformations for {\it two} fermion modes.
The original papers of Bogolyubov \citep{bogo1,bogo2}
and Valatin \citep{vala1} study the application of quasifermion
operators constructed linearly (with certain real coefficients $u$ and $v$)
from {\it two} fermionic modes in the most simple way. Our study
can, therefore, be considered as the widest possible (take this with 
a grain of salt) generalization
of the work of Bogolyubov and Valatin for two fermion modes. For example,
we will encounter the generalization(s) of the condition 
(\citep{bogo1}, p.\ 59, (p.\ 42 of the English transl.),
\citep{bogo2}, p.\ 796)
\begin{equation}
\label{origbv}
u^2\ +\ v^2\ =\ 1
\end{equation}
which is the well-known condition for the original (linear) 
Bogolyubov-Valatin transformation to be canonical (We display this condition
here symbolically without precisely defining its meaning.).\\

The study of nonlinear Bogolyubov-Valatin transformations requires 
the study of a new mathematical problem not considered previously --
nonlinear basis transformations in a Clifford algebra. This involves 
transformations among multivectors of different grades which can
be considered as a mathematical realization of the physical concept
of a supersymmetry. Furthermore, nonlinear Bogolyubov-Valatin 
transformations of two fermion modes have connections to various 
concrete physical and mathematical problems. Via Jordan-Wigner
transformations two-fermion systems are related to 
two-spin-$\frac{1}{2}$ and single-spin-$\frac{3}{2}$ systems which
have been studied from various points of view in the past and have received 
broad attention in the context of quantum information
and computation theory recently. Nonlinear Bogolyubov-Valatin transformations
correspond to nonlinear spin transformations in these systems.
As two-fermion systems are closely related to the Clifford algebra
$C(0,4)$ (and their cousins of different signature), the study of 
nonlinear Bogolyubov-Valatin transformations is closely connected
to Dirac matrices and the Dirac equation in general. Consequently, 
early studies of Dirac matrices and the Dirac equation performed by
Eddington feature prominently in the list of references accompanying
the article. And finally, via the relation of two-fermion systems
to $4\times 4$ matrices the study of nonlinear Bogolyubov-Valatin 
transformations has close connections to the representation and
parametrization of the group $SU(4)$, and its related orthogonal group
$SO(6;\mathbb{R})$ to which it is the double cover.\\

As in our earlier paper \citep{vanh1}, the main focus lies on 
structural and methodical aspects of the problem of 
nonlinear Bogolyubov-Valatin transformations. We start in 
Sec.\ \ref{prelim} with a discussion of certain algebraic
aspects of a (fixed) set of fermion creation and annihilation operators for
two fermion modes. In particular, we make contact with 
Clifford algebras (Subsec.\ \ref{prelimferm})
and the concept of paravectors (Subsec.\ \ref{structures}). 
In the main 
section \ref{NBV} of our article we first relate the concept of 
nonlinear Bogolyubov-Valatin transformations to the mathematical
concept of nonlinear basis transformations in a Clifford algebra
(Subsec.\ \ref{cliffform})
and then study those in mathematical detail for the Clifford
algebra $C(0,4)$ (Subsec.\ \ref{basis}). 
We find that nonlinear Bogolyubov-Valatin 
transformations have the structure of the group
$SO(6;\mathbb{R})/\mathbb{Z}_2$ [eq.\ (\ref{ntwofurtheradd78})]. 
In section \ref{diagham} we then apply the obtained insight to 
the problem of diagonalizing fermion (Subsec.\ \ref{twoferm})
and spin (Subsec.\ \ref{twospin}) Hamiltonians. As the whole 
discussion in the sections \ref{NBV} and \ref{diagham} is closely linked
to the intricate relationship between the groups $SO(6;\mathbb{R})$
and $SU(4)$ in section \ref{param} we study explicitly and in detail 
the relationship between their corresponding $6\times 6$ and 
$4\times 4$ representation matrices. Discussions of various aspects 
of the present work and their relation with the work of other authors
are collected in Sec.\ \ref{disc}. Short conclusions are 
presented in Sec.\ \ref{concl}. To facilitate the readability 
of the article, a number of technical details are deferred to six Appendices.  
The article is accompanied by a comprehensive list of references 
which may serve as a guide to the relevant literature.\\

\section{\label{prelim}Algebraic preliminaries}

\subsection{\label{prelimferm}Fermion creation and annihilation operators
and the associated Clifford algebra}

We consider two pairs of fermion creation and annihilation
operators $\hat{a}_k^+$, $\hat{a}_k$ ($k = 1,2$). We regard the
creation operators $\hat{a}_k^+$ as the hermitian conjugates of the
annihilation operators $\hat{a}_k$: $\hat{a}_k^+ = \hat{a}_k^\dagger$
(we will use the latter notation throughout). They are acting on
the vacuum state $\vert 0\rangle$ the following way
\begin{eqnarray}
\label{ntwoadd1a}
\hat{a}_k\vert 0\rangle&=& 0\ ,\\[0.3cm]
\label{ntwoadd1b}
\hat{a}_k^\dagger\vert 0\rangle&=&\vert 1\rangle_k\ .
\end{eqnarray}
The four-dimensional (complex) Hilbert space $\mathbb{C}_4$
(Sometimes, we will refer to it as spin space.)
of the two-fermion system
is spanned by the vacuum state $\vert 0\rangle$, 
the two one-particle states $\vert 1\rangle_k$,
and the two-particle state $\vert 2\rangle_{(2,1)}$
[be aware of the index convention in eq.\ (\ref{ntwoadd1d})]
\begin{eqnarray}
\label{ntwoadd1c}
\hat{a}_2^\dagger\hat{a}_1^\dagger\vert 0\rangle&=&\vert 2\rangle_{(2,1)}\ ,
\ =\ - \vert 2\rangle_{(1,2)}\ ,\\[0.3cm]
\label{ntwoadd1d}
\langle 0\vert\hat{a}_1\hat{a}_2&=&\langle 2\vert_{(2,1)}
\ =\ - \langle 2\vert_{(1,2)}\ .
\end{eqnarray}
The fermion creation and annihilation
operators obey the canonical anticommutation relations (CAR), 
($k,l = 1,2$; $\openone_4$ is the $4\times 4$ unit 
matrix)\footnote{As we will emphasize in this paper on many occasions
the concrete aspects of the problem under consideration we have decided 
to indicate the presence of unit operators explicitly.},
\begin{eqnarray}
\label{p1a}
\{\hat{a}_k^\dagger,\hat{a}_l\}&=&\hat{a}_k^\dagger\ \hat{a}_l\ +\
\hat{a}_k\ \hat{a}_l^\dagger\ =\ \delta_{kl}\ \openone_4, \\
\label{p1b}
\{\hat{a}_k,\hat{a}_l\}&=&\hat{a}_k\ \hat{a}_l\ +\
\hat{a}_k\ \hat{a}_l\ =\ 0, 
\end{eqnarray}
the latter equation entailing
\begin{eqnarray}
\label{p1c}
\{\hat{a}_k^\dagger,\hat{a}_l^\dagger\}
&=&\hat{a}_k^\dagger\ \hat{a}_l^\dagger\ +\
\hat{a}_k^\dagger\ \hat{a}_l^\dagger\ =\ 0\ .
\end{eqnarray}
It is now useful to consider the following
pairs of anti-hermitian operators.
\begin{eqnarray}
\label{p2a}
\hat{a}_k^{[1]}&=&- \hat{a}_k^{[1]\dagger}\ =
i\left(\hat{a}_k+\hat{a}_k^\dagger\right)\\
\label{p2b}
\hat{a}_k^{[2]}&=&- \hat{a}_k^{ [2]\dagger}\ =
\hat{a}_k-\hat{a}_k^\dagger
\end{eqnarray}
As a consequence of the CAR, eqs.\ (\ref{p1a})-(\ref{p1c}),
these operators obey the equation ($p,q = 1,2$)
\begin{eqnarray}
\label{p3}
\{\hat{a}_k^{[p]},\hat{a}_l^{[q]}\}&=&
-2\ \delta_{pq}\ \delta_{kl}\ \openone_4.
\end{eqnarray}
We introduce now the following useful notation:
\begin{eqnarray}
\label{ntwoadd30a} 
\hat{c}_{2k-1}&=&\hat{a}^{[1]}_k\ ,\ \ \ k\ =\ 1,
\ldots ,n\ (=2)\ ,\nonumber\\[0.3cm]
\label{ntwoadd30b} 
\hat{c}_{2k}&=&\hat{a}^{[2]}_k\ ,\ \ \ k\ =\ 1,\ldots ,n\ (=2)\ .
\end{eqnarray}
The anti-Hermitian operators $\hat{c}_k = - \hat{c}_k^\dagger$ 
obey the Clifford algebra relation [$k,l = 1,\ldots ,2n\ (=4)$]
\begin{eqnarray}
\label{ntwoadd31b}
\left\{\hat{c}_k,\hat{c}_l\right\}&=&2\ \hat{c}_k\cdot\hat{c}_l\ =\
2\ g_{kl}\ \openone_4\ =\ -2\ \delta_{kl}\ \openone_4\ , \\[0.3cm]
\label{ntwoadd31c}
g_{kl}&=&- \delta_{kl}\ ,
\end{eqnarray}
where $\hat{c}_k\cdot\hat{c}_l$ denotes the inner product in the Clifford 
algebra. The operators $\hat{c}_k$ generate 
the (real) Clifford algebra $C(0,4)$ which
is isomorphic to the twofold tensor product of the algebra 
of quaternions $\mathbb{H}$ (cf., e.g., \citep{port2}, Chap.\ 15, p.\ 123,
\citep{loun}, Chap.\ 16, p.\ 205).
To simplify the further calculations and to minimize sign errors,
in the following we will raise and lower indices by means of $\delta_{kl}$
and not by means of $g_{kl}$.\\

\subsection{\label{structures}Some useful algebraic structures in the 
Clifford algebra \texorpdfstring{$C(0,4)$}{C(0,4)}} 

Having related the two-mode pairs of 
fermion creation and annihilation operators to the Clifford algebra
$C(0,4)$ we will focus now our attention onto this algebraic
structure. In this subsection, we will discuss certain algebraic objects
in the Clifford algebra $C(0,4)$ that we will encounter in the
course of the further investigation. Their significance will become
clear in the further sections only. The somewhat unconventional 
index convention (including an index -1) to be used in the following
has been chosen with an eye to possible future generalizations of it to 
a larger number of fermionic modes.\\

\paragraph{Paravectors}\hfill\ \\
\nopagebreak
Define now 
\begin{eqnarray}
\label{ntwofurther63pa}
\hat{\hat{c}\hspace{0.5mm}}_k&=&\hat{c}_k\ ,\hspace{2cm} k = 1,\ldots , 4\ ,\\
\label{ntwofurther63pb}
\hat{\hat{c}\hspace{0.5mm}}_{(-1)}&=&i\hat{c}_1\hat{c}_2\hat{c}_3\hat{c}_4\ .
\end{eqnarray}
These operators obey the equation ($k,l = -1, 1,\ldots, 4$)
\begin{eqnarray}
\label{ntwofurther63pc}
\left\{\hat{\hat{c}\hspace{0.5mm}}_k,\hat{\hat{c}\hspace{0.5mm}}_l\right\}
&=&-2\ \delta_{kl}\ \openone_4.
\end{eqnarray}
They are generators of the (non-universal) Clifford algebra $C(0,5)$.
Going one step further we define
\begin{eqnarray}
\label{ntwofurther63pd}
\hat{\hat{\hat{c}\hspace{0.3mm}}}_k&=&\hat{\hat{c}\hspace{0.5mm}}_k
\ ,\hspace{2cm} k = -1, 1,\ldots , 4\ ,\\
\label{ntwofurther63pe}
\hat{\hat{\hat{c}\hspace{0.2mm}}}_0&=&-\openone_4\ .
\end{eqnarray}
These operators $\hat{\hat{\hat{c}\hspace{0.3mm}}}_k$ span the 
{\it paravector} space\footnote{For paravector space see, e.g.,  
\citep{port1}, Chap.\ 13, pp.\ 254-259, \citep{port2}, Chap.\ 16, p.\ 140, 
\citep{loun}, 1.\ and 2.\ ext.\ ed., Sec.\ 19.3, p.\ 247,
\citep{bayl1}. The term {\it paravector} has been introduced 
in \citep{maks2}, p.\ 22. In a context related to our study, 
paravectors have been used earlier 
in \citep{baru2}, p.\ 14, eq.\ (5.3), (in some
disguise) in \citep{lord2}, p.\ 340, eq.\ (1.2), and 
in \citep{lord4}, Sec.\ 2, p.\ 92; also note \citep{lord3}. 
For an early, related discussion see \citep{miln1}, in particular, p.\ 3,
eq.\ (5).}  $V_6$ associated with the Clifford algebra $C(0,5)$
the same way as the operators $\hat{c}_k$ span the 
vector space that is associated with the Clifford algebra $C(0,4)$.
It should be mentioned here that the 
choice in sign for $\hat{\hat{\hat{c}\hspace{0.3mm}}}_0 = -\openone_4$ 
is a matter of convenience. Our choice has 
been suggested by eq.\ (\ref{ntwofurtheradd65a}) further 
below\footnote{A different choice 
($\hat{\hat{\hat{c}\hspace{0.3mm}}}_0 = \openone_4$) 
would have entailed to modify eq.\ (\ref{ntwoadd32a})
accordingly.}.\\

As a consequence of eq.\ (\ref{ntwofurther63pc}) and the
anti-Hermiticity of the operators $\hat{\hat{c}\hspace{0.5mm}}_k$,
the operators $\hat{\hat{\hat{c}\hspace{0.3mm}}}_k$ [i.e.,
the basis elements of the paravector 
space associated with the Clifford algebra $C(0,5)$] 
obey the relation ($k,l=-1,\ldots , 4$)
\begin{eqnarray}
\label{ntwofurtheradd63x}
\hat{\hat{\hat{c}\hspace{0.3mm}}}^\dagger_k \hat{\hat{\hat{c}\,}}_l\ +\
\hat{\hat{\hat{c}\hspace{0.3mm}}}^\dagger_l \hat{\hat{\hat{c}\,}}_k\ =\ 
\hat{\hat{\hat{c}\,}}_k\ \hat{\hat{\hat{c}\hspace{0.3mm}}}^\dagger_l\ +\
\hat{\hat{\hat{c}\,}}_l\ \hat{\hat{\hat{c}\hspace{0.3mm}}}^\dagger_k
&=& 2\ \delta_{kl}\ \openone_4.
\end{eqnarray}
Note, that eq.\ (\ref{ntwofurtheradd63x}) has the form of the 
{\it unitary Hurwitz-Radon matrix problem}
(\citep{eckm1}, Subsec.\
1.4, p.\ 25, also see \citep{wong1}, part II, p.\ 67 and \citep{wong4}).
One can convince oneself that the relation
\begin{eqnarray}
\label{ntwofurtheradd63y}
\hat{\hat{\hat{c}\,}}_m&=&-\frac{i}{5!}\
\epsilon_m^{\ \ npqrs}\ \hat{\hat{\hat{c}\,}}_n\;
\hat{\hat{\hat{c}\hspace{0.3mm}}}^\dagger_p\; \hat{\hat{\hat{c}\,}}_q\;
\hat{\hat{\hat{c}\hspace{0.3mm}}}^\dagger_r\; \hat{\hat{\hat{c}\,}}_s
\end{eqnarray}
holds ($\epsilon_{mnpqrs}$
is the 6-dimensional totally antisymmetric tensor,
$\epsilon_{(-1)01234} = 1$;
we have raised here the indices
by means of $\delta_{kl}$ and not using $g_{kl}$),
and consequently also ($n\neq p\neq q\neq r\neq s$)
\begin{eqnarray}
\label{ntwofurtheradd63z}
\hat{\hat{\hat{c}\,}}_n\;
\hat{\hat{\hat{c}\hspace{0.3mm}}}^\dagger_p\; \hat{\hat{\hat{c}\,}}_q\;
\hat{\hat{\hat{c}\hspace{0.3mm}}}^\dagger_r\; \hat{\hat{\hat{c}\,}}_s
&=&-i\ \epsilon^{\ \ \ \ \ \ m}_{npqrs}\ \hat{\hat{\hat{c}\,}}_m
\ .
\end{eqnarray}

\paragraph{Biparavectors}\hfill\ \\
\nopagebreak
We define now the antisymmetric objects
\begin{eqnarray}
\label{ntwofurtheradd65aa}
\hat{\hat{\hat{c}\,}}_{m_1 m_2}&=&
- \hat{\hat{\hat{c}\,}}_{m_2 m_1}\ =\ \frac{1}{2}
\left(\hat{\hat{\hat{c}\hspace{0.3mm}}}^\dagger_{m_1}
\hat{\hat{\hat{c}\,}}_{m_2}
- \hat{\hat{\hat{c}\hspace{0.3mm}}}^\dagger_{m_2}\hat{\hat{\hat{c}\,}}_{m_1}
\right)\ =\ -\ \hat{\hat{\hat{c}\hspace{0.3mm}}}_{m_1 m_2}^\dagger\ ,
\end{eqnarray}
the decomposable/simple 
biparavectors of the paravector space $V_6$ associated with the Clifford
algebra $C(0,5)$ (cf.\ \citep{bayl1}, Sec.\ 3.4, p.\ 12). 
They span the 15-dimensional biparavector 
space $\bigwedge^2 (V_6)$. To simplify the notation we will often
write $\hat{\hat{\hat{c}\hspace{0.3mm}}}_M$ instead of
$\hat{\hat{\hat{c}\hspace{0.3mm}}}_{m_1 m_2}$ [$M = (m_1,m_2)$; 
if capital letters, e.g., $M$, are used for summations, $m_1 < m_2$ is 
understood].\\

It seems worth noting here that the set of biparavectors 
$\hat{\hat{\hat{c}\hspace{0.2mm}}}_M$ (supplemented by the unit operator
$\openone_4$) is equivalent to the $E$-numbers 
introduced by Eddington \citep{eddi1}\footnote{Also see 
\citep{eddi2,eddi3,eddi4};
a systematic presentation is given 
in \citep{eddi5}, Part I, Chaps.\ 2-4, pp.\ 20-61, \citep{eddi6}, 
Chaps. VI/VII, pp.\ 106-158, and
a discussion of Eddington's $E$-numbers from a modern perspective
can be found in \citep{sali1}.}. It has been recognized from early
on\footnote{\citep{scho3,iwan1,litt1,kwal1,lema2}, 
\citep{eddi5}, \S 3.8, p.\ 47,
\citep{mccr1}, \citep{eddi6}, \S 60, p.\ 120, \citep{miln2,depr1}.} 
that the $E$-numbers of Eddington form 
an algebraic system\footnote{One can convince oneself that the 
biparavector space $\bigwedge^2 (V_6)$ supplemented by the unit operator
$\openone_4$ spans the operator space of the Clifford algebra $C(0,4)$.
Initially, biparavectors (supplemented by the unit operator)  have been named 
`{\it sedenions}' (\citep{scho3}, p.\ 105, footnote 3)
or `{\it quadriquaternions}' \citep{comb1}. Note, that these sedenions are
different from the weird algebraic objects presently called 
'sedenions' and obtained from the octonions in the following step of 
the Cayley-Dickson process.}
which can be obtained by taking
the tensor product of two (commuting) copies, 
${\bf I}_1$, ${\bf J}_1$, ${\bf K}_1$ 
(${\bf I}_1 {\bf J}_1 = {\bf K}_1$, etc.)
and ${\bf I}_2$, ${\bf J}_2$, ${\bf K}_2$, 
(${\bf I}_2 {\bf J}_2 = {\bf K}_2$, etc.),
of the system of 
quaternions $\mathbb{H}$
(cf.\ also the comment at the end of Subsec.\ \ref{prelimferm}). 
To be more specific, one possible 
(symmetric) choice for 
the generators of the Clifford algebra $C(0,4)$ is (for a different
choice see Appendix \ref{appc})
\begin{eqnarray}
\label{ntwofurtheraddz66a}
\hat{\hat{\hat{c}\,}}_1&=&\hat{c}_1\ =\ \frac{1}{\sqrt{2}}
\left[\openone_4 - i\ {\bf K}_1{\bf K}_2\right]\ {\bf I}_1\ =\
\frac{1}{\sqrt{2}}\left[{\bf I}_1 - i\ {\bf J}_1{\bf K}_2\right]\ ,\\[0.3cm]
\label{ntwofurtheraddz66b}
\hat{\hat{\hat{c}\,}}_2&=&\hat{c}_2\ =\ \frac{1}{\sqrt{2}}
\left[\openone_4 - i\ {\bf K}_1{\bf K}_2\right]\ {\bf J}_1\ =\
\frac{1}{\sqrt{2}}\left[{\bf J}_1 + i\ {\bf I}_1{\bf K}_2\right]\ ,\\[0.3cm]
\label{ntwofurtheraddz66c}
\hat{\hat{\hat{c}\,}}_3&=&\hat{c}_3\ =\ \frac{1}{\sqrt{2}}
\left[\openone_4 + i\ {\bf K}_1{\bf K}_2\right]\ {\bf I}_2\ =\
\frac{1}{\sqrt{2}}\left[{\bf I}_2 + i\ {\bf K}_1{\bf J}_2\right]\ ,\\[0.3cm]
\label{ntwofurtheraddz66d}
\hat{\hat{\hat{c}\,}}_4&=&\hat{c}_4\ =\ \frac{1}{\sqrt{2}}
\left[\openone_4 + i\ {\bf K}_1{\bf K}_2\right]\ {\bf J}_2\ =\
\frac{1}{\sqrt{2}}\left[{\bf J}_2 - i\ {\bf K}_1{\bf I}_2\right]\ ,
\end{eqnarray}
and, consequently,
\begin{eqnarray}
\label{ntwofurtheraddz66e}
\hat{\hat{\hat{c}\,}}_{(-1)}&=&i\ {\bf K}_1{\bf K}_2\ .
\end{eqnarray}
Inversely, the two commuting copies of the system of quaternions can be
given in terms of the generators of the Clifford algebra $C(0,4)$ by
means of the following equations.
\begin{eqnarray}
\label{ntwofurtheraddz67a}
{\bf I}_1&=&\frac{1}{\sqrt{2}}
\left[\openone_4 + \hat{\hat{\hat{c}\,}}_{(-1)}\right]\ \hat{\hat{\hat{c}\,}}_1
\ =\ - \frac{1}{\sqrt{2}}
\left[\hat{\hat{\hat{c}\,}}_{01} + \hat{\hat{\hat{c}\,}}_{(-1)1}\right]
\ =\ \frac{1}{\sqrt{2}}
\left[\hat{c}_1 + i\ \hat{c}_2 \hat{c}_3 \hat{c}_4\right]\ \ \ \\[0.3cm]
\label{ntwofurtheraddz67b}
{\bf J}_1&=&\frac{1}{\sqrt{2}}
\left[\openone_4 + \hat{\hat{\hat{c}\,}}_{(-1)}\right]\ \hat{\hat{\hat{c}\,}}_2
\ =\ - \frac{1}{\sqrt{2}}
\left[\hat{\hat{\hat{c}\,}}_{02} + \hat{\hat{\hat{c}\,}}_{(-1)2}\right]
\ =\ \frac{1}{\sqrt{2}}
\left[\hat{c}_2 - i\ \hat{c}_1 \hat{c}_3 \hat{c}_4\right]\ \ \ \\[0.3cm]
\label{ntwofurtheraddz67c}
{\bf K}_1&=&\hat{\hat{\hat{c}\,}}_1 \hat{\hat{\hat{c}\,}}_2 
\ =\ -\ \hat{\hat{\hat{c}\,}}_{12}\ =\ \hat{c}_1 \hat{c}_2\\[0.3cm]
\label{ntwofurtheraddz67d}
{\bf I}_2&=&\frac{1}{\sqrt{2}}
\left[\openone_4 - \hat{\hat{\hat{c}\,}}_{(-1)}\right]\ \hat{\hat{\hat{c}\,}}_3
\ =\ - \frac{1}{\sqrt{2}}
\left[\hat{\hat{\hat{c}\,}}_{03} - \hat{\hat{\hat{c}\,}}_{(-1)3}\right]
\ =\ \frac{1}{\sqrt{2}}
\left[\hat{c}_3 - i\ \hat{c}_1 \hat{c}_2 \hat{c}_4\right]\ \ \ \\[0.3cm]
\label{ntwofurtheraddz67e}
{\bf J}_2&=&\frac{1}{\sqrt{2}}
\left[\openone_4 - \hat{\hat{\hat{c}\,}}_{(-1)}\right]\ \hat{\hat{\hat{c}\,}}_4
\ =\ - \frac{1}{\sqrt{2}}
\left[\hat{\hat{\hat{c}\,}}_{04} - \hat{\hat{\hat{c}\,}}_{(-1)4}\right] 
\ =\ \frac{1}{\sqrt{2}}
\left[\hat{c}_4 + i\ \hat{c}_1 \hat{c}_2 \hat{c}_3\right]\ \ \ \\[0.3cm]
\label{ntwofurtheraddz67f}
{\bf K}_2&=&\hat{\hat{\hat{c}\,}}_3 \hat{\hat{\hat{c}\,}}_4
\ =\ -\ \hat{\hat{\hat{c}\,}}_{34}\ =\ \hat{c}_3 \hat{c}_4
\end{eqnarray}
As the system of quaternions ${\bf I}$, ${\bf J}$, ${\bf K}$ is related to the 
group SU(2) and, therefore, to spin variables $S^x$, $S^y$, $S^z$
($S^x = i {\bf I}/2$, $S^y = i {\bf J}/2$, $S^z = i {\bf K}/2$), a 
quadriquaternion representation of the biparavectors 
$\hat{\hat{\hat{c}\,}}_M$ is related to a 
two-spin-$\frac{1}{2}$ system\footnote{In fact, 
the equations (\ref{ntwofurtheraddz66a})-(\ref{ntwofurtheraddz67f})
are a modified version of a Jordan-Wigner 
transformation used in \citep{sire1}, p.\ 13835, above of eq.\ (5).}.
For example, this fact is relied on in the product operator
formalism widely used in the field of nuclear magnetic resonance (NMR) 
\citep{sore1,ven1,pack1}. Two-spin-$\frac{1}{2}$ systems
will be discussed in Subsection \ref{twospinone}. Another approach relating
quaternions to Eddington's $E$-numbers has been put forward
by Conway \citep{conw1} (also see \citep{syng1}, Sec.\ 9, p.\ 48).\\

\paragraph{Algebra of biparavectors}\hfill\ \\
\nopagebreak
Using the notation introduced in eq.\ (\ref{ntwofurtheradd65aa}),
the eqs.\ (\ref{ntwofurtheradd63y}), (\ref{ntwofurtheradd63z})
can be transformed to 
read\footnote{Related equations can be found in:\\[-0.7cm]
\begin{tabbing}
\citep{hris3}, p.\ 203, eq.\ (2),\hskip2.9cm \=
\citep{hris1}\ p.\ 756, eq.\ (8),\hskip2.cm \=
\citep{hris4}, p.\ 27, eq.\ (14),\\
\citep{buch1}, p.\ 359, eq.\ (2.8),
\>\citep{hris10}, p.\ 118, eq.\ (5), and 
\>\citep{nash1}, p.\ 205, eq.\ (9).
\end{tabbing} }
\begin{eqnarray}
\label{ntwofurtheradd63ya}
\hat{\hat{\hat{c}\hspace{0.3mm}}}_M&=&\frac{i}{6}\
\epsilon_M^{\ \ PQ}\ 
\hat{\hat{\hat{c}\hspace{0.3mm}}}_P\;\hat{\hat{\hat{c}\hspace{0.3mm}}}_Q
\end{eqnarray}
and
\begin{eqnarray}
\label{ntwofurtheradd63za}
\hat{\hat{\hat{c}\hspace{0.3mm}}}_P\;\hat{\hat{\hat{c}\hspace{0.3mm}}}_Q&=&
-\ \delta_{PQ}\ \openone_4\
\ -\ \delta_{p_1 q_1}\ \hat{\hat{\hat{c}\hspace{0.3mm}}}_{(p_2 q_2)}
\ -\ \delta_{p_2 q_2}\ \hat{\hat{\hat{c}\hspace{0.3mm}}}_{(p_1 q_1)}
\nonumber\\[0.3cm]
&&\ \ +\ \delta_{p_1 q_2}\ \delta_{p_2 q_1}\ \openone_4\
\ +\ \delta_{p_1 q_2}\ \hat{\hat{\hat{c}\hspace{0.3mm}}}_{(p_2 q_1)}
\ +\ \delta_{p_2 q_1}\ \hat{\hat{\hat{c}\hspace{0.3mm}}}_{(p_1 q_2)}
\nonumber\\[0.3cm]
&&\ \ -\ i\ \epsilon_{PQ}^{\ \ \ \ N}\ \hat{\hat{\hat{c}\hspace{0.3mm}}}_N\ .
\end{eqnarray}
In the last equation we have partially lifted
the condition $p_1\neq p_2\neq q_1\neq q_2$, it still applies
$p_1\neq p_2$, $q_1\neq q_2$.
Equation (\ref{ntwofurtheradd63za}) can be found in a number 
of studies\footnote{
\citep{scho12}, p.\ 656, \S 8, eqs.\ (8.6), (8.7) (here, 
a five-dimensional formulation is still being used), 

\noindent
\citep{hari1}, p.\ 32 (reprint \# 1: p.\ 50, reprint \# 2: p.\ 943), eq.\ (9),
\\[-0.7cm]
\begin{tabbing}
\citep{hris3}, p.\ 203, eq.\ (2),\hskip2.9cm \=
\citep{hris1}, p.\ 756, eq.\ (7),\hskip2.cm \=
\citep{hris4}, p.\ 27, eq.\ (12),\\
\citep{popo5}, p.\ 3121, eq.\ (F),
\>\citep{macf1}, p.\ 136, eqs.\ (2.4), (2.10)
\>\citep{fedo3}, p.\ 127, eqs.\ (2), (3),\\
\citep{buch1}, p.\ 359, eq.\ (2.6), 
\>\citep{tenk1}, p.\ 183, eq.\ (8a),
\>\citep{hris10}, p.\ 118, eq.\ (4),\\
\citep{buch2}, p.\ 140, eq.\ (2.15),
\>\citep{klot1}, p.\ 2244, eq.\ (V),
\>\citep{nash1}, p.\ 204, eq.\ (1),\\
\citep{tsel1}, Appendix C, p.\ 105021-19, eq.\ (C.13).
\end{tabbing} }.
The commutator $[\hat{\hat{\hat{c}\,}}_P,\hat{\hat{\hat{c}\,}}_Q]$
reads\footnote{The commutator can be found in the literature in:\\[-0.7cm]
\begin{tabbing}
\citep{vala2} p.\ 97, eq.\ ($1^a$),\hskip3cm \=
\citep{hris1}, p.\ 756, eq.\ (7),\hskip2.1cm \=
\citep{hris4}, p. 27, eq.\ (12),\\
\citep{hepn1}, p.\ 354, eq.\ (2), 
\>\citep{baru1}, p.\ B841, eq.\ (3.2),
\>\citep{macf1}, p.\ 136, eq.\ (2.11),\\
\citep{dewe1}, Subsec.\ 2d, p.\ 21,
\>\citep{buch1}, p.\ 360, eq.\ (3.2),
\>\citep{tenk1}, p.\ 182, eq.\ (6),
\end{tabbing}
\ \\[-0.7cm]
\citep{kono1}, p.\ 368 (p.\ 1212 of the English transl.), below from eq.\ (7),
\\[-0.7cm]
\begin{tabbing}
\citep{hris10}, p.\ 118, eq.\ (3),\hskip2.9cm \= 
\citep{klot1}, p.\ 2244, eq.\ (III),\hskip1.7cm \= 
\citep{nash1}, p.\ 204, eq.\ (3),\\
\citep{lauf1}, Sec.\ 5.3, p.\ 168,
\>\citep{tsel1}, Appendix C, p.\ 105021-19, eq.\ (C.12).
\end{tabbing} }:
\begin{eqnarray}
\label{ntwofurtheradd63zb}
[\hat{\hat{\hat{c}\,}}_P,\hat{\hat{\hat{c}\,}}_Q]\ =\
\hat{\hat{\hat{c}\,}}_P\;\hat{\hat{\hat{c}\,}}_Q\ -\ 
\hat{\hat{\hat{c}\,}}_Q\;\hat{\hat{\hat{c}\,}}_P
&=&-\ 2\ \delta_{p_1 q_1}\ \hat{\hat{\hat{c}\,}}_{(p_2 q_2)}
\ -\ 2\ \delta_{p_2 q_2}\ \hat{\hat{\hat{c}\hspace{0.2mm}}}_{(p_1 q_1)}
\nonumber\\[0.3cm]
&&\ \ +\ 2\ \delta_{p_1 q_2}\ \hat{\hat{\hat{c}\hspace{0.3mm}}}_{(p_2 q_1)}
\ +\ 2\ \delta_{p_2 q_1}\ \hat{\hat{\hat{c}\hspace{0.4mm}}}_{(p_1 q_2)}\ ,
\end{eqnarray}
while the anticommutator
$\{\hat{\hat{\hat{c}\,}}_P,\hat{\hat{\hat{c}\,}}_Q\}$ is given
by\footnote{The anticommutator can be found in the literature in:

\noindent
\citep{hari1}, p.\ 33 (reprint \# 1: p.\ 50, 
reprint \# 2: p.\ 944), eq.\ (10a),\\[-0.7cm]
\begin{tabbing}
\citep{hris1}, p.\ 756, eq.\ (7),\hskip2.9cm \=
\citep{hris4}, p. 27, eq.\ (12),\hskip2.1cm \=
\citep{buch1}, p.\ 360, eq.\ (3.3),\\
\citep{tenk1}, p.\ 182, eq.\ (7),
\>\citep{hris10}, p.\ 122, eq.\ (40),
\>\citep{lord2}, p.\ 341, eq.\ (1.11).
\end{tabbing} }:
\begin{eqnarray}
\label{ntwofurtheradd63zc}
\{\hat{\hat{\hat{c}\,}}_P,\hat{\hat{\hat{c}\,}}_Q\}&=&
\hat{\hat{\hat{c}\,}}_P\;\hat{\hat{\hat{c}\,}}_Q\ +\ 
\hat{\hat{\hat{c}\,}}_Q\;\hat{\hat{\hat{c}\,}}_P\nonumber\\[0.3cm]
&=&\left(-\ 2\ \delta_{PQ}\ 
\ +\ 2\ \delta_{p_1 q_2}\ \delta_{p_2 q_1}\right)\ \openone_4\
-\ 2 i\ \epsilon_{PQ}^{\ \ \ \ N}\ \hat{\hat{\hat{c}\,}}_N\ .
\end{eqnarray}
Equation (\ref{ntwofurtheradd63zb}) displays the structure of 
the Lie algebra of $\mathsf{su}(4) \simeq \mathsf{so}(6)$. Note that 
this Lie algebra (i.e., its generators) is realized here in terms of 
products of one to four fermion creation and annihilation operators.
This representation of the Lie algebra
$\mathsf{su}(4) \simeq \mathsf{so}(6)$ has been discussed earlier in
\citep{pale1,grub3} (In general, for $n$ fermionic modes
the Lie algebra generated by all possible products of fermion 
creation and annihilation operators is $\mathsf{su}(2^n)$ \citep{judd1},
Sec.\ 5.2, p.\ 21, \citep{pale2}.). 
Most discussions, in the literature, of Lie algebras 
in terms of fermion creation and annihilation operators
rely on their bilinear combinations only (see, e.g., 
\citep{dora1}, \citep{frap1}, Subsecs.\ 1.59, 1.60, pp.\ 88-93) or,
at best, supplement these combinations by the fermion 
creation and annihilation operators themselves 
\citep{wybo1,fuku2,fuku1}, \citep{bose1}, App.\ 1, p.\ 919 (p.\ 548
of the reprint), \citep{bulg1,buza1}, \citep{zhan3}, p.\ 907,
\citep{vanh1}. Cases that go beyond this linear/bilinear 
framework have been discussed in \citep{judd1},
Sec.\ 5.2, p.\ 21, \citep{pale1,pale2,grub3,kell1,zhan2}.
Finally, we mention that certain graphical representations 
of the commutator/anticommutator relations 
(\ref{ntwofurtheradd63zb}), (\ref{ntwofurtheradd63zc})
have appeared in the literature \citep{good2,sani1,plan1,plan2,sani2,rau4}. 
To relate the works of Saniga and 
collaborators \citep{sani1,plan1,plan2,sani2} 
and Rau \citep{rau4} to the equations 
(\ref{ntwofurtheradd63zb}), (\ref{ntwofurtheradd63zc}) one must
rely on the representation of the bipa\-ra\-vectors 
$\hat{\hat{\hat{c}\hspace{0.3mm}}}_M$
in terms of the (twofold) tensor product of quaternions $\mathbb{H}$ discussed 
at the end of Appendix \ref{appc}.\\

\section{\label{NBV}Nonlinear Bogolyubov-Valatin transformations}

\subsection{\label{cliffform}Clifford algebra formulation 
of canonical fermion transformations}
\subsubsection{The ansatz for nonlinear Bogolyubov-Valatin transformations}

We proceed now to writing down an ansatz for the most general
Bogolyubov-Valatin transformation for two fermionic modes.
In view of the eqs.\ (\ref{p1a}), (\ref{p1b}), the new 
fermion annihilation and creation operators $\hat{b}_k$, 
$\hat{b}_k^\dagger$ read
\begin{eqnarray}
\label{ntwo3a}
\hat{b}_k&=&{\sf B}_k\left(\{\lambda\};\{\hat{a}\}\right)\ =\ 
{\sf U}\,\hat{a}_k\,{\sf U}^\dagger\nonumber\\[0.3cm]
&=&\lambda_k^{(0\vert 0)}\ \openone_4\ +\
\lambda_k^{(1\vert 0)}\ \hat{a}_1^\dagger\ +\
\lambda_k^{(2\vert 0)}\ \hat{a}_2^\dagger\ +\
\lambda_k^{(0\vert 1)}\ \hat{a}_1\ +\
\lambda_k^{(0\vert 2)}\ \hat{a}_2\nonumber\\[0.3cm]
&&\ +\ \lambda_k^{(1,2\vert 0)}\ \hat{a}_1^\dagger\hat{a}_2^\dagger\ +\
\lambda_k^{(1\vert 1)}\ \hat{a}_1^\dagger\hat{a}_1\ +\
\lambda_k^{(1\vert 2)}\ \hat{a}_1^\dagger\hat{a}_2\nonumber\\[0.3cm]
&&\ +\ \lambda_k^{(2\vert 1)}\ \hat{a}_2^\dagger\hat{a}_1
\ +\ \lambda_k^{(2\vert 2)}\ \hat{a}_2^\dagger\hat{a}_2
\ +\ \lambda_k^{(0\vert 1,2)}\ \hat{a}_1\hat{a}_2\nonumber\\[0.3cm]
&&\ +\ \lambda_k^{(1,2\vert 1)}\ \hat{a}_1^\dagger\hat{a}_2^\dagger\hat{a}_1
\ +\ \lambda_k^{(1,2\vert 2)}\ \hat{a}_1^\dagger\hat{a}_2^\dagger\hat{a}_2
\nonumber\\[0.3cm]
&&
\ +\ \lambda_k^{(1\vert 1,2)}\ \hat{a}_1^\dagger\hat{a}_1\hat{a}_2
\ +\ \lambda_k^{(2\vert 1,2)}\ \hat{a}_2^\dagger\hat{a}_1\hat{a}_2
\nonumber\\[0.3cm]
&&\ +\ \lambda_k^{(1,2\vert 1,2)}\ 
\hat{a}_1^\dagger\hat{a}_2^\dagger\hat{a}_1\hat{a}_2\ ,\\[0.3cm]
\label{ntwo3b}
\hat{b}_k^\dagger&=&{\sf B}_k\left(\{\lambda\};\{\hat{a}\}\right)^\dagger
\ =\ {\sf U}\,\hat{a}_k^\dagger\, {\sf U}^\dagger\nonumber\\[0.3cm]
&=&\overline{\lambda_k^{(0\vert 0)}}\ \openone_4\ +\
\overline{\lambda_k^{(1\vert 0)}}\ \hat{a}_1\ +\
\overline{\lambda_k^{(2\vert 0)}}\ \hat{a}_2\ +\
\overline{\lambda_k^{(0\vert 1)}}\ \hat{a}_1^\dagger\ +\
\overline{\lambda_k^{(0\vert 2)}}\ \hat{a}_2^\dagger
\nonumber\\[0.3cm]
&&\ -\ \overline{\lambda_k^{(1,2\vert 0)}}\ 
\hat{a}_1\hat{a}_2\ +\
\overline{\lambda_k^{(1\vert 1)}}\ \hat{a}_1^\dagger\hat{a}_1
\ +\ \overline{\lambda_k^{(1\vert 2)}}\ \hat{a}_2^\dagger\hat{a}_1
\nonumber\\[0.3cm]
&&\ +\ \overline{\lambda_k^{(2\vert 1)}}\ 
\hat{a}_1^\dagger\hat{a}_2
\ +\ \overline{\lambda_k^{(2\vert 2)}}\ \hat{a}_2^\dagger\hat{a}_2
\ -\ \overline{\lambda_k^{(0\vert 1,2)}}\ \hat{a}_1^\dagger\hat{a}_2^\dagger
\nonumber\\[0.3cm]
&&\ -\ \overline{\lambda_k^{(1,2\vert 1)}}\ 
\hat{a}_1^\dagger\hat{a}_1\hat{a}_2
\ -\ \overline{\lambda_k^{(1,2\vert 2)}}\ 
\hat{a}_2^\dagger\hat{a}_1\hat{a}_2
\nonumber\\[0.3cm]
&&\ -\ \overline{\lambda_k^{(1\vert 1,2)}}\ 
\hat{a}_1^\dagger\hat{a}_2^\dagger\hat{a}_1
\ -\ \overline{\lambda_k^{(2\vert 1,2)}}\ 
\hat{a}_1^\dagger\hat{a}_2^\dagger\hat{a}_2
\nonumber\\[0.3cm]
&&\ +\ \overline{\lambda_k^{(1,2\vert 1,2)}}\ 
\hat{a}_1^\dagger\hat{a}_2^\dagger\hat{a}_1\hat{a}_2\ .
\end{eqnarray}
We assume the coefficients to be complex numbers:
$\lambda^{(\ldots)}\in\mathbb{C}$.
$\{\lambda\}$ denotes the set of all 16 coefficients 
$\lambda^{(\ldots)}$, and $\{\hat{a}\}$ the set of the creation and
annihilation operators.
${\sf U} = {\sf U}\left(\{\lambda\};\{\hat{a}\}\right)$ is an 
unitary operator belonging to the group $SU(4)$ that implements the 
nonlinear Bogolyubov-Valatin transformation.
The new annihilation operators $\hat{b}_k$ act on the new vacuum state
\begin{eqnarray}
\label{ntwoadd4a}
\vert 0^\prime\rangle&=&\vert 0,\{\lambda\}\rangle\ =\ 
{\sf U}\left(\{\lambda\};\{\hat{a}\}\right)\vert 0\rangle
\end{eqnarray}
the usual way
\begin{eqnarray}
\label{ntwoadd4b}
\hat{b}_k\vert 0^\prime\rangle&=& 0\ .
\end{eqnarray}
The new fermion creation and annihilation
operators also must obey the canonical anticommutation 
relations [CAR, cf.\ eqs.\ (\ref{p1a}), (\ref{p1b})], ($k,l = 1,2$),
\begin{eqnarray}
\label{p1ba}
\{\hat{b}_k^\dagger,\hat{b}_l\}&=&\hat{b}_k^\dagger\ \hat{b}_l\ +\
\hat{b}_k\ \hat{b}_l^\dagger\ =\ \delta_{kl}\ \openone_4\ , \\
\label{p1bb}
\{\hat{b}_k,\hat{b}_l\}&=&\hat{b}_k\ \hat{b}_l\ +\
\hat{b}_k\ \hat{b}_l\ =\ 0, 
\end{eqnarray}
the latter equation entailing
\begin{eqnarray}
\label{p1bc}
\{\hat{b}_k^\dagger,\hat{b}_l^\dagger\}
&=&\hat{b}_k^\dagger\ \hat{b}_l^\dagger\ +\
\hat{b}_k^\dagger\ \hat{b}_l^\dagger\ =\ 0\ .
\end{eqnarray}

Before proceeding further, it is useful to consider the expression 
for the trace of any annihilation (or creation) operator.
\begin{eqnarray}
\label{ntwoadd6}
&&\hspace{-1.5cm}{\rm tr}\;\hat{b}_k\nonumber\\[0.3cm]
&=&\langle 0^\prime\vert \hat{b}_k\vert 0^\prime\rangle
\ +\ \langle 1^\prime\vert_1\; \hat{b}_k\vert 1^\prime\rangle_1
\ +\ \langle 1^\prime\vert_2\; \hat{b}_k\vert 1^\prime\rangle_2
\ +\ \langle 2^\prime\vert_{(1,2)}\; \hat{b}_k\vert 2^\prime\rangle_{(1,2)}
\ =\ 0 
\end{eqnarray}
The trace has to vanish.
From eq.\ (\ref{ntwo3a}),  for eq.\ (\ref{ntwoadd6}) it then follows:
\begin{eqnarray}
\label{ntwoadd7}
{\rm tr}\;\hat{b}_k&=&
4 \lambda^{(0\vert 0)}_k\ +\ 2 \lambda^{(1\vert 1)}_k
\ +\ 2 \lambda^{(2\vert 2)}_k\ -\ \lambda^{(1,2\vert 1,2)}_k\ =\ 0
\end{eqnarray}
(In the one-mode case studied earlier \citep{vanh1}, 
the corresponding equation is eq.\ (13), p.\ 10247.).\\

To facilitate the further discussion, we now introduce 
the antihermitian operators
\begin{eqnarray}
\label{ntwoadd13a}
\hat{b}_k^{[1]}&=&- \hat{b}_k^{[1]\dagger}\ =
i\left(\hat{b}_k+\hat{b}_k^\dagger\right)\nonumber\\[0.3cm]
&=&{\rm Re}\;\kappa^{(1\vert 0)}_k\ \hat{a}_1^{[1]}\ +\ 
{\rm Re}\;\kappa^{(0\vert 1)}_k\ \hat{a}_1^{[2]}\ +\ 
{\rm Re}\;\kappa^{(2\vert 0)}_k\ \hat{a}_2^{[1]}\ +\
{\rm Re}\;\kappa^{(0\vert 2)}_k\ \hat{a}_2^{[2]}\nonumber\\[0.3cm]
&&\ +\ {\rm Re}\;\kappa^{(1\vert 1)}_k\ \ \hat{a}_1^{[1]}\hat{a}_1^{[2]} 
\ +\ {\rm Re}\;\kappa^{(1,2\vert 0)}_k\ \hat{a}_1^{[1]}\hat{a}_2^{[1]}
\ +\ {\rm Re}\;\kappa^{(1\vert 2)}_k\ \hat{a}_1^{[1]}\hat{a}_2^{[2]}
\nonumber\\[0.3cm]
&&\ +\ {\rm Re}\;\kappa^{(2\vert 1)}_k\ \hat{a}_1^{[2]}\hat{a}_2^{[1]}
\ +\ {\rm Re}\;\kappa^{(0\vert 1,2)}_k\ \hat{a}_1^{[2]}\hat{a}_2^{[2]}
\ +\ {\rm Re}\;\kappa^{(2\vert 2)}_k\ \ \hat{a}_2^{[1]}\hat{a}_2^{[2]} 
\nonumber\\[0.3cm]
&&\ +\ {\rm Re}\;\kappa^{(1,2\vert 1)}_k\
i\hat{a}_1^{[1]}\hat{a}_1^{[2]}\hat{a}_2^{[1]}
\ +\ {\rm Re}\;\kappa^{(1\vert 1,2)}_k\
i\hat{a}_1^{[1]}\hat{a}_1^{[2]}\hat{a}_2^{[2]}\nonumber\\[0.3cm]
&&\ +\ {\rm Re}\;\kappa^{(1,2\vert 2)}_k\
i\hat{a}_1^{[1]}\hat{a}_2^{[1]}\hat{a}_2^{[2]}
\ +\ {\rm Re}\;\kappa^{(2\vert 1,2)}_k\
i\hat{a}_1^{[2]}\hat{a}_2^{[1]}\hat{a}_2^{[2]}\nonumber\\[0.3cm]
&&\ +\ {\rm Re}\;\kappa^{(1,2\vert 1,2)}_k\
i\hat{a}_1^{[1]}\hat{a}_1^{[2]}\hat{a}_2^{[1]}\hat{a}_2^{[2]}
\end{eqnarray}
and
\begin{eqnarray}
\label{ntwoadd13b}
\hat{b}_k^{[2]}&=&- \hat{b}_k^{[2]\dagger}\ =
\hat{b}_k-\hat{b}_k^\dagger\nonumber\\[0.3cm]
&=&{\rm Im}\;\kappa^{(1\vert 0)}_k\ \hat{a}_1^{[1]}\ +\ 
{\rm Im}\;\kappa^{(0\vert 1)}_k\ \hat{a}_1^{[2]}\ +\ 
{\rm Im}\;\kappa^{(2\vert 0)}_k\ \hat{a}_2^{[1]}\ +\
{\rm Im}\;\kappa^{(0\vert 2)}_k\ \hat{a}_2^{[2]}\nonumber\\[0.3cm]
&&\ +\ {\rm Im}\;\kappa^{(1\vert 1)}_k\ \ \hat{a}_1^{[1]}\hat{a}_1^{[2]} 
\ +\ {\rm Im}\;\kappa^{(1,2\vert 0)}_k\ \hat{a}_1^{[1]}\hat{a}_2^{[1]}
\ +\ {\rm Im}\;\kappa^{(1\vert 2)}_k\ \hat{a}_1^{[1]}\hat{a}_2^{[2]}
\nonumber\\[0.3cm]
&&\ +\ {\rm Im}\;\kappa^{(2\vert 1)}_k\ \hat{a}_1^{[2]}\hat{a}_2^{[1]}
\ +\ {\rm Im}\;\kappa^{(0\vert 1,2)}_k\ \hat{a}_1^{[2]}\hat{a}_2^{[2]}
\ +\ {\rm Im}\;\kappa^{(2\vert 2)}_k\ \ \hat{a}_2^{[1]}\hat{a}_2^{[2]} 
\nonumber\\[0.3cm]
&&\ +\ {\rm Im}\;\kappa^{(1,2\vert 1)}_k\
i\hat{a}_1^{[1]}\hat{a}_1^{[2]}\hat{a}_2^{[1]}
\ +\ {\rm Im}\;\kappa^{(1\vert 1,2)}_k\
i\hat{a}_1^{[1]}\hat{a}_1^{[2]}\hat{a}_2^{[2]}\nonumber\\[0.3cm]
&&\ +\ {\rm Im}\;\kappa^{(1,2\vert 2)}_k\
i\hat{a}_1^{[1]}\hat{a}_2^{[1]}\hat{a}_2^{[2]}
\ +\ {\rm Im}\;\kappa^{(2\vert 1,2)}_k\
i\hat{a}_1^{[2]}\hat{a}_2^{[1]}\hat{a}_2^{[2]}\nonumber\\[0.3cm]
&&\ +\ {\rm Im}\;\kappa^{(1,2\vert 1,2)}_k\
i\hat{a}_1^{[1]}\hat{a}_1^{[2]}\hat{a}_2^{[1]}\hat{a}_2^{[2]}\ .
\end{eqnarray}
With some hindsight, the primary ordering
of the operators on the r.h.s.\ is being done 
in accordance with the mode number.
In deriving the eqs.\ (\ref{ntwoadd13a}), (\ref{ntwoadd13b}) we
have taken into account eq.\ (\ref{ntwoadd7}). The explicit relation of
the coefficients $\kappa^{(\ldots )}$ in the above two equations to the 
coefficients in the eqs.\ (\ref{ntwo3a}), (\ref{ntwo3b})
is given in Appendix \ref{appa}. 
As a consequence of the CAR, eqs.\ (\ref{p1ba})-(\ref{p1bc}),
the operators $\hat{b}^{[p]}_k$ must obey the equation:
\begin{eqnarray}
\label{ntwoadd10c}
\left\{\hat{b}^{[p]}_k,\hat{b}^{[q]}_l\right\}&=&
-2\ \delta_{pq}\ \delta_{kl}\ \openone_4\ .
\end{eqnarray}

\subsubsection[Clifford algebra counterpart of 
nonlinear Bogolyubov-Valatin transformations]{Clifford 
algebra counterpart of nonlinear Bogolyubov-Valatin\\
transformations}

In analogy to eq.\ (\ref{ntwoadd30a}),
we introduce the following useful notation:
\begin{eqnarray}
\hat{d}_{2k-1}&=&\hat{b}^{[1]}_k\ ,\ \ \ k\ =\ 1,
\ldots ,n\ (=2)\ ,\nonumber\\[0.3cm]
\hat{d}_{2k}&=&\hat{b}^{[2]}_k\ ,\ \ \ k\ =\ 1,\ldots ,n\ (=2)\ .
\end{eqnarray}
The operators $\hat{d}_k$ must obey the $C(0,2n = 4)$ Clifford algebra 
relation [$k,l = 1,\ldots ,2n\ (=4)$]
\begin{eqnarray}
\label{ntwoadd31a}
\left\{\hat{d}_k,\hat{d}_l\right\}&=&2\ \hat{d}_k\cdot\hat{d}_l\ =\
2\ g_{kl}\ \openone_4\ =\ -2\ \delta_{kl}\ \openone_4\ .
\end{eqnarray}
This relation is the Clifford algebra analogue of the canonical
anticommutation relations, (\ref{p1ba})-(\ref{p1bc}), for fermions.
The study of nonlinear Bogolyubov-Valatin transformations 
can consequently be performed by studying
nonlinear basis transformations in the Clifford algebra $C(0,4)$.\\ 

To prepare ourselves for the further investigation, 
the equations (\ref{ntwoadd13a}), (\ref{ntwoadd13b}) now can be compactly 
written as [The factor of $i$ in the last two terms can also be 
deduced, in principle, from the condition 
$\hat{b}_k^{[p]} =- \hat{b}_k^{[p]\dagger}$, see eqs.\ (\ref{ntwoadd13a}),
(\ref{ntwoadd13b}).]:
\begin{eqnarray}
\label{ntwoadd32a}
\hat{d}_k&=&\chi^{[1]\ (m)}_k\ \hat{c}_m\ +\ 
\frac{1}{2!}\chi^{[2]\ (m,n)}_k\ \hat{c}_m\hat{c}_n\ +\
\frac{1}{3!}\chi^{[3]\ (m,n,p)}_k\ i\ \hat{c}_m\hat{c}_n\hat{c}_p
\nonumber\\[0.3cm]
&&\ \ +\
\frac{1}{4!}\chi^{[4]\ (m,n,p,q)}_k\ i\ \hat{c}_m\hat{c}_n\hat{c}_p\hat{c}_q
\\[0.3cm]
\label{ntwoadd32b}
&=&\chi^{[1]\ (m)}_k\ \hat{c}_m\ +\ 
\frac{1}{2!}\chi^{[2]\ (m,n)}_k\ \hat{c}_m\wedge\hat{c}_n\ +\
\frac{1}{3!}\chi^{[3]\ (m,n,p)}_k\ i\ \hat{c}_m\wedge\hat{c}_n\wedge\hat{c}_p
\nonumber\\[0.3cm]
&&\ \ +\
\frac{1}{4!}\chi^{[4]\ (m,n,p,q)}_k\ i\ 
\hat{c}_m\wedge\hat{c}_n\wedge\hat{c}_p\wedge\hat{c}_q
\\[0.3cm]
\label{ntwoadd32c}
&=&\chi^{[1]\ (1)}_k\ \hat{c}_1\ +\ \chi^{[1]\ (2)}_k\ \hat{c}_2\ +\
\chi^{[1]\ (3)}_k\ \hat{c}_3\ +\ \chi^{[1]\ (4)}_k\ \hat{c}_4
\nonumber\\[0.3cm]
&&\ +\
\chi^{[2]\ (1,2)}_k\ \hat{c}_1\hat{c}_2\ +\
\chi^{[2]\ (1,3)}_k\ \hat{c}_1\hat{c}_3\ +\
\chi^{[2]\ (1,4)}_k\ \hat{c}_1\hat{c}_4\nonumber\\[0.3cm]
&&\ +\
\chi^{[2]\ (2,3)}_k\ \hat{c}_2\hat{c}_3\ +\
\chi^{[2]\ (2,4)}_k\ \hat{c}_2\hat{c}_4\ +\
\chi^{[2]\ (3,4)}_k\ \hat{c}_3\hat{c}_4
\nonumber\\[0.3cm]
&&\ +\
\chi^{[3]\ (1,2,3)}_k\ i\ \hat{c}_1\hat{c}_2\hat{c}_3\ +\
\chi^{[3]\ (1,2,4)}_k\ i\ \hat{c}_1\hat{c}_2\hat{c}_4\nonumber\\[0.3cm]
&&\ +\
\chi^{[3]\ (1,3,4)}_k\ i\ \hat{c}_1\hat{c}_3\hat{c}_4\ +\
\chi^{[3]\ (2,3,4)}_k\ i\ \hat{c}_2\hat{c}_3\hat{c}_4
\nonumber\\[0.3cm]
&&\ +\
\chi^{[4]\ (1,2,3,4)}_k\ i\ \hat{c}_1\hat{c}_2\hat{c}_3\hat{c}_4
\end{eqnarray}
The tensors $\chi^{[1]\ (m)}_k$, $\chi^{[2]\ (m,n)}_k$, 
$\chi^{[3]\ (m,n,p)}_k$, $\chi^{[4]\ (m,n,p,q)}_k$
are real and the last three tensors are totally antisymmetric with respect
to their upper indices. For the relation of the tensors 
$\chi^{[l]\ (\ldots )}_k$ to the coefficients $\kappa^{(\ldots )}$
see Appendix \ref{appa}.\\

In principle, already here one could go one step further and define
the operators $\hat{d}_k$ in terms of biparavectors (with some
coefficients $\chi_k^{\ \ M}$):
\begin{eqnarray}
\label{ntwofurtheradd65aintro}
\hat{d}_k&=&-\ \hat{d}^\dagger_k
\ = \ -\ \chi_k^{\ \ M}\ \hat{\hat{\hat{c}\,}}_M\ .
\end{eqnarray}
However, this step can much better be understood in course
of the further investigation and we postpone it, therefore,
to Part II of Subsec.\ \ref{global}, eq.\ (\ref{ntwofurtheradd65a}).\\

\subsection{\label{basis}Nonlinear basis transformations 
in the Clifford algebra 
\texorpdfstring{$C(0,4)$}{C(0,4)}} 
\subsubsection{\label{antico}Anticommutators \texorpdfstring{--}{-} 
Conditions for canonical transformations} 

From the eqs.\ (\ref{ntwoadd32a}), (\ref{ntwoadd32b}),
we find for the anticommutator, eq.\ (\ref{ntwoadd31a}),
\begin{eqnarray}
\label{ntwofurther44}
\left\{\hat{d}_k,\hat{d}_l\right\}&=&
\hat{d}_k \hat{d}_l\ +\ \hat{d}_l \hat{d}_k
\ \stackrel{\rm def}{=}\ 
2\ \hat{d}_k\cdot\hat{d}_l\nonumber\\[0.3cm]
&=&-2\left[\chi^{[1]}_{k\ (m)}\ \chi^{[1]\ (m)}_l
\ +\ \frac{1}{2!}\ \chi^{[2]}_{k\ (m,n)}\ \chi^{[2]\ (m,n)}_l
\right.\nonumber\\[0.3cm]
&&\ \ \ \ +\ \left.\frac{1}{3!}\ \chi^{[3]}_{k\ (m,n,p)}\ \chi^{[3]\ (m,n,p)}_l
\ +\ \frac{1}{4!}\ \chi^{[4]}_{k\ (m,n,p,q)}\ \chi^{[4]\ (m,n,p,q)}_l
\right]\ \openone_4\nonumber\\[0.3cm]
&&-\ i\ \left[\chi^{[2]}_{k\ (m,n)}\ \chi^{[3]\ (m,n,q)}_l
\ +\ \chi^{[3]\ (m,n,q)}_k\ \chi^{[2]}_{l\ (m,n)}\right] \hat{c}_q
\nonumber\\[0.3cm]
&&-\ i\ \Bigg[\chi^{[1]}_{k\ (m)}\ \chi^{[3]\ (m,p,q)}_l
\ +\ \chi^{[3]\ (m,p,q)}_k\ \chi^{[1]}_{l\ (m)}\nonumber\\[0.3cm]
&&\ \ \ \ \ \left. +\frac{1}{2}\ \chi^{[2]}_{k\ (m,n)}\ \chi^{[4]\ (m,n,p,q)}_l
\ +\ \frac{1}{2}\ \chi^{[4]\ (m,n,p,q)}_k\ \chi^{[2]}_{l\ (m,n)}\right]
\hat{c}_p\wedge\hat{c}_q\nonumber\\[0.3cm]
&&+\ \left[\chi^{[1]\ (n)}_k\ \chi^{[2]\ (p,q)}_l
\ +\ \chi^{[2]\ (p,q)}_k\ \chi^{[1]\ (n)}_l\right] 
\hat{c}_n\wedge\hat{c}_p\wedge\hat{c}_q
\nonumber\\[0.3cm]
&&+\ \frac{1}{2}\ \chi^{[2]\ (m,n)}_k\ \chi^{[2]\ (p,q)}_l\
\hat{c}_m\wedge\hat{c}_n\wedge\hat{c}_p\wedge\hat{c}_q\ .
\end{eqnarray}
The above result has been derived 
by means of the eqs.\ (\ref{ntwofurther40a})-(\ref{ntwofurther40d})
given in Appendix \ref{appb} [In the product
$\hat{d}_k\hat{d}_l$, use for the first factor the representation  
(\ref{ntwoadd32a}) and for the second factor (\ref{ntwoadd32b}).].
Now, imposing on the anticommutator (\ref{ntwofurther44})
the Clifford algebra analogue of the 
canonical anticommutation relation (CAR)
\begin{eqnarray}
\label{ntwofurther44bb}
\left\{\hat{d}_k,\hat{d}_l\right\}
&\stackrel{\displaystyle !}{=}&2\ g_{kl}\ \openone_4
\ =\ -2\ \delta_{kl}\ \openone_4
\end{eqnarray}
the equations (\ref{ntwofurther45a})-(\ref{ntwofurther45e}) follow:\\

\begin{itemize}
\item scalar part of eq.\ (\ref{ntwofurther44bb}):
\begin{eqnarray}
\label{ntwofurther45a}
&&\hspace{-1.cm}\chi^{[1]}_{k\ (m)}\ \chi^{[1]\ (m)}_l
\ +\ \frac{1}{2!}\ \chi^{[2]}_{k\ (m,n)}\ \chi^{[2]\ (m,n)}_l 
\ +\ \frac{1}{3!}\ \chi^{[3]}_{k\ (m,n,p)}\ \chi^{[3]\ (m,n,p)}_l
\nonumber\\[0.3cm]
&& +\ \frac{1}{4!}\ \chi^{[4]}_{k\ (m,n,p,q)}\ \chi^{[4]\ (m,n,p,q)}_l
\nonumber\\[0.3cm]
&=& \left[\chi^{[1]\ (1)}_k\ \chi^{[1]\ (1)}_l
\ +\ \chi^{[1]\ (2)}_k\ \chi^{[1]\ (2)}_l\right.
\ +\ \chi^{[1]\ (3)}_k\ \chi^{[1]\ (3)}_l
\ +\ \chi^{[1]\ (4)}_k\ \chi^{[1]\ (4)}_l\nonumber\\[0.3cm]
&&\ \ \ \ \ +\ \chi^{[2]\ (1,2)}_k\ \chi^{[2]\ (1,2)}_l
\ +\ \chi^{[2]\ (1,3)}_k\ \chi^{[2]\ (1,3)}_l
\ +\ \chi^{[2]\ (1,4)}_k\ \chi^{[2]\ (1,4)}_l
\nonumber\\[0.3cm]
&&\ \ \ \ \ +\ \chi^{[2]\ (2,3)}_k\ \chi^{[2]\ (2,3)}_l
\ +\ \chi^{[2]\ (2,4)}_k\ \chi^{[2]\ (2,4)}_l
\ +\ \chi^{[2]\ (3,4)}_k\ \chi^{[2]\ (3,4)}_l
\nonumber\\[0.3cm]
&&\ \ \ \ \ +\ \chi^{[3]\ (1,2,3)}_k\ \chi^{[3]\ (1,2,3)}_l
\ +\ \chi^{[3]\ (1,2,4)}_k\ \chi^{[3]\ (1,2,4)}_l
\ +\  \chi^{[3]\ (1,3,4)}_k\ \chi^{[3]\ (1,3,4)}_l
\nonumber\\[0.3cm]
&&\ \ \ \ \ +\ \chi^{[3]\ (2,3,4)}_k\ \chi^{[3]\ (2,3,4)}_l
\ +\ \left. \chi^{[4]\ (1,2,3,4)}_k\ \chi^{[4]\ (1,2,3,4)}_l\right]
\ =\ \delta_{kl}\hspace{2cm}
\end{eqnarray}
(The above equation is the orthogonality/orthonormality condition for 
4 vectors in a 15-dimensional Euclidean vector space.),

\item vector part of eq.\ (\ref{ntwofurther44bb}):
\begin{eqnarray}
\label{ntwofurther45b}
\left[\chi^{[2]}_{k\ (m,n)}\ \chi^{[3]\ (m,n,q)}_l
\ +\ \chi^{[3]\ (m,n,q)}_k\ \chi^{[2]}_{l\ (m,n)}\right] \hat{c}_q&=&0\ ,
\end{eqnarray}

\item bivector part of eq.\ (\ref{ntwofurther44bb}):
\begin{eqnarray}
\label{ntwofurther45c}
&&\hspace{-1.cm}\Bigg[\chi^{[1]}_{k\ (m)}\ \chi^{[3]\ (m,p,q)}_l
\ +\ \chi^{[3]\ (m,p,q)}_k\ \chi^{[1]}_{l\ (m)}\nonumber\\[0.3cm]
&&\hspace{-1.cm}
\ \ \ \ \ \left. +\frac{1}{2}\ \chi^{[2]}_{k\ (m,n)}\ \chi^{[4]\ (m,n,p,q)}_l
\ +\ \frac{1}{2}\ \chi^{[4]\ (m,n,p,q)}_k\ \chi^{[2]}_{l\ (m,n)}\right]
\hat{c}_p\wedge\hat{c}_q\ =\ 0\ ,
\end{eqnarray}

\item trivector part of eq.\ (\ref{ntwofurther44bb}):
\begin{eqnarray}
\label{ntwofurther45d}
\left[\chi^{[1]\ (n)}_k\ \chi^{[2]\ (p,q)}_l
\ +\ \chi^{[2]\ (p,q)}_k\ \chi^{[1]\ (n)}_l\right] 
\hat{c}_n\wedge\hat{c}_p\wedge\hat{c}_q&=&0\ ,
\end{eqnarray}

\item quadrivector part of eq.\ (\ref{ntwofurther44bb}):
\begin{eqnarray}
\label{ntwofurther45e}
\chi^{[2]\ (m,n)}_k\ \chi^{[2]\ (p,q)}_l\
\hat{c}_m\wedge\hat{c}_n\wedge\hat{c}_p\wedge\hat{c}_q&=&0\ .
\end{eqnarray}
\end{itemize}
These equations have to be studied in order to investigate
basis transformations in the Clifford algebra $C(0,4)$ and,
consequently, nonlinear Bogolyubov-Valatin transformations for
two fermionic modes.\\

For the sake of completeness, here we also give the expression for 
the commutator of the operators $\hat{d}_k$ (which, incidentally,
also defines their wedge product).
\begin{eqnarray}
\label{ntwofurther43}
\left[\hat{d}_k,\hat{d}_l\right]&=&
\hat{d}_k \hat{d}_l\ -\ \hat{d}_l \hat{d}_k
\ \stackrel{\rm def}{=}\ 
2\ \hat{d}_k\stackrel{d}{\displaystyle \wedge}\hat{d}_l
\nonumber\\[0.3cm]
&=&2 \left[ \chi^{[2]\ (m,q)}_k \chi^{[1]}_{l\ (m)}
\ -\ \chi^{[1]}_{k\ (m)}\ \chi^{[2]\ (m,q)}_l
\ +\ \frac{1}{3!}\ \chi^{[4]\ (m,n,p,q)}_k \chi^{[3]}_{l\ (m,n,p)}\right.
\nonumber\\[0.3cm]
&&\ \ \ \left. -\ 
\frac{1}{3!}\ \chi^{[3]}_{k\ (m,n,p)}\  \chi^{[4]\ (m,n,p,q)}_l\right]\hat{c}_q
\nonumber\\[0.3cm]
&&+\ 2\ \Bigg[ \chi^{[1]}_{k\ (p)}\ \chi^{[1]\ (q)}_l
\ +\ \chi^{[2]}_{k\ (m,p)}\ \chi^{[2]\ (m,q)}_l
\nonumber\\[0.3cm]
&&\ \ \ \left. +\ \frac{1}{2}\  \chi^{[3]}_{k\ (m,n,p)}\
\chi^{[3]\ (m,n,q)}_l\right] \hat{c}^{\ p}\wedge\hat{c}_q\nonumber\\[0.3cm]
&&+\ i\ \left[ \frac{1}{3}\ \chi^{[4]\ (m,n,p,q)}_k\ \chi^{[1]}_{l\ (m)}
\ -\ \frac{1}{3}\ \chi^{[1]}_{k\ (m)}\ \chi^{[4]\ (m,n,p,q)}_l\right.
\nonumber\\[0.3cm]
&&\ \ \ 
\ +\ \chi^{[2]\ (r,n)}_k\ \delta_{rs}\ \chi^{[3]\ (s,p,q)}_l
- \chi^{[3]\ (s,p,q)}_k\ \delta_{rs}\ \chi^{[2]\ (r,n)}_l\Bigg]
\hat{c}_n\wedge\hat{c}_p\wedge\hat{c}_q\nonumber\\[0.3cm]
&&+\ \frac{i}{3}\ \left[\chi^{[1]\ (m)}_k \chi^{[3]\ (n,p,q)}_l
\ -\ \chi^{[3]\ (n,p,q)}_k \chi^{[1]\ (m)}_l\ \right]
\hat{c}_m\wedge\hat{c}_n\wedge\hat{c}_p\wedge\hat{c}_q
\end{eqnarray}

At the end of this subsection, a conceptual comment is in order.
The first two lines of eq.\ (\ref{ntwofurther44})
define the inner product in the new set of vectors $\hat{d}_k$.
As the vectors $\hat{d}_k$ are not linear combinations of the original
vectors $\hat{c}_k$, the vector space within the Clifford algebra
$C(0,4)$ spanned by the vectors $\hat{d}_k$ differs from the 
vector space spanned by the vectors $\hat{c}_k$ \footnote{Certain 
special types of such nonlinear basis transformations in a 
Clifford algebras have been
considered previously in \citep{berg2,berg1,abe1}
and in generality for the Clifford algebra $C(0,2)$ in \citep{vanh1}.
For related aspects see \citep{mosn1}.}. Consequently, 
also the wedge products built over the respective vector spaces 
differ conceptually, as do concepts related to them such as the 
grade of a multivector. Consequently, the first line of eq.\
(\ref{ntwofurther43}) defines the wedge product related to the 
vector space spanned by the vectors $\hat{d}_k$ (We indicate this
fact by means of the letter '$d$' atop of the wedge symbol: 
$\stackrel{d}{\displaystyle \wedge}$.). To see 
this explicitly compare eq.\ (\ref{ntwofurther43}) with the following
equation which has been evaluated using the linearity and associativity
of the wedge product related to the vectors $\hat{c}_k$.
\begin{eqnarray}
\label{ntwofurther43b} 
\hat{d}_k\wedge\hat{d}_l
&=&\ \chi^{[1]}_{k\ (p)}\ \chi^{[1]\ (q)}_l
\ \hat{c}^{\ p}\wedge\hat{c}_q\nonumber\\[0.3cm]
&&+\ \frac{1}{2!}\ 
\left[\chi^{[1]\ (n)}_k\ \chi^{[2]\ (p,q)}_l
\ +\ \chi^{[2]\ (p,q)}_k\ \chi^{[1]\ (n)}_l\right]
\hat{c}_n\wedge\hat{c}_p\wedge\hat{c}_q\nonumber\\[0.3cm]
&&+\ \left[\frac{i}{3!}\ \chi^{[1]\ (m)}_k \chi^{[3]\ (n,p,q)}_l
\ -\ \frac{i}{3!}\ \chi^{[3]\ (n,p,q)}_k \chi^{[1]\ (m)}_l\right.
\nonumber\\[0.3cm]
&&\left. 
\ \ \ \ \ \ +\ \frac{1}{4}\ \chi^{[2]\ (m,n)}_k\ \chi^{[2]\ (p,q)}_l\right]
\hat{c}_m\wedge\hat{c}_n\wedge\hat{c}_p\wedge\hat{c}_q
\end{eqnarray}
Contrary to what naively one might expect,
the product $\hat{d}_k\wedge\hat{d}_l$ fails to be antisymmetric 
with respect to the indices $k$ and $l$ unless the condition 
$\hat{d}_k\cdot\hat{d}_l = - \delta_{kl}\ \openone_4$ 
[eq.\ (\ref{ntwofurther44bb})]
is imposed.\\

\subsubsection{Solution of the Clifford algebra analogue of the CAR 
\texorpdfstring{--}{-} Infinitesimal method} 

We are now going to solve the eqs.\ 
(\ref{ntwofurther45a})-(\ref{ntwofurther45e}) by means of two 
different methods, an infinitesimal and a global one. 
We start with the infinitesimal method in the 
vicinity of the identical Bogolyubov-Valatin transformation.
We apply the relations $\chi^{[1]}_{k\ (m)} = \delta_{km}
+ \Delta\chi^{[1]}_{k\ (m)}$,
$\chi^{[2]}_{k\ (m,n)} = \Delta\chi^{[2]}_{k\ (m,n)}$,
$\chi^{[3]}_{k\ (m,n,p)} = \Delta\chi^{[3]}_{k\ (m,n,p)}$,
$\chi^{[4]}_{k\ (m,n,p,q)} = \Delta\chi^{[4]}_{k\ (m,n,p,q)}$, where
all quantities preceeded by $\Delta$ are infinitesimally small.
Neglecting higher order terms the eqs.\ (\ref{ntwofurther45a}),
(\ref{ntwofurther45c}), (\ref{ntwofurther45d}) yield the following
conditions, respectively:
\begin{eqnarray}
\label{ntwofurther46a}
\Delta\chi^{[1]}_{k\ (l)}\ +\ \Delta\chi^{[1]}_{l\ (k)}&=&0\ ,\\[0.3cm]
\label{ntwofurther46c}
\Delta\chi^{[3]}_{k\ (m,n,l)}\ +\ \Delta\chi^{[3]}_{l\ (m,n,k)}&=&0\ ,\\[0.3cm]
\label{ntwofurther46d}
\Delta\chi^{[2]\ast}_{l\ (k,m)}\ +\ 
\Delta\chi^{[2]\ast}_{k\ (l,m)}&=&0\ .
\end{eqnarray}
Here, $\Delta\chi^{[2]\ast}_{k\ (l,m)} = 
\epsilon_{lmpq}\ \Delta\chi^{[2]\ (p,q)}_k$
($\epsilon_{lmpq}$ is the completely antisymmetric tensor in four dimensions
with $\epsilon_{1234} = 1$.). 
To arrive at this 
form of the eq.\ (\ref{ntwofurther46d}) we have made use of the relation
\begin{eqnarray}
\label{ntwofurther46da}
\epsilon_{mnpq}\ \epsilon^{\ \ \ q}_{rst}
&=&\delta_{mr}\delta_{ns}\delta_{pt}
\ +\ \delta_{ms}\delta_{nt}\delta_{pr}
\ +\ \delta _{mt}\delta_{nr}\delta_{ps}\nonumber\\[0.3cm]
&&\ -\ \delta_{ms}\delta_{nr}\delta_{pt}
\ -\ \delta_{mr}\delta_{nt}\delta_{ps}
\ -\ \delta _{mt}\delta_{ns}\delta_{pr}\ .
\end{eqnarray}
The eqs.\ (\ref{ntwofurther45b}) and (\ref{ntwofurther45e}) do not 
give rise to any condition for the infinitesimal parameters.
From the eqs.\ (\ref{ntwofurther46a})-(\ref{ntwofurther46d})
we can conclude that $\Delta\chi^{[1]}_{k\ (l)}$,
$\Delta\chi^{[2]\ast}_{k\ (l,m)}$,
$\Delta\chi^{[3]}_{k\ (m,n,l)}$ are objects which are 
completely antisymmetric in \underline{all} of their lower indices.   
Consequently, we find that the infinitesimal (nonlinear)
Bogolyubov-Valatin transformations depend on 15 parameters:
6 parameters given by $\Delta\chi^{[1]}_{k\ (l)}$, 4 parameters given
by $\Delta\chi^{[2]\ast}_{k\ (l,m)}$, 1 parameter given by the totally 
antisymmetric object $\Delta\chi^{[3]}_{k\ (m,n,l)}$ and 
4 more parameters given by 
$\Delta\chi^{[4]}_{k\ (m,n,p,q)}$ (The latter object does not receive any
restrictions within our infinitesimal consideration.). 
The number of parameters for each object is in agreement
with the global solution to be discussed further below 
[cf.\ eqs.\ (\ref{ntwofurther52a}), (\ref{ntwofurther49a}), 
(\ref{ntwofurther55a}), (\ref{ntwofurther57a})]
and the total number of 
15 parameters stands in correspondence to insight coming from the
implementation of the Bogolyubov-Valatin transformation
by means of an unitary transformation 
${\sf U}\left(\{\lambda\};\hat{c}\right)$ 
[cf.\ eq.\ (\ref{ntwo3a})]
\begin{eqnarray}
\label{ntwofurther47}
\hat{d}_k&=& {\sf B}\left(\{\lambda\};\hat{c}\right)
\ =\ {\sf U}\left(\{\lambda\};\hat{c}\right)\ \hat{c}_k\
{\sf U}\left(\{\lambda\};\hat{c}\right)^\dagger.
\end{eqnarray}
${\sf U}\left(\{\lambda\};\hat{c}\right)$ is for two ($n=2$) fermionic modes
an element of $SU(2^n = 4)$ which is a 15-parametric group and the 
double cover of the group $SO(6;\mathbb{R})$.\\

\subsubsection{\label{global}Solution of the Clifford algebra analogue
of the CAR \texorpdfstring{--}{-} Global method} 
\paragraph{I.\ \ \ Structural analysis of the Clifford algebra analogue 
of the CAR}\hspace{8cm}\\
\nopagebreak
\addcontentsline{toc}{subsubsection}{\ \ \ \ \ \ \ \ \ \  
I.\ \ Structural analysis of the Clifford algebra analogue of the CAR}
We turn our attention now to the global study of the eqs.\ 
(\ref{ntwofurther45a})-(\ref{ntwofurther45e}). The best strategy
appears to be to start with the consideration of the 
eq.\ (\ref{ntwofurther45e}), the quadrivector part of these equations.
The reason for this lies in the fact that in difference to the other
eqs.\ (\ref{ntwofurther45a})-(\ref{ntwofurther45d}), 
eq.\ (\ref{ntwofurther45e}) depends on one type of coefficients
[$\chi^{[2]\ (m,n)}_k$, i.e., the bivector coefficients in eq.\
(\ref{ntwoadd32a})] only.\\

\subparagraph{\sf (A) Quadrivector part}\hspace{8cm}\\
\nopagebreak
For the case $k = l$, eq.\ (\ref{ntwofurther45e}) reads
\begin{eqnarray}
\label{ntwofurther48a}
\hat{\chi}^{[2]}_k\wedge\hat{\chi}^{[2]}_k&=&0
\end{eqnarray}
where we have applied the notation
$\hat{\chi}^{[2]}_k = \chi^{[2]\ (m,n)}_k\ 
\hat{c}_m\wedge\hat{c}_n$ for the bivector $\hat{\chi}^{[2]}_k$.
Equivalently, this condition can be formulated as 
(\citep{kozl1}, \S 7, p.\ 101,
eq.\ (3) [p.\ 2250, eq.\ (7.3) of the English transl.])
\begin{eqnarray}
\label{ntwofurther48b}
\langle\hat{\chi}^{[2]\ast}_k,\hat{\chi}^{[2]}_k\rangle&=&
\epsilon^{mnpq} \chi^{[2]}_{k\ (m,n)}\chi^{[2]}_{k\ (p,q)}\ =\ 0\ .
\end{eqnarray}
Here, $\langle A,B\rangle$ denotes the 
scalar product of the bivectors $A$, $B$.
For its definition see, e.g., \citep{kozl1}, \S\ 1, p.\ 85, eq.\ (3)
[eq.\ (1.3), p.\ 2240 of the English translation].
$\hat{\chi}^{[2]\ast}_k$ denotes the bivector that is related 
to the Hodge dual of $\chi^{[2]}_k$:
$\chi^{[2]\ast\ (m,n)}_k = \epsilon^{mnpq} \chi^{[2]}_{k\ (p,q)}$.
Finally, a third, equivalent formulation for the 
condition (\ref{ntwofurther48a}) is: 
\begin{eqnarray}
\label{ntwofurther48c}
{\rm Pf}\;\chi^{[2]}_k&=&0
\end{eqnarray}
where $\rm Pf$ denotes the Pfaffian of the matrix $\chi^{[2]}_k$.
The eqs.\ (\ref{ntwofurther48a})-(\ref{ntwofurther48c}) each are
the condition for the decomposability of the bivector $\hat{\chi}^{[2]}_k$
\ \footnote{
\citep{cart1}, Vol.\ I, p.\ 26 (p.\ 19 of the English transl.),

\citep{scho1}, all eds., Chap.\ II, \S\ 6, p.\ 27,

\citep{ster1}, p.\ 25, Exercise 5.1,

\citep{penr1}, Vol.\ 1, Chap.\ 3, end of \S\ 5, p.\ 165,

\citep{kozl2}, p.\ 69, 

\citep{kozl1}, p.\ 101 
(p.\ 2250 of the English transl.),

\citep{vinb1}, Chap.\ 8, end of \S\ 4, {\cyrrm Primer 3} 
[Primer 3]/[Example 3], p.\ 341, 2.\ and 3.\ ed. 
(p.\ 324 of the English transl.),

\citep{kost1}, Vol.\ 2 ({\cyrrm \protect{Ch}ast\cprime }\ II: 
{\cyrrm Line\u inaya
Algebra}\ [Chast' II: Line\u\i naya Algebra]), {\cyrrm gl.} 6 [Chap.\ 6], 
\S\ 5, pp.\ 293-295.}. 
Consequently, we can write for the matrix
$\chi^{[2]\ (m,n)}_k$ and the corresponding bivector [For the last
part of eq.\ (\ref{ntwofurther49c}) use eq.\ (\ref{ntwofurther41}).]
\begin{eqnarray}
\label{ntwofurther49a}
\chi^{[2]\ (m,n)}_k&=&\beta^{(m)}\ {\cal H}^{(n)}_k
\ -\ {\cal H}^{(m)}_k\ \beta^{(n)}\ ,\\[0.3cm]
\label{ntwofurther49b}
\chi^{[2]}_k&=&2\ \beta\wedge{\cal H}_k\ ,\\[0.3cm]
\label{ntwofurther49c}
\hat{\chi}^{[2]}_k&=&2\ \hat{\beta}\wedge\hat{\cal H}_k
\ =\ 2\ \hat{\beta} \hat{\cal H}_k\ -\ 2\ {\cal H}_k \beta^T\ \openone_4\ .
\end{eqnarray}
Here, $\hat{\cal H}_k = {\cal H}^{(m)}_k\ \hat{c}_m$. 
${\cal H}_k$, $\beta$ denote row vectors 
($1\times 4$ matrices)\footnote{Their elements will finally 
turn out to be real and can,
for simplicity, immediately be chosen to be real in view of the fact
that $\chi^{[2]\ (m,n)}_k\in\mathbb{R}$.}. 
The ansatz (\ref{ntwofurther49a}), (\ref{ntwofurther49c})
immediately also solves eq.\ (\ref{ntwofurther45e})
for $k\neq l$.
\begin{eqnarray}
\label{ntwofurther50}
\hat{\chi}^{[2]}_k\wedge\hat{\chi}^{[2]}_l&=&0
\end{eqnarray}

\subparagraph{\sf (B) Trivector part}\hspace{8cm}\\
\nopagebreak
Eq.\ (\ref{ntwofurther45d}) can be written as follows.
\begin{eqnarray}
\label{ntwofurther51a}
\hat{\chi}^{[1]}_k\wedge\hat{\chi}^{[2]}_l
\ +\ \hat{\chi}^{[2]}_k\wedge\hat{\chi}^{[1]}_l&=&0
\end{eqnarray}
Introducing the ansatz (\ref{ntwofurther49c})
into eq.\ (\ref{ntwofurther51a}), it can be transformed to read
\begin{eqnarray}
\label{ntwofurther51b}
\hat{\beta}\wedge\left[
\hat{\chi}^{[1]}_k\wedge\hat{\cal H}_l
\ -\ \hat{\cal H}_k\wedge\hat{\chi}^{[1]}_l\right]&=&0
\end{eqnarray}
and its solution can immediately be read off to be 
(${\cal H}^{(0)}_k$, $\beta^{(0)}$ are real numbers)
\begin{eqnarray}
\label{ntwofurther52a}
\chi^{[1]\ (m)}_k&=&\beta^{(0)}\ {\cal H}^{(m)}_k
\ -\ {\cal H}^{(0)}_k\ {\beta}^{(m)}\ ,\\[0.3cm]
\label{ntwofurther52b}
\hat{\chi}^{[1]}_k&=&\beta^{(0)}\ \hat{\cal H}_k
\ -\ {\cal H}^{(0)}_k\ \hat{\beta}\ .
\end{eqnarray}
The significance of the chosen notation will become evident in a moment.\\

\subparagraph{\sf (C) Vector part}\hspace{8cm}\\
\nopagebreak
We can now also study eq.\ (\ref{ntwofurther45b}). Introducing
the Hodge dual of $\chi^{[3]}_k$ by writing 
$\chi^{[3]\ (m,n,p)}_k = \epsilon^{mnpq}\ \chi^{[3]\ast}_{k\ (q)}$
it reads 
\begin{eqnarray}
\label{ntwofurther53}
\chi^{[2]}_{k\ (m,n)}\ \epsilon^{mnpq}\ \chi^{[3]\ast}_{l\ (q)}
\ +\ \chi^{[3]\ast}_{k\ (q)}\ \epsilon^{mnpq}\ 
\chi^{[2]}_{l\ (m,n)}&=&0\ .
\end{eqnarray}
Operating on it with the expression $\epsilon_{rstp}\
\hat{c}^{\ r}\wedge\hat{c}^{\ s}\wedge\hat{c}^{\ t}$ and summing over
repeated indices, eq.\ (\ref{ntwofurther53}) can be brought to the 
form [taking into account eq.\ (\ref{ntwofurther46da})]
\begin{eqnarray}
\label{ntwofurther54}
\hat{\chi}^{[2]}_k\wedge\hat{\chi}^{[3]\ast}_l
\ +\ \hat{\chi}^{[3]\ast}_k\wedge\hat{\chi}^{[2]}_l&=&0\ .
\end{eqnarray}
This equation is analogous to eq.\ (\ref{ntwofurther51a}),
and immediately its solution can be found to be
(${\cal H}^{(-1)}_k$, $\beta^{(-1)}$ are real numbers,
the sign is chosen with hindsight)
\begin{eqnarray}
\label{ntwofurther55a}
\chi^{[3]\ast\ (m)}_k&=&-\left(\beta^{(-1)}\ {\cal H}^{(m)}_k
\ -\ {\cal H}^{(-1)}_k \beta^{(m)}\right)\ ,\\[0.3cm]
\label{ntwofurther55b}
\hat{\chi}^{[3]\ast}_k&=&-\left(\beta^{(-1)}\ \hat{\cal H}_k
\ -\ {\cal H}^{(-1)}_k \hat{\beta}\right)\ ,
\end{eqnarray}
or 
\begin{eqnarray}
\label{ntwofurther55c}
\chi^{[3]\ (m,n,p)}_k&=& -\ \epsilon^{mnpq}\ 
\left(\beta^{(-1)}\ {\cal H}_{k\ (q)}
\ -\ {\cal H}^{(-1)}_k \beta_{(q)}\right)\ ,\\[0.3cm]
\label{ntwofurther55d}
\hat{\chi}^{[3]}_k&=&-\ \left(\beta^{(-1)}\ \hat{\cal H}^\ast_k
\ -\ {\cal H}^{(-1)}_k \hat{\beta}^\ast\right)
\end{eqnarray}
where we have applied the notation
$\hat{\chi}^{[3]}_k = \chi^{[3]\ (m,n,p)}_k\ 
\hat{c}_m\wedge\hat{c}_n\wedge\hat{c}_p$ for the 
trivector $\hat{\chi}^{[3]}_k$, and
${\cal H}^{\ast\ (m,n,p)}_k = \epsilon^{mnpq}\ {\cal H}_{k\ (q)}$,
$\beta^{\ast\ (m,n,p)} = \epsilon^{mnpq}\ \beta_{(q)}$.
Again, the significance of the chosen notation will become 
evident in a moment.\\

\subparagraph{\sf (D) Bivector part}\hspace{8cm}\\
\nopagebreak
We turn our attention now to eq.\ (\ref{ntwofurther45c}).
Introducing the Hodge dual of $\chi^{[4]}_k$ by writing 
$\chi^{[4]\ (m,n,p,q)}_k = \epsilon^{mnpq}\ \chi^{[4]\ast}_k$
and making use of the eqs.\ (\ref{ntwofurther49a}), 
(\ref{ntwofurther52a}), (\ref{ntwofurther45d}) it reads:
\begin{eqnarray}
\label{ntwofurther56}
\left[\beta_{(m)}\ {\cal H}_{k\ (n)}\ 
\left(\beta^{(0)}\ {\cal H}^{(-1)}_l 
-  \beta^{(-1)}\ {\cal H}^{(0)}_l\right)\ \ \ \ \ \ \ \ \ \ 
\right.&&\nonumber\\[0.3cm]
\ +\ \left(\beta^{(0)}\ {\cal H}^{(-1)}_k
- \beta^{(-1)}\ {\cal H}^{(0)}_k\right)\ 
\beta_{(m)}\ {\cal H}_{l\ (n)}\ \ \ \ \ &&\nonumber\\[0.3cm]
\left.\ -\ \beta_{(m)}\ {\cal H}_{k\ (n)}\ \chi^{[4]\ast}_l
\ -\ \chi^{[4]\ast}_k\ \beta_{(m)}\ {\cal H}_{l\ (n)}\right]\
&\epsilon^{mnpq}\ \hat{c}_p\wedge\hat{c}_q\ =&0\ .
\end{eqnarray}
And the solution of this equation is found to be:
\begin{eqnarray}
\label{ntwofurther57a}
\chi^{[4]\ast}_k&=&\beta^{(0)}\ {\cal H}^{(-1)}_k
- \beta^{(-1)}\ {\cal H}^{(0)}_k
\\[0.3cm]
\label{ntwofurther57b}
\chi^{[4]\ (m,n,p,q)}_k&=&\epsilon^{mnpq}\ 
\left(\beta^{(0)}\ {\cal H}^{(-1)}_k
- \beta^{(-1)}\ {\cal H}^{(0)}_k
\right)\ .
\end{eqnarray}

\subparagraph{\sf (E) Scalar part}\hspace{8cm}\\
\nopagebreak
The remaining task consists in studying eq.\ 
(\ref{ntwofurther45a}). Inserting the eqs.\ 
(\ref{ntwofurther52a}), (\ref{ntwofurther49a}), 
(\ref{ntwofurther55c}), (\ref{ntwofurther57b}) into it, it can be transformed
to read:
\begin{eqnarray}
\label{ntwofurther58b}
\left[\sum_{m=-1}^4\left(\beta^{(m)}\right)^2\right]
\left[\sum_{n=-1}^4{\cal H}^{(n)}_k\ {\cal H}^{(n)}_l\right]
\hspace{2cm}&&\nonumber\\[0.3cm]
\ -\ \left[\sum_{m=-1}^4{\cal H}^{(m)}_k\ \beta^{(m)}\right]
\left[\sum_{n=-1}^4{\cal H}^{(n)}_l\ \beta^{(n)}\right]
&=&\delta_{kl}\ .\ \ 
\end{eqnarray}
This equation obviates the usefulness of the  
notation chosen earlier in the subparagraphs (A)-(D). From now on we denote
by the symbol ${\cal H}$ the $4\times 6$ matrix ${\cal H}$
with row number $k$ and column number $m$ matrix elements 
${\cal H}^{(m)}_k$ and by $\beta$ the row vector ($1\times 6$ matrix)
$\beta$. Eq.\ (\ref{ntwofurther58b}) can then compactly be written as
\begin{eqnarray}
\label{ntwofurther59}
\vert\beta\vert^2\ {\cal H}{\cal H}^T\ -\ 
\left({\cal H}\beta^T\right)\left(\beta{\cal H}^T\right)&=&
\openone_4\ .
\end{eqnarray}
It is immediately obvious from eq.\ (\ref{ntwofurther59}) that it
is invariant under $(S)O(6;\mathbb{R})$ transformations 
[${\cal H}^\prime = {\cal H} L$, $\beta^\prime = \beta L$,
$L$ is a $6\times 6$ matrix with $L\in (S)O(6;\mathbb{R})$].
Consequently, any two-mode nonlinear Bogolyubov-Valatin transformation
can be parametrized by means of the $(S)O(6;\mathbb{R})$ transformed
parameters of any particular (fixed) Bogolyubov-Valatin transformation.
The most natural starting point is the identical Bogolyubov-Valatin 
transformation $\hat{d}_k = \hat{c}_k$, i.e., in view of 
eq.\ (\ref{ntwofurther52a}) we can choose 
\begin{eqnarray}
\label{ntwofurther60c}
\beta &=&\beta_I\ =\ (0,1,0,0,0,0)\ ,\\[0.3cm]
\label{ntwofurther60d}
{\cal H}^{(m)}_k&=&{\cal H}^{\ \ \ (m)}_{I\ k}\ =\ \delta_k^m\ ,\ \ \ 
k,m = 1,\ldots,4,\\[0.3cm]
\label{ntwofurther60e}
{\cal H}^{(-1)}_k&=&{\cal H}^{(0)}_k\ =\ 0\ . 
\end{eqnarray}
We immediately find that always
\begin{eqnarray}
\label{ntwofurther60a}
{\cal H}\beta^T&=&0\ ,\\[0.3cm]
\label{ntwofurther60b}
\beta \beta^T\ =\ \vert\beta\vert^2&=&1
\end{eqnarray}
applies. Consequently, eq.\ (\ref{ntwofurther59})
finally reads
\begin{eqnarray}
\label{ntwofurther61}
{\cal H}{\cal H}^T&=&\openone_4\ .
\end{eqnarray}
We can now go one step further in streamlining the notation by
writing $\beta = {\cal H}_0$ (i.e., $\beta^{(m)} = {\cal H}_0^{(m)}$).
From now on we denote
by the symbol ${\cal H}$ the $5\times 6$ matrix ${\cal H}$
with row number $k = 0,\ldots,4$ and column number $m = -1,\ldots,4$ 
matrix elements ${\cal H}^{(m)}_k$. The equations 
(\ref{ntwofurther60a})-(\ref{ntwofurther61}) 
can then compactly be written 
in a single line as
\begin{eqnarray}
\label{ntwofurther61b}
{\cal H}{\cal H}^T&=&\openone_5\ .
\end{eqnarray}
Any nonlinear Bogolyubov-Valatin transformation can conveniently be 
parametrized by means of a $(S)O(6;\mathbb{R})$ matrix $L$
according to the equation
\begin{eqnarray}
\label{ntwofurther62}
{\cal H}&=& {\cal H}_I\ L\ .
\end{eqnarray}

The above structural analysis of the Clifford algebra analogue of the 
CAR has now paved the way for a systematic discussion of nonlinear
basis transformations in the Clifford algebra $C(0,4)$ (and the 
nonlinear Bogolyubov-Valatin transformations related to them).\\ 

\paragraph{II.\ Biparavector structure of basis transformations in 
the Clifford algebra \texorpdfstring{$C(0,4)$}{C(0,4)} 
}\hspace{8cm}\\
\addcontentsline{toc}{subsubsection}{\ \ \ \ \ \ \ \ \ \
II.\ \ Biparavector structure of basis transformations in the
\texorpdfstring{\hfill\ \\
Clifford algebra $C(0,4)$}{Clifford algebra C(0,4)}} 
To structure the results obtained in the previous subsection, 
let us define the antisymmetric $6\times 6$ matrix 
$\chi_{0k}$ with matrix elements
$\chi^{(m_1,m_2)}_{0k}$ by means of the equation[$M = (m_1,m_2)$]
\begin{eqnarray}
\label{ntwofurther63a}
\chi^{(m_1,m_2)}_{0k}&=&\chi^{\ \ M}_{0k}\ =\ 
{\cal H} ^{(m_1)}_0\ {\cal H}^{(m_2)}_k\ -\
{\cal H}^{(m_2)}_0\ {\cal H}^{(m_1)}_k\ =\
C_2({\cal H})^{\ \ \ m_1m_2}_{0k}
\end{eqnarray}
or, symbolically,
\begin{eqnarray}
\label{ntwofurther63b}
\chi_{0k}&=&\left(
\begin{array}{*{3}{c}}
0&-\chi^{[4]\ast}_k&
-\chi^{[3]\ast}_k\\
\chi^{[4]\ast}_k&
0&
\chi^{[1]}_k\\
\left(\chi^{[3]\ast}_k\right)^T&
-\left(\chi^{[1]}_k\right)^T&
\chi^{[2]}_k
\end{array}
\right)\ = \ -\ \chi_{0k}^T\\[0.3cm]
\label{ntwofurther63c}
&=&2\ {\cal H}_0\wedge{\cal H}_k
\ = \ 2\ \left[({\cal H}_I\wedge{\cal H}_I)\ C_2(L)\right]_{0k}\ .
\end{eqnarray}
Here, $\chi^{[1]}_k$, $\chi^{[3]\ast }_k$ denote the row vectors 
($1\times 6$ matrices)
$\chi^{[1]}_k$, $\chi^{[3]\ast }_k$ with components $\chi^{[1]\ (m)}_k$,
$\chi^{[3]\ast\ (m)}_k$, respectively, and $C_l(L)$ denotes the order $l$
compound matrix of the matrix $L$
(cf.\ Appendix \ref{appcomp}).
Writing eq.\ (\ref{ntwoadd32a}) as
\begin{eqnarray}
\label{ntwofurtheradd64}
\hat{d}_k&=&\chi^{[1]\ (m)}_k\ \hat{c}_m\ +\ 
\frac{1}{2!}\chi^{[2]\ (m,n)}_k\ \hat{c}_m\hat{c}_n\ -\
\chi^{[3]\ast\ (q)}_k\ \hat{\hat{c}\,}_{(-1)}\hat{c}_q\ +\
\chi^{[4]\ast}_k\ \hat{\hat{c}\,}_{(-1)}\ \ \ \ \ \
\end{eqnarray}
one can convince oneself that the following representation applies
[here, $M = (m_1,m_2)$, $m_1 < m_2$].
\begin{eqnarray}
\label{ntwofurtheradd65a}
\hat{d}_k&=&-\ \hat{d}^\dagger_k\ =\ -\ \frac{1}{2}\ 
\hat{\hat{\hat{c}\hspace{0.45mm}}}^\dagger\chi_{0k}\ 
\hat{\hat{\hat{c}\hspace{0.45mm}}}
\ = \ -\ \chi_{0k}^{\ \ M}\ \hat{\hat{\hat{c}\hspace{0.45mm}}}_M\\[0.3cm]
\label{ntwofurtheradd65asec}
&=&- 2\ \left({\cal H}_0\wedge{\cal H}_k\right)^M
\ \hat{\hat{\hat{c}\hspace{0.4mm}}}_M
\end{eqnarray}
We use the notation
\begin{eqnarray}
\label{ntwofurtheradd65aaa}
 \hat{\hat{\hat{c}\hspace{0.45mm}}}&=&\left(
\begin{array}{*{1}{c}}
 \hat{\hat{\hat{c}\hspace{0.5mm}}}_{-1}\\
 \hat{\hat{\hat{c}\hspace{0.5mm}}}_0\\
 \hat{\hat{\hat{c}\hspace{0.5mm}}}_1\\
 \hat{\hat{\hat{c}\hspace{0.5mm}}}_2\\
 \hat{\hat{\hat{c}\hspace{0.5mm}}}_3\\
 \hat{\hat{\hat{c}\hspace{0.5mm}}}_4
\end{array}
\right)\ =\ 
\left(
\begin{array}{*{1}{c}}
i \hat{c}_1 \hat{c}_2 \hat{c}_3 \hat{c}_4\\
 -\openone_4\\
\hat{c}_1\\
\hat{c}_2\\
\hat{c}_3\\
\hat{c}_4
\end{array}
\right)\ ,\ \ \ \ \\[0.3cm]
\label{ntwofurtheradd65aab}
\hat{\hat{\hat{c}\hspace{0.45mm}}}^\dagger&=&
\left(\hat{\hat{\hat{c}\hspace{0.3mm}}}_{-1}^\dagger,
\hat{\hat{\hat{c}\hspace{0.3mm}}}_0^\dagger,
\hat{\hat{\hat{c}\hspace{0.3mm}}}_1^\dagger,
\hat{\hat{\hat{c}\hspace{0.3mm}}}_2^\dagger,
\hat{\hat{\hat{c}\hspace{0.3mm}}}_3^\dagger,
\hat{\hat{\hat{c}\hspace{0.3mm}}}_4^\dagger\right)\nonumber\\[0.3cm]
&=&
\left(-i \hat{c}_1 \hat{c}_2 \hat{c}_3 \hat{c}_4,-\openone_4,
-\hat{c}_1,-\hat{c}_2,-\hat{c}_3,-\hat{c}_4\right)\ .
\end{eqnarray}
$\hat{\hat{\hat{c}\,}}_M$ used in 
eq.\ (\ref{ntwofurtheradd65a}) is a biparavector [cf.\ eq.\
(\ref{ntwofurtheradd65aa})]. Taking into account 
the eqs.\ (\ref{ntwofurther60c})-(\ref{ntwofurther60e}), 
(\ref{ntwofurther63c}),  eq.\ (\ref{ntwofurtheradd65asec}) can also
be written as
\begin{eqnarray}
\label{ntwofurtheradd65b}
\hat{d}_k&=& -\ C_2(L)_{0k}^{\ \ \ M}\ \hat{\hat{\hat{c}\,}}_M\ .
\end{eqnarray}
This equation describes the structure of nonlinear basis transformations 
in the Clifford algebra $C(0,4)$. However, as these transformations
must have a group theoretical structure the 
eqs.\ (\ref{ntwofurtheradd65asec}), (\ref{ntwofurtheradd65b}) fail
to reflect this structural, group theoretical, aspect because the 
new (transformed) generators $\hat{d}_k$ of the Clifford algebra $C(0,4)$
are given in terms of other (however, related) objects --
the biparavectors $\hat{\hat{\hat{c}\,}}_M$.\\

We will now derive a
more symmetric representation of the nonlinear basis transformations
in the Clifford algebra $C(0,4)$.
Taking into account eq.\ (\ref{ntwofurtheradd65aa}), we find (with some
hindsight) for eq.\ (\ref{ntwofurtheradd65asec})
\begin{eqnarray}
\label{ntwofurtheradd65c}
-\ \hat{d}_k\ =\ \hat{\hat{\hat{d}\ }}_{\hspace{-1mm}0k}
&=&\frac{1}{2}\left(
\hat{\hat{\hat{d}\ }}^\dagger_{\hspace{-1mm}0}\; 
\hat{\hat{\hat{d}\ }}_{\hspace{-1mm}k}
- \hat{\hat{\hat{d}\ }}^\dagger_{\hspace{-1mm}k}\; 
\hat{\hat{\hat{d}\ }}_{\hspace{-1mm}0}\right)
\nonumber\\[0.3cm]
&=&2\ \left({\cal H}_0\wedge{\cal H}_k\right)^M 
\hat{\hat{\hat{c}\hspace{0.3mm}}}_M\ =\ 
C_2({\cal H})_{0k}^{\ \ \ M}\ \hat{\hat{\hat{c}\,}}_M\nonumber\\[0.3cm]
&=&\frac{1}{2}\left(\hat{\hat{\hat{\cal H}\;}}_{\hspace{-1mm}0}^\dagger
\hat{\hat{\hat{\cal H}\;}}_{\hspace{-1mm}k}\ -\
\hat{\hat{\hat{\cal H}\;}}^{\hspace{-1mm}\dagger}_{\hspace{-1mm}k}\;
\hat{\hat{\hat{\cal H}\;}}_{\hspace{-1mm}0}\right)\ .
\end{eqnarray}
Here, $\hat{\hat{\hat{\cal H}\;}}_{\hspace{-1mm}k} = 
{\cal H}_k \hat{\hat{\hat{c}\,}} = 
\sum^4_{m=-1}{\cal H}^{(m)}_k\ \hat{\hat{\hat{c}\,}}_{\hspace{-1mm}m}$ is a 
paravector ,
$\hat{\hat{\hat{\cal H}\;}}^{\hspace{-1mm}\dagger}_{\hspace{-1mm}k}
= - 2 {\cal H}^{(0)}_k\ \openone_4 
- \hat{\hat{\hat{\cal H}\;}}_{\hspace{-1mm}k}$.
From the eqs.\ (\ref{ntwofurtheradd63x}),
(\ref{ntwofurther61b}) we recognize that for the paravectors
$\hat{\hat{\hat{\cal H}\;}}_{\hspace{-1mm}k}$ applies ($k,l = 0,\ldots , 4$)
\begin{eqnarray}
\label{ntwofurtheradd73}
\hat{\hat{\hat{\cal H}\;}}^{\hspace{-1mm}\dagger}_{\hspace{-1mm}k}
\hat{\hat{\hat{\cal H}\;}}_{\hspace{-1mm}l}\ +\ 
\hat{\hat{\hat{\cal H}\;}}^{\hspace{-1mm}\dagger}_{\hspace{-1mm}l}
\hat{\hat{\hat{\cal H}\;}}_{\hspace{-1mm}k}\ = \ 
\hat{\hat{\hat{\cal H}\;}}_{\hspace{-1mm}k}
\hat{\hat{\hat{\cal H}\;}}^{\hspace{-1mm}\dagger}_{\hspace{-1mm}l}\ +\ 
\hat{\hat{\hat{\cal H}\;}}_{\hspace{-1mm}l}
\hat{\hat{\hat{\cal H}\;}}^{\hspace{-1mm}\dagger}_{\hspace{-1mm}k}
&=&2\ \delta_{kl}\ \openone_4\ .
\end{eqnarray}
Consequently, we obtain
\begin{eqnarray}
\label{ntwofurtheradd69}
-\ \hat{d}_k&=&\hat{\hat{\hat{d}\ }}_{\hspace{-1mm}0k}
\ =\ \hat{\hat{\hat{d}\ }}^\dagger_{\hspace{-1mm}0}\; 
\hat{\hat{\hat{d}\ }}_{\hspace{-1mm}k}\ =\ 
-\ \hat{\hat{\hat{d}\ }}^\dagger_{\hspace{-1mm}k}\; 
\hat{\hat{\hat{d}\ }}_{\hspace{-1mm}0}\ =\
\hat{\hat{\hat{\cal H}\;}}^{\hspace{-1mm}\dagger}_{\hspace{-1mm}0}\;
\hat{\hat{\hat{\cal H}\;}}_{\hspace{-1mm}k}\ =\
-\ \hat{\hat{\hat{\cal H}\;}}^{\hspace{-1mm}\dagger}_{\hspace{-1mm}k}\;
\hat{\hat{\hat{\cal H}\;}}_{\hspace{-1mm}0}\ .
\end{eqnarray}
Using the relation [cf.\ eqs.\ (\ref{ntwofurther60b}),
(\ref{ntwofurtheradd73})]
\begin{eqnarray}
\label{ntwofurtheradd70}
\hat{\hat{\hat{\cal H}\;}}^{\hspace{-1mm}\dagger}_{\hspace{-1mm}0}\;
\hat{\hat{\hat{\cal H}\;}}_{\hspace{-1mm}0}
&=&
\hat{\hat{\hat{\cal H}\;}}_{\hspace{-1mm}0}\;
\hat{\hat{\hat{\cal H}\;}}^{\hspace{-1mm}\dagger}_{\hspace{-1mm}0}
\ =\ {\cal H}_0 {\cal H}^T_0\ \openone_4\ =\ \openone_4
\end{eqnarray}
we find for $\hat{d}_k\stackrel{d}{\displaystyle \wedge}\hat{d}_l$, 
($k\neq l$, $k,l = 1,\ldots , 4$)
\begin{eqnarray}
\label{ntwofurtheradd72c}
-\ \hat{d}_k\stackrel{d}{\displaystyle \wedge}\hat{d}_l
\ =\ - \ \hat{d}_k \hat{d}_l 
&=&\hat{\hat{\hat{d}\ }}_{\hspace{-1mm}kl}
\ =\ \frac{1}{2}\left(
\hat{\hat{\hat{d}\ }}^\dagger_{\hspace{-1mm}k}\; 
\hat{\hat{\hat{d}\ }}_{\hspace{-1mm}l}
- \hat{\hat{\hat{d}\ }}^\dagger_{\hspace{-1mm}l}\; 
\hat{\hat{\hat{d}\ }}_{\hspace{-1mm}k}\right)
\nonumber\\[0.3cm]
&=&\frac{1}{2}\left(\hat{\hat{\hat{\cal H}\;}}_{\hspace{-1mm}k}^\dagger
\hat{\hat{\hat{\cal H}\;}}_{\hspace{-1mm}l}\ -\
\hat{\hat{\hat{\cal H}\;}}^{\hspace{-1mm}\dagger}_{\hspace{-1mm}l}\;
\hat{\hat{\hat{\cal H}\;}}_{\hspace{-1mm}k}\right)
\ = \ -\ \hat{\hat{\hat{\cal H}\;}}^{\hspace{-1mm}\dagger}_{\hspace{-1mm}k}
\hat{\hat{\hat{\cal H}\;}}_{\hspace{-1mm}l}
\ = \ \hat{\hat{\hat{\cal H}\;}}^{\hspace{-1mm}\dagger}_{\hspace{-1mm}l}
\hat{\hat{\hat{\cal H}\;}}_{\hspace{-1mm}k}\nonumber\\[0.3cm]
&=&2\ \left({\cal H}_k\wedge{\cal H}_l\right)^M \hat{\hat{\hat{c}\ }}_M\ =\ 
C_2({\cal H})_{kl}^{\ \ \ M}\ \hat{\hat{\hat{c}\,}}_M\ \ . \ \ \ 
\end{eqnarray}

Finally, only studying objects with index $k = -1$ remains to be done.
In view of eq.\ (\ref{ntwofurtheradd63y}) 
it turns out to be useful to define and to study the object
\begin{eqnarray}
\label{ntwofurtheradd74a}
\hat{\hat{\hat{\cal H}\;}}_{\hspace{-1mm}(-1)}&=&
-\ i\ \hat{\hat{\hat{\cal H}\;}}_{\hspace{-1mm}0}
\hat{\hat{\hat{\cal H}\;}}^{\hspace{-1mm}\dagger}_{\hspace{-1mm}1}
\hat{\hat{\hat{\cal H}\;}}_{\hspace{-1mm}2}
\hat{\hat{\hat{\cal H}\;}}^{\hspace{-1mm}\dagger}_{\hspace{-1mm}3}
\hat{\hat{\hat{\cal H}\;}}_{\hspace{-1mm}4}\ =\ 
-\ i\ \hat{\hat{\hat{\cal H}\;}}_{\hspace{-1mm}4}
\hat{\hat{\hat{\cal H}\;}}^{\hspace{-1mm}\dagger}_{\hspace{-1mm}3}
\hat{\hat{\hat{\cal H}\;}}_{\hspace{-1mm}2}
\hat{\hat{\hat{\cal H}\;}}^{\hspace{-1mm}\dagger}_{\hspace{-1mm}1}
\hat{\hat{\hat{\cal H}\;}}_{\hspace{-1mm}0}\ .
\end{eqnarray}
${\cal H}_{(-1)}$ implicitly defined by means of eq.\ (\ref{ntwofurtheradd74a})
is a row vector ($1\times 6$ matrix). For the identical Bogolyubov-Valatin
transformation one quickly finds ${\cal H}_{I (-1)} = (1,0,0,0,0,0)$.
From now on we denote
by the symbol ${\cal H}$ the (quadratic) $6\times 6$ matrix ${\cal H}$
with row number $k = -1,\ldots,4$ and column number $m = -1,\ldots,4$ 
matrix elements ${\cal H}^{(m)}_k$. 
Consequently, for the identical Bogolyubov-Valatin
transformation we have
\begin{eqnarray}
\label{ntwofurtheradd74b}
{\cal H}_I&=&\openone_6\ .
\end{eqnarray}
For eq.\ (\ref{ntwofurtheradd74a}), we can also write
($\epsilon_{mnpqrs}$
is the 6-dimensional totally antisymmetric tensor,
$\epsilon_{(-1)01234} = 1$;
we have raised here the indices
by means of $\delta_{kl}$ and not using $g_{kl}$)
\begin{eqnarray}
\label{ntwofurtheradd74c}
\hat{\hat{\hat{\cal H}\;}}_{\hspace{-1mm}(-1)}
&=&-\frac{i}{5!}\ \epsilon^{\ \ \ \ npqrs}_{(-1)}\ 
\hat{\hat{\hat{\cal H}\;}}_{\hspace{-1mm}n}
\hat{\hat{\hat{\cal H}\;}}^{\hspace{-1mm}\dagger}_{\hspace{-1mm}p}
\hat{\hat{\hat{\cal H}\;}}_{\hspace{-1mm}q}
\hat{\hat{\hat{\cal H}\;}}^{\hspace{-1mm}\dagger}_{\hspace{-1mm}r}
\hat{\hat{\hat{\cal H}\;}}_{\hspace{-1mm}s}\ .
\end{eqnarray}
Eq.\ (\ref{ntwofurtheradd74c}) can be transformed 
[taking into account eq.\ (\ref{ntwofurther61b})] to read
\begin{eqnarray}
\label{ntwofurtheradd75a}
\hat{\hat{\hat{\cal H}\;}}_{\hspace{-1mm}(-1)}
&=&-\frac{i}{5!}\ \epsilon^{\ \ \ \ npqrs}_{(-1)}\ \nonumber\\[0.3cm]
&&\ \times\ \sum_{n^\prime , p^\prime ,
q^\prime , r^\prime , s^\prime= -1\atop
n^\prime\neq p^\prime\neq
q^\prime\neq r^\prime\neq s^\prime}^4
{\cal H}_n^{(n^\prime)} {\cal H}_p^{(p^\prime)}
{\cal H}_q^{(q^\prime)} {\cal H}_r^{(r^\prime)}
{\cal H}_s^{(s^\prime)}\ 
\hat{\hat{\hat{c}\,}}_{n^\prime}\;\hat{\hat{\hat{c}\,}}^\dagger_{p^\prime}\;
\hat{\hat{\hat{c}\,}}_{q^\prime}\;\hat{\hat{\hat{c}\,}}^\dagger_{r^\prime}\;
\hat{\hat{\hat{c}\,}}_{s^\prime}\ \ . \ \ 
\end{eqnarray}
Using eq.\ (\ref{ntwofurtheradd63z}), taking into account
eq.\ (\ref{ntwofurtheradd75c}), and applying the Laplace expansion
of a determinant [cf.\ eq.\ (\ref{A2}) in our Appendix \ref{appcomp}]
we find
\begin{eqnarray}
\label{ntwofurtheradd75b}
\hat{\hat{\hat{\cal H}\;}}_{\hspace{-1mm}(-1)}
&=&\frac{1}{5!}\ \epsilon^{\ \ \ \ npqrs}_{(-1)}\  
{\cal H}_n^{(n^\prime)} {\cal H}_p^{(p^\prime)}
{\cal H}_q^{(q^\prime)} {\cal H}_r^{(r^\prime)}
{\cal H}_s^{(s^\prime)}\ 
\epsilon^{\ \ \ \ \ \ \ \ \ m}_{n^\prime p^\prime q^\prime r^\prime s^\prime}\ 
\hat{\hat{\hat{c}\,}}_m\\[0.3cm]
\label{ntwofurtheradd75d}
&=&\left[
C_5\left(L^T\right)^\star\right]_{(-1)}^{\ \ \ \ m}\ 
\hat{\hat{\hat{c}\,}}_m\ =\
\frac{L_{(-1)}^{\ \ \ \ (m)}}{\det L}\ \hat{\hat{\hat{c}\,}}_m
\end{eqnarray}
Consistency  with eq.\ (\ref{ntwofurtheradd74b}) now requires that 
$\det L = 1$, i.e., $L\in SO(6;\mathbb{R})$.  This is closely related
to the fact that the unitary group implementing the 
nonlinear Bogolyubov-Valatin transformations
[i.e., $SU(4)$] is simply connected.
From the eqs.\ (\ref{ntwofurther62}), (\ref{ntwofurtheradd75d})
we find in total
\begin{eqnarray}
\label{ntwofurtheradd75c}
{\cal H}&=& L\ ,\ \  L L^T\ =\ \openone_6\ .
\end{eqnarray}
One can convince oneself that eq.\ (\ref{ntwofurtheradd73})
also applies for the whole index range $k,l = -1,\ldots , 4$.\\

In analogy to eq.\ (\ref{ntwofurtheradd65c})
we can write ($\hat{\hat{d}\ }_{\hspace{-1.5mm}(-1)}
\ =\ i\hat{d}_1\hat{d}_2\hat{d}_3\hat{d}_4\ =\ 
i\hat{d}_1\stackrel{d}{\displaystyle
 \wedge}\hat{d}_2\stackrel{d}{\displaystyle
 \wedge}\hat{d}_3\stackrel{d}{\displaystyle \wedge}\hat{d}_4$)
\begin{eqnarray}
\label{ntwofurtheradd76}
\hat{\hat{d}\ }_{\hspace{-1mm}(-1)}
&=&\hat{\hat{\hat{d}\ }}_{\hspace{-1mm}(-1)0}
\ =\ \frac{1}{2}\left(
\hat{\hat{\hat{d}\ }}^\dagger_{\hspace{-1mm}(-1)}\; 
\hat{\hat{\hat{d}\ }}_{\hspace{-1mm}0}
- \hat{\hat{\hat{d}\ }}^\dagger_0\; \hat{\hat{\hat{d}\ }}_{(-1)}\right)
\nonumber\\[0.3cm]
&=&\frac{1}{2}\left(\hat{\hat{\hat{\cal H}\;}}_{\hspace{-1mm}(-1)}^\dagger
\hat{\hat{\hat{\cal H}\;}}_{\hspace{-1mm}0}\ -\
\hat{\hat{\hat{\cal H}\;}}^{\hspace{-1mm}\dagger}_{\hspace{-1mm}0}\;
\hat{\hat{\hat{\cal H}\;}}_{\hspace{-1mm}(-1)}\right)
\ = \ i\ \hat{\hat{\hat{\cal H}\;}}^{\hspace{-1mm}\dagger}_{\hspace{-1mm}1}
\hat{\hat{\hat{\cal H}\;}}_{\hspace{-1mm}2}
\hat{\hat{\hat{\cal H}\;}}^{\hspace{-1mm}\dagger}_{\hspace{-1mm}3}
\hat{\hat{\hat{\cal H}\;}}_{\hspace{-1mm}4}\nonumber\\[0.3cm]
&=&2\ \left({\cal H}_{(-1)}\wedge{\cal H}_0\right)^M \hat{\hat{\hat{c}\, }}_M
\ =\ C_2({\cal H})_{(-1)0}^{\ \ \ \ \ \ M}\ \hat{\hat{\hat{c}\,}}_M\ .
\end{eqnarray}
Furthermore, we have
\begin{eqnarray}
\label{ntwofurtheradd77a}
- i\hat{d}_2\hat{d}_3\hat{d}_4&=& 
- i\hat{d}_2\stackrel{d}{\displaystyle 
 \wedge}\hat{d}_3\stackrel{d}{\displaystyle \wedge}\hat{d}_4
\ =\ -\ \hat{\hat{d}\ }_{\hspace{-1mm}(-1)}
\hat{\hat{d}\ }_{\hspace{-1mm}1}\nonumber\\[0.3cm]
&=&\hat{\hat{\hat{d}\ }}_{\hspace{-1mm}(-1)1}
\ =\ \frac{1}{2}\left(
\hat{\hat{\hat{d}\ }}^\dagger_{\hspace{-1mm}(-1)}\; 
\hat{\hat{\hat{d}\ }}_{\hspace{-1mm}1}
- \hat{\hat{\hat{d}\ }}^\dagger_{\hspace{-1mm}1}\; 
\hat{\hat{\hat{d}\ }}_{\hspace{-1mm}(-1)}\right)
\nonumber\\[0.3cm]
&=&\frac{1}{2}\left(\hat{\hat{\hat{\cal H}\;}}_{\hspace{-1mm}(-1)}^\dagger
\hat{\hat{\hat{\cal H}\;}}_{\hspace{-1mm}1}\ -\
\hat{\hat{\hat{\cal H}\;}}^{\hspace{-1mm}\dagger}_{\hspace{-1mm}1}\;
\hat{\hat{\hat{\cal H}\;}}_{\hspace{-1mm}(-1)}\right)
\ = \ -\ i\ \hat{\hat{\hat{\cal H}\;}}^{\hspace{-1mm}\dagger}_{\hspace{-1mm}0}
\hat{\hat{\hat{\cal H}\;}}_{\hspace{-1mm}2}
\hat{\hat{\hat{\cal H}\;}}^{\hspace{-1mm}\dagger}_{\hspace{-1mm}3}
\hat{\hat{\hat{\cal H}\;}}_{\hspace{-1mm}4}\nonumber\\[0.3cm]
&=&2\ \left({\cal H}_{(-1)}\wedge{\cal H}_1\right)^M \hat{\hat{\hat{c}\, }}_M
\ =\ C_2({\cal H})_{(-1)1}^{\ \ \ \ \ \ M}\ \hat{\hat{\hat{c}\,}}_M
\ ,\\[0.3cm]
\label{ntwofurtheradd77b}
i\hat{d}_1\hat{d}_3\hat{d}_4&=& 
i\hat{d}_1\stackrel{d}{\displaystyle 
\wedge}\hat{d}_3\stackrel{d}{\displaystyle \wedge}\hat{d}_4
\ =\ -\ \hat{\hat{d}\ }_{\hspace{-1mm}(-1)}
\hat{\hat{d}\ }_{\hspace{-1mm}2}\nonumber\\[0.3cm]
&=&\hat{\hat{\hat{d}\ }}_{\hspace{-1mm}(-1)2}
\ =\ \frac{1}{2}\left(
\hat{\hat{\hat{d}\ }}^\dagger_{\hspace{-1mm}(-1)}\;
\hat{\hat{\hat{d}\ }}_{\hspace{-1mm}2}
- \hat{\hat{\hat{d}\ }}^\dagger_{\hspace{-1mm}2}\; 
\hat{\hat{\hat{d}\ }}_{\hspace{-1mm}(-1)}\right)
\nonumber\\[0.3cm]
&=&\frac{1}{2}\left(\hat{\hat{\hat{\cal H}\;}}_{\hspace{-1mm}(-1)}^\dagger
\hat{\hat{\hat{\cal H}\;}}_{\hspace{-1mm}2}\ -\
\hat{\hat{\hat{\cal H}\;}}^{\hspace{-1mm}\dagger}_{\hspace{-1mm}2}\;
\hat{\hat{\hat{\cal H}\;}}_{\hspace{-1mm}(-1)}\right)
\ = \ i\ \hat{\hat{\hat{\cal H}\;}}^{\hspace{-1mm}\dagger}_{\hspace{-1mm}0}
\hat{\hat{\hat{\cal H}\;}}_{\hspace{-1mm}1}
\hat{\hat{\hat{\cal H}\;}}^{\hspace{-1mm}\dagger}_{\hspace{-1mm}3}
\hat{\hat{\hat{\cal H}\;}}_{\hspace{-1mm}4}\nonumber\\[0.3cm]
&=&2\ \left({\cal H}_{(-1)}\wedge{\cal H}_2\right)^M \hat{\hat{\hat{c}\, }}_M
\ =\ C_2({\cal H})_{(-1)2}^{\ \ \ \ \ \ M}\ \hat{\hat{\hat{c}\,}}_M
\ ,\\[0.3cm]
\label{ntwofurtheradd77c}
- i\hat{d}_1\hat{d}_2\hat{d}_4&=& 
- i\hat{d}_1\stackrel{d}{\displaystyle 
\wedge}\hat{d}_2\stackrel{d}{\displaystyle \wedge}\hat{d}_4
\ =\ -\ \hat{\hat{d}\ }_{\hspace{-1mm}(-1)}
\hat{\hat{d}\ }_{\hspace{-1mm}3}\nonumber\\[0.3cm]
&=&\hat{\hat{\hat{d}\ }}_{\hspace{-1mm}(-1)3}
\ =\ \frac{1}{2}\left(
\hat{\hat{\hat{d}\ }}^\dagger_{\hspace{-1mm}(-1)}\; 
\hat{\hat{\hat{d}\ }}_{\hspace{-1mm}3}
- \hat{\hat{\hat{d}\ }}^\dagger_{\hspace{-1mm}3}\; 
\hat{\hat{\hat{d}\ }}_{\hspace{-1mm}(-1)}\right)
\nonumber\\[0.3cm]
&=&\frac{1}{2}\left(\hat{\hat{\hat{\cal H}\;}}_{\hspace{-1mm}(-1)}^\dagger
\hat{\hat{\hat{\cal H}\;}}_{\hspace{-1mm}3}\ -\
\hat{\hat{\hat{\cal H}\;}}^{\hspace{-1mm}\dagger}_{\hspace{-1mm}3}\;
\hat{\hat{\hat{\cal H}\;}}_{\hspace{-1mm}(-1)}\right)
\ = \ -\ i\ \hat{\hat{\hat{\cal H}\;}}^{\hspace{-1mm}\dagger}_{\hspace{-1mm}0}
\hat{\hat{\hat{\cal H}\;}}_{\hspace{-1mm}1}
\hat{\hat{\hat{\cal H}\;}}^{\hspace{-1mm}\dagger}_{\hspace{-1mm}2}
\hat{\hat{\hat{\cal H}\;}}_{\hspace{-1mm}4}\nonumber\\[0.3cm]
&=&2\ \left({\cal H}_{(-1)}\wedge{\cal H}_3\right)^M 
\hat{\hat{\hat{c}\,}}_M
\ =\ C_2({\cal H})_{(-1)3}^{\ \ \ \ \ \ M}\ \hat{\hat{\hat{c}\,}}_M
\ ,\\[0.3cm]
\label{ntwofurtheradd77d}
i\hat{d}_1\hat{d}_2\hat{d}_3&=& 
i\hat{d}_1\stackrel{d}{\displaystyle 
\wedge}\hat{d}_2\stackrel{d}{\displaystyle \wedge}\hat{d}_3
\ =\ -\ \hat{\hat{d}\ }_{\hspace{-1mm}(-1)}
\hat{\hat{d}\ }_{\hspace{-1mm}4}\nonumber\\[0.3cm]
&=&\hat{\hat{\hat{d}\ }}_{\hspace{-1mm}(-1)4}
\ =\ \frac{1}{2}\left(
\hat{\hat{\hat{d}\ }}^\dagger_{\hspace{-1mm}(-1)}\; 
\hat{\hat{\hat{d}\ }}_{\hspace{-1mm}4}
- \hat{\hat{\hat{d}\ }}^\dagger_{\hspace{-1mm}4}\; 
\hat{\hat{\hat{d}\ }}_{\hspace{-1mm}(-1)}\right)
\nonumber\\[0.3cm]
&=&\frac{1}{2}\left(\hat{\hat{\hat{\cal H}\;}}_{\hspace{-1mm}(-1)}^\dagger
\hat{\hat{\hat{\cal H}\;}}_{\hspace{-1mm}4}\ -\
\hat{\hat{\hat{\cal H}\;}}^{\hspace{-1mm}\dagger}_{\hspace{-1mm}4}\;
\hat{\hat{\hat{\cal H}\;}}_{\hspace{-1mm}(-1)}\right)
\ = \ i\ \hat{\hat{\hat{\cal H}\;}}^{\hspace{-1mm}\dagger}_{\hspace{-1mm}0}
\hat{\hat{\hat{\cal H}\;}}_{\hspace{-1mm}1}
\hat{\hat{\hat{\cal H}\;}}^{\hspace{-1mm}\dagger}_{\hspace{-1mm}2}
\hat{\hat{\hat{\cal H}\;}}_{\hspace{-1mm}3}\nonumber\\[0.3cm]
&=&2\ \left({\cal H}_{(-1)}\wedge{\cal H}_4\right)^M \hat{\hat{\hat{c}\ }}_M
\ =\ C_2({\cal H})_{(-1)4}^{\ \ \ \ \ \ M}\ \hat{\hat{\hat{c}\,}}_M\ .
\end{eqnarray}

\paragraph{III.\ Final result}\hspace{8cm}\\
\nopagebreak
\addcontentsline{toc}{subsubsection}{\ \ \ \ \ \ \ \ \ \  
III.\ Final result}
Summarizing the above study of two-mode nonlinear Bogolyubov-Valatin 
transformations we can say that these can be studied
in terms of nonlinear basis transformations of the Clifford algebra
$C(0,4)$. The 15-dimensional space of nontrivial operators 
(the unit operator omitted) in the fermionic Fock space can by spanned
in terms of biparavectors. In terms of biparavectors of the Clifford
algebra $C(0,5)$, nonlinear Bogolyubov-Valatin transformations
can be described
by means of the equation [$L_m = {\cal H}_m$ are the row vectors 
with row number $m$ of the matrix $L = {\cal H}\in SO(6;\mathbb{R})$.]
\begin{eqnarray}
\label{ntwofurtheradd78}
\hat{\hat{\hat{d}\ }}_M&=&
{\sf U}\; \hat{\hat{\hat{c}\,}}_M\; {\sf U}^\dagger\ =\ 
\chi_M^{\ \ N}\ \hat{\hat{\hat{c}\,}}_N
\nonumber\\[0.3cm]
&=&2\left({\cal H}_{m_1}\wedge{\cal H}_{m_2} \right)^N
\ \hat{\hat{\hat{c}\,}}_N\ =\
2\left(L_{m_1}\wedge L_{m_2} \right)^N
\ \hat{\hat{\hat{c}\,}}_N
\ =\ C_2(L)_M^{\ \ N}\ \hat{\hat{\hat{c}\,}}_N\ .\ \ \ \
\end{eqnarray}
In a more compact notation we can write
\begin{eqnarray}
\label{ntwofurtheradd78b}
\hat{\hat{\hat{D}\, }}&=&
{\sf U}\; \hat{\hat{\hat{C}\,}}\; {\sf U}^\dagger\ =\ 
\chi\ \hat{\hat{\hat{C}\,}}\ ,
\end{eqnarray}
where $\hat{\hat{\hat{D}\ }}$ and $\hat{\hat{\hat{C}\, }}$
are column vectors with biparavector components (To achieve a
more compact graphical display, we give here the 
Hermitian conjugate of $\hat{\hat{\hat{C}\, }}$.):
\begin{eqnarray}
\label{ntwofurtheradd79}
\hat{\hat{\hat{C}\hspace{0.5mm}}}^\dagger
&=&\left(\hat{\hat{\hat{c}\hspace{0.3mm}}}^\dagger_{(-1)0},
\hat{\hat{\hat{c}\hspace{0.3mm}}}^\dagger_{(-1)1},
\hat{\hat{\hat{c}\hspace{0.3mm}}}^\dagger_{(-1)2},
\hat{\hat{\hat{c}\hspace{0.3mm}}}^\dagger_{(-1)3},
\hat{\hat{\hat{c}\hspace{0.3mm}}}^\dagger_{(-1)4},\right.\nonumber\\[0.3cm]
&&\ \ \ \left.\hat{\hat{\hat{c}\hspace{0.3mm}}}^\dagger_{01},
\hat{\hat{\hat{c}\hspace{0.3mm}}}^\dagger_{02},
\hat{\hat{\hat{c}\hspace{0.3mm}}}^\dagger_{03},
\hat{\hat{\hat{c}\hspace{0.3mm}}}^\dagger_{04},
\hat{\hat{\hat{c}\hspace{0.3mm}}}^\dagger_{12},
\hat{\hat{\hat{c}\hspace{0.3mm}}}^\dagger_{13},
\hat{\hat{\hat{c}\hspace{0.3mm}}}^\dagger_{14},
\hat{\hat{\hat{c}\hspace{0.3mm}}}^\dagger_{23},
\hat{\hat{\hat{c}\hspace{0.3mm}}}^\dagger_{24},
\hat{\hat{\hat{c}\hspace{0.3mm}}}^\dagger_{34}\right)
\nonumber\\[0.3cm]
&=&\left(i\hat{c}_1\hat{c}_2\hat{c}_3\hat{c}_4,
i\hat{c}_2\hat{c}_3\hat{c}_4,-i\hat{c}_1\hat{c}_3\hat{c}_4,
i\hat{c}_1\hat{c}_2\hat{c}_4,-i\hat{c}_1\hat{c}_2\hat{c}_3,
\right.\nonumber\\[0.3cm]
&&\ \ \ \left.
\hat{c}_1,\hat{c}_2,\hat{c}_3,\hat{c}_4,
\hat{c}_1\hat{c}_2,\hat{c}_1\hat{c}_3,\hat{c}_1\hat{c}_4,
\hat{c}_2\hat{c}_3,\hat{c}_2\hat{c}_4,\hat{c}_3\hat{c}_4\right)
\end{eqnarray}
and $\chi$ denotes the $15\times 15$-matrix
$\chi =  C_2(L)$. In component notation, we have
\begin{eqnarray}
\label{ntwofurtheradd78c}
\chi_{M\; N}\ =\  C_2(L)_{M\; N}
&=&L_{m_1 n_1}\; L_{m_2 n_2}\ -\ L_{m_1 n_2}\; L_{m_2 n_1}\ .
\end{eqnarray}
Finally, as a special case of eq.\ (\ref{ntwofurtheradd78})
we display again eq.\ (\ref{ntwofurtheradd65b}):
\begin{eqnarray}
\label{ntwofurtheradd65bfinal}
\hat{d}_k&=&-\ \chi_ {0k}^{\ \ N}\ \hat{\hat{\hat{c}\,}}_N\ =\
 -\ C_2(L)_{0k}^{\ \ \ N}\ \hat{\hat{\hat{c}\,}}_N
\end{eqnarray}
expressing the new (transformed) generators of the Clifford algebra 
$C(0,4)$ in terms of the original (untransformed) biparavectors.\\

Eq.\ (\ref{ntwofurtheradd78}) can be reformulated in a different way.
Consider now $\hat{\hat{\hat{D}\ }}$ and $\hat{\hat{\hat{C}\, }}$ not 
as column vectors in the 15-dimensional biparavector space
$\bigwedge^2 (V_6)$ but rather as 
antisymmetric $6\times 6$ matrices with biparavector entries for which
we now write $\hat{\hat{\hat{\cal D}\ }}$ and $\hat{\hat{\hat{\cal C}\, }}$,
respectively. For example,
the antisymmetric matrix $\hat{\hat{\hat{\cal D}\ }}$ has the matrix elements
\begin{eqnarray}
\label{ntwofurtheradd81}
\hat{\hat{\hat{d}\ }}_{\hspace{-1mm}kl}&=&
\frac{1}{2}\ \left(\hat{\hat{\hat{d}\ }}^\dagger_{\hspace{-1mm}k} 
\hat{\hat{\hat{d}\ }}_{\hspace{-1mm}l}
- \hat{\hat{\hat{d}\ }}^\dagger_{\hspace{-1mm}l} 
\hat{\hat{\hat{d}\ }}_{\hspace{-1mm}k}\right).
\end{eqnarray}
Eq.\ (\ref{ntwofurtheradd78}) then equivalently reads
\begin{eqnarray}
\label{ntwofurtheradd82}
\hat{\hat{\hat{\cal D}\ }}&=&L\ \hat{\hat{\hat{\cal C}\, }} L^T\ .
\end{eqnarray}
Results corresponding to eq.\ (\ref{ntwofurtheradd82})
have been obtained earlier by Hristev, \citep{hris5}, ext.\ French
version in Rev.\ Roum.\ Math.\ Pures Appl., p.\ 637, eq.\ (15.7), 
\citep{hris6}, p.\ 176, eq.\ (4.1), 
ten Kate \citep{tenk1}, p.\ 183, eq.\ (9), and 
Buchdahl \citep{buch1}, p.\ 363, Sec.\ 8.\\

As the matrices $L$ and $-L$ describe the same Bogolyubov-Valatin 
transformation (i.e., the same matrix $\chi$) the group of two-mode
nonlinear Bogolyubov-Valatin transformations is equivalent to
the group $SO(6;\mathbb{R})/\mathbb{Z}_2$ 
(cf.\ also \citep{tenk1}, specifically Sec.\ II and Theorem 1 therein;
for related considerations see
\citep{penr1}, Vol.\ 2, Sec.\ 9.4, p.\ 316, \citep{klot1}, 
\citep{lauf1}, Sec.\ 5.3, p.\ 162).\\

\paragraph{IV.\ Further analysis}\hspace{8cm}\\
\nopagebreak
\addcontentsline{toc}{subsubsection}{\ \ \ \ \ \ \ \ \ \  
IV.\ Further analysis}
On the basis of the equation [cf.\ the eqs.\ (\ref{ntwofurtheradd65a}),
(\ref{ntwofurtheradd65bfinal})] 
\begin{eqnarray}
\label{ntwofurtheradd65athi}
\hat{d}_k&=&-\ \chi_{0k}^{\ \ M}\ \hat{\hat{\hat{c}\,}}_M
\end{eqnarray}
and taking into account the eq.\ (\ref{ntwofurtheradd63za}), 
we can quickly rederive and reformulate
the expression for the anticommutator (\ref{ntwofurther44}). We find
\begin{eqnarray}
\label{ntwofurther44b}
\hat{d}_k \hat{d}_l&=&
- \chi_{0k}^{\ \ M}\ \chi_{0l\ M}\ \openone_4\ -\
\sum_{m=-1}^4 \chi_{0k}^{\ \ m n_1} \chi_{0l}^{\ \ m n_2}\
\hat{\hat{\hat{c}\,}}_{n_1 n_2}
\nonumber\\[0.3cm]
&&\ -\ i\ \chi_{0k}^{\ \ M} \chi_{0l}^{\ \ N}\ 
\epsilon_{MN}^{\ \ \ \ P}\ 
\hat{\hat{\hat{c}\,}}_P
\end{eqnarray}
and, consequently,
\begin{eqnarray}
\label{ntwofurther44c}
\left\{\hat{d}_k,\hat{d}_l\right\}&=&
\hat{d}_k \hat{d}_l\ +\ \hat{d}_l \hat{d}_k
\stackrel{\displaystyle !}{=}\ 2\ g_{kl}\ \openone_4
\ =\ -2\ \delta_{kl}\ \openone_4\nonumber\\[0.3cm]
&=&-\ 2\ \chi_{0k}^{\ \ M}\ \chi_{0l\ M}\ \openone_4
\ -\ 2 i\ \chi_{0k}^{\ \ M} \chi_{0l}^{\ \ N}\ 
\epsilon_{MN}^{\ \ \ \ P}\ 
\hat{\hat{\hat{c}\,}}_P\ .
\end{eqnarray}
The eqs.\ (\ref{ntwofurther45a})-(\ref{ntwofurther45e})
can be written compactly as
\begin{eqnarray}
\label{ntwofurther44d}
\chi_{0k}^{\ \ M} \chi_{0l\ M}&=&\delta_{kl}\ ,\\[0.3cm]
\label{ntwofurther44e}
\chi_{0k}^{\ \ M} \chi_{0l}^{\ \ N}\ \epsilon_{MNP}&=&0\ .
\end{eqnarray}
Eq.\ (\ref{ntwofurther44b}) now reads
\begin{eqnarray}
\label{ntwofurther44f}
\hat{d}_k \hat{d}_l&=&\ -\
\sum_{m=-1}^4 \chi_{0k}^{\ \ m n_1} \chi_{0l}^{\ \ m n_2}\
\hat{\hat{\hat{c}\,}}_{n_1 n_2}\ -\ \delta_{kl}\ \openone_4\\[0.3cm]
\label{ntwofurther44g}
&=&- \chi_{kl}^{\ \ N}\ \hat{\hat{\hat{c}\,}}_N
\ -\ \delta_{kl}\ \openone_4\ ,\\[0.3cm]
\label{ntwofurther44h}
\chi_{kl}^{\ \ N}&=&
\sum_{m=-1}^4 \left(\chi_{0k}^{\ \ m n_1} \chi_{0l}^{\ \ m n_2}
- \chi_{0k}^{\ \ m n_2} \chi_{0l}^{\ \ m n_1}\right)\ .
\end{eqnarray}
Introducing the complex vectors 
$\left({\bf e}^\prime_k\right)^T = \left(
\kappa^{(1\vert 0)}_k,
\kappa^{(0\vert 1)}_k,
\kappa^{(2\vert 0)}_k,
\kappa^{(0\vert 2)}_k,
\kappa^{(1\vert 1)}_k,
\kappa^{(1,2\vert 0)}_k,\right.$\hfill\  \linebreak 
$\left.
\kappa^{(1\vert 2)}_k,
\kappa^{(2\vert 1)}_k,
\kappa^{(0\vert 1,2)}_k,
\kappa^{(2\vert 2)}_k,
\kappa^{(1,2\vert 1)}_k,
\kappa^{(1\vert 1,2)}_k,
\kappa^{(1,2\vert 2)}_k,
\kappa^{(2\vert 1,2)}_k,
\kappa^{(1,2\vert 1,2)}_k\right)$, ($k = 1,2$), i.e., \hfill

\noindent
${\rm Re}\; {\bf e}^\prime_{k\ M} = \chi_{0(2k-1)\ M}$,
${\rm Im}\; {\bf e}^\prime_{k\ M} = \chi_{0(2k)\ M}$)
the eqs.\ (\ref{ntwofurther44d}), (\ref{ntwofurther44e}) can be 
written as
\begin{eqnarray}
\label{ntwofurther544d}
\left({\bf e}^\prime\right)^T_k\ \overline{{\bf e}^\prime}_l
&=&2\ \delta_{kl}\ ,\\[0.3cm]
\label{ntwofurther644d}
\left({\bf e}^\prime\right)^T_k\ {\bf e}^\prime_l
&=&0\ ,\\[0.3cm]
\label{ntwofurther544e}
{\bf e}_k^{\prime\ M}\ \overline{{\bf e}_l^{\prime\ N}}
\ \epsilon_{MNP}&=&0\ ,\\[0.3cm]
\label{ntwofurther644e}
{\bf e}_k^{\prime\ M}\ {\bf e}_l^{\prime\ N}\ \epsilon_{MNP}&=&0\ .
\end{eqnarray}
Consequently, the 15-component complex 
vectors ${\bf e}^\prime_1$, ${\bf e}^\prime_2$
should be two orthogonal isotropic vectors of length $\sqrt{2}$ fulfilling the
additional conditions (\ref{ntwofurther544e}), (\ref{ntwofurther644e}).\\

Eq.\ (\ref{ntwofurther44d}) which is the generalization of the 
condition (\ref{origbv}) characteristic
for the original (linear) Bogolyubov-Valatin transformations
can be understood as an orthonormality
condition in the 15-dimensional biparavector space $\bigwedge^2 (V_6)$ and 
eq.\ (\ref{ntwofurther44e}) is (for $k=l$) a decomposability condition.
Consequently, any set of four operators $\hat{d}_k$ fulfilling the 
Clifford algebra analogue of the canonical anticommutation 
relations should be a set of four 
decomposable orthonormal biparavectors. A related result has been 
obtained earlier (in a somewhat more general form, i.e., for pentades) by
Haantjes, \citep{haan1}, p.\ 51, stelling 5 [proposition 5], and 
a corresponding comment can also be found in ref.\ \citep{buch1}, \S 14,
p.\ 269, above of eq.\ (14.8). For a further discussion of these
aspects see Subsec.\ \ref{bibidis}.\\

The eqs.\ (\ref{ntwofurther44d}), (\ref{ntwofurther44e}) can
be further generalized. Inserting the equation 
[cf.\ eq.\ (\ref{ntwofurtheradd78})] 
\begin{eqnarray}
\label{ntwofurtheradd78dpre}
\hat{\hat{\hat{d}\ }}_M&=&\chi_M^{\ \ N}\ \hat{\hat{\hat{c}\,}}_N
\end{eqnarray}
into the anticommutator  (\ref{ntwofurtheradd63zc}) we find
in generalization of the eq.\ (\ref{ntwofurther44d})
\begin{eqnarray}
\label{ntwofurtheradd78d}
\chi_{p_1p_2}^{\ \ \ \ M} \chi_{q_1q_2\ M}&=&
\delta_{p_1 q_1}\delta_{p_2 q_2}\ -\
\delta_{p_1 q_2}\delta_{p_2 q_1}\ ,
\end{eqnarray}
and also 
\begin{eqnarray}
\label{ntwofurtheradd78db}
\chi_{\ \ p_1p_2}^M\; \chi_{M\ q_1q_2}&=&
\delta_{p_1 q_1}\delta_{p_2 q_2}\ -\
\delta_{p_1 q_2}\delta_{p_2 q_1}\ .
\end{eqnarray}
The generalizations of eq.\ (\ref{ntwofurther44e}) read
\begin{eqnarray}
\label{ntwofurtheradd78e}
\chi_J^{\ \ M}\; \chi_K^{\ \ N}\; \chi_L^{\ \ P}\ \epsilon_{MNP}
&=&\epsilon_{JKL}\ ,\\[0.3cm]
\label{ntwofurtheradd78f}
\epsilon_{MNP}\ \chi^M_{\ J}\; \chi^N_{\ K}\; \chi^P_{\ L}
&=&\epsilon_{JKL}\ .
\end{eqnarray}
If all the six indices $j_1$, $j_2$, $k_1$, $k_2$, $l_1$, and $l_2$ are 
chosen pairwise differently, the eqs.\ (\ref{ntwofurtheradd78e}),
(\ref{ntwofurtheradd78f}) can be transformed to read
\begin{eqnarray}
\label{ntwofurtheradd78g}
\det\chi&=&1\ .
\end{eqnarray}
This sharpens the condition $\det\chi = \pm 1$ that can be derived 
from the eqs.\  (\ref{ntwofurtheradd78d}), (\ref{ntwofurtheradd78db}).
If two of the six indices $j_1$, $j_2$, $k_1$, $k_2$, $l_1$, and $l_2$ are
equal, say $k_1 = l_1 = j$ (The cases $r_1 = r_2$ are trivial and need
not to be discussed.), the eqs.\ (\ref{ntwofurtheradd78e}),
(\ref{ntwofurtheradd78f}) can be transformed to read 
[using the orthogonality condition (\ref{ntwofurtheradd78d}),
(\ref{ntwofurtheradd78db})]
\begin{eqnarray}
\label{ntwofurther44efin}
\chi_{jk}^{\ \ M} \chi_{jl}^{\ \ N}\ \epsilon_{MNP}&=&0\ .
\end{eqnarray}

Finally, we study the connection between the conditions 
(\ref{ntwofurtheradd78d}), (\ref{ntwofurtheradd78db}),  
and (\ref{ntwofurtheradd78e}), (\ref{ntwofurtheradd78f}).
Using the known expression for $\chi$ in terms of $L$ 
[eq.\ (\ref{ntwofurtheradd78c})], the 
eq.\ (\ref{ntwofurtheradd78d}) can be written as
\begin{eqnarray}
\label{ntwofurtheradd78dc}
C_2\left(L\right)C_2\left(L^T\right)&=&\openone_{15}\ .
\end{eqnarray}
Taking into account the compound matrix relation (Laplace expansion of
a determinant; for the notation see Appendix \ref{appcomp})
\begin{eqnarray}
\label{ntwofurtheradd78dd}
C_2\left(L\right)C_4\left(L\right)^\star&=&\openone_{15}
\end{eqnarray}
we can immediately conclude that 
\begin{eqnarray}
\label{ntwofurtheradd78de}
C_2\left(L\right)&=&C_4\left(L^T\right)^\star\ .
\end{eqnarray}
This equation can also be written in the following form
\begin{eqnarray}
\label{ntwofurtheradd78df}
\chi_{PQ}&=&\frac{1}{3!}\ \epsilon_P^{\ \ KM}\ \epsilon_Q^{\ \ LN}\
\chi_{KL}\ \chi_{MN}
\end{eqnarray}
which can also be derived from the eqs.\ (\ref{ntwofurtheradd78e}),
(\ref{ntwofurtheradd78f}) using the eqs.\ (\ref{ntwofurtheradd78d}),
(\ref{ntwofurtheradd78db}).\\

\section{\label{diagham}Diagonalizing Hamiltonians}

Relying on the insight obtained in the previous
section into the structure of nonlinear
Bogolyubov-Valatin transformations, in this section we will study
the diagonalization of certain fermion and spin Hamiltonians. 
In principle, the methods of this section can also be applied, with 
certain modifications, to other quantities of physical interest, for example,
the density matrix which is widely being investigated in quantum information
theory.\\

\subsection{\label{twoferm}Diagonalizing two-fermion Hamiltonians}

Let us now look at an arbitrary two-fermion Hamiltonian $H^\prime$. 
\begin{eqnarray}
\label{ntwofurtheradd100}
H^\prime&=&h^{(0\vert 0)}\ \openone_4\ +\
h^{(1\vert 0)}\ \hat{a}_1^\dagger\ +\
h^{(2\vert 0)}\ \hat{a}_2^\dagger\ +\
h^{(0\vert 1)}\ \hat{a}_1\ +\
h^{(0\vert 2)}\ \hat{a}_2\nonumber\\[0.3cm]
&&\ +\ h^{(1,2\vert 0)}\ \hat{a}_1^\dagger\hat{a}_2^\dagger\ +\
h^{(1\vert 1)}\ \hat{a}_1^\dagger\hat{a}_1\ +\
h^{(1\vert 2)}\ \hat{a}_1^\dagger\hat{a}_2\nonumber\\[0.3cm]
&&\ +\ h^{(2\vert 1)}\ \hat{a}_2^\dagger\hat{a}_1
\ +\ h^{(2\vert 2)}\ \hat{a}_2^\dagger\hat{a}_2
\ +\ h^{(0\vert 1,2)}\ \hat{a}_1\hat{a}_2\nonumber\\[0.3cm]
&&\ +\ h^{(1,2\vert 1)}\ \hat{a}_1^\dagger\hat{a}_2^\dagger\hat{a}_1
\ +\ h^{(1,2\vert 2)}\ \hat{a}_1^\dagger\hat{a}_2^\dagger\hat{a}_2
\nonumber\\[0.3cm]
&&
\ +\ h^{(1\vert 1,2)}\ \hat{a}_1^\dagger\hat{a}_1\hat{a}_2
\ +\ h^{(2\vert 1,2)}\ \hat{a}_2^\dagger\hat{a}_1\hat{a}_2
\ +\ h^{(1,2\vert 1,2)}\ 
\hat{a}_1^\dagger\hat{a}_2^\dagger\hat{a}_1\hat{a}_2
\end{eqnarray}
We assume it to be hermitian ($H^\prime =H^{\prime\dagger}$). 
From the hermiticity condition the following 10 relations derive.
\begin{eqnarray}
\label{ntwofurtheradd101a}
h^{(0\vert 0)}&=&\overline{h^{(0\vert 0)}}\\[0.3cm]
\label{ntwofurtheradd101b}
h^{(1\vert 0)}&=&\overline{h^{(0\vert 1)}}\\[0.3cm]
\label{ntwofurtheradd101c}
h^{(2\vert 0)}&=&\overline{h^{(0\vert 2)}}\\[0.3cm]
\label{ntwofurtheradd101d}
h^{(1,2\vert 0)}&=&-\ \overline{h^{(0\vert 1,2)}}\\[0.3cm]
\label{ntwofurtheradd101e}
h^{(1\vert 1)}&=&\overline{h^{(1\vert 1)}}\\[0.3cm]
\label{ntwofurtheradd101f}
h^{(1\vert 2)}&=&\overline{h^{(2\vert 1)}}\\[0.3cm]
\label{ntwofurtheradd101g}
h^{(2\vert 2)}&=&\overline{h^{(2\vert 2)}}\\[0.3cm]
\label{ntwofurtheradd101h}
h^{(1,2\vert 1)}&=&-\ \overline{h^{(1\vert 1,2)}}\\[0.3cm]
\label{ntwofurtheradd101i}
h^{(1,2\vert 2)}&=&-\ \overline{h^{(2\vert 1,2)}}\\[0.3cm]
\label{ntwofurtheradd101j}
h^{(1,2\vert 1,2)}&=&\overline{h^{(1,2\vert 1,2)}}
\end{eqnarray}
Taking into account the equations 
(\ref{ntwofurtheradd101a})-(\ref{ntwofurtheradd101j})
(in particular, that $h^{(0\vert 0)}$, $h^{(1\vert 1)}$,
$h^{(2\vert 2)}$, $h^{(1,2\vert 1,2)}$ are real), eq.\
(\ref{ntwofurtheradd100}) reads
\begin{eqnarray}
\label{ntwofurtheradd102a}
H^\prime&=&\left(h^{(0\vert 0)}\ +\ \frac{1}{2}\ h^{(1\vert 1)}
\ +\ \frac{1}{2}\ h^{(2\vert 2)}
\ -\ \frac{1}{4}\ h^{(1,2\vert 1,2)}\right)\ \openone_4\nonumber\\[0.3cm]
&&\ \ -\ i\ {\rm Re}\left( h^{(1\vert 0)} +
\frac{1}{2}\ h^{(1,2\vert 2)}\right)\ \hat{c}_1
\ -\ i\ {\rm Im}\left( h^{(1\vert 0)} +
\frac{1}{2}\ h^{(1,2\vert 2)}\right)\ \hat{c}_2\nonumber\\[0.3cm]
&&\ \ -\ i\ {\rm Re}\left( h^{(2\vert 0)} -
\frac{1}{2}\ h^{(1,2\vert 1)}\right)\ \hat{c}_3
\ -\ i\ {\rm Im}\left( h^{(2\vert 0)} -
\frac{1}{2}\ h^{(1,2\vert 1)}\right)\ \hat{c}_4\nonumber\\[0.3cm]
&&\ \ -\ \frac{i}{2}\ h^{(1\vert 1)}\ \hat{c}_1\hat{c}_2
\ -\ \frac{i}{2}\ {\rm Im}\left( h^{(1,2\vert 0)} +
h^{(1\vert 2)}\right)\ \hat{c}_1\hat{c}_3
\nonumber\\[0.3cm]
&&\ \ +\ \frac{i}{2}\ {\rm Re}\left( h^{(1,2\vert 0)} -
h^{(1\vert 2)}\right)\ \hat{c}_1\hat{c}_4
\ +\ \frac{i}{2}\ {\rm Re}\left(  h^{(1,2\vert 0)} +
h^{(1\vert 2)}\right)\ \hat{c}_2\hat{c}_3\nonumber\\[0.3cm]
&&\ \ +\ \frac{i}{2}\ {\rm Im}\left( h^{(1,2\vert 0)} -
h^{(1\vert 2)}\right)\ \hat{c}_2\hat{c}_4
\ -\ \frac{i}{2}\ h^{(2\vert 2)}\ \hat{c}_3\hat{c}_4
\nonumber\\[0.3cm]
&&\ \ -\ \frac{i}{2}\ {\rm Re}\;h^{(1,2\vert 1)}\
i\hat{c}_1\hat{c}_2\hat{c}_3
\ -\ \frac{i}{2}\ {\rm Im}\;h^{(1,2\vert 1)}\
i\hat{c}_1\hat{c}_2\hat{c}_4\nonumber\\[0.3cm]
&&\ \ +\ \frac{i}{2}\ {\rm Re}\;h^{(1,2\vert 2)}\
i\hat{c}_1\hat{c}_3\hat{c}_4
\ +\ \frac{i}{2}\ {\rm Im}\;h^{(1,2\vert 2)}\
i\hat{c}_2\hat{c}_3\hat{c}_4\nonumber\\[0.3cm]
&&\ \ -\ \frac{i}{4}\ h^{(1,2\vert 1,2)}\ 
i\hat{c}_1\hat{c}_2\hat{c}_3\hat{c}_4\\[0.3cm]
\label{ntwofurtheradd102b}
&=&\left(h^{(0\vert 0)}\ +\ \frac{1}{2}\ h^{(1\vert 1)}
\ +\ \frac{1}{2}\ h^{(2\vert 2)}
\ -\ \frac{1}{4}\ h^{(1,2\vert 1,2)}\right)\ \openone_4\nonumber\\[0.3cm]
&&\ -\ \frac{i}{4}\ 
{\rm tr}_{\, V_6}\left( Y \hat{\hat{\hat{\cal C}\, }}\right)\ .
\end{eqnarray}
Here, the subscript $V_6$ indicates that the trace operation is
carried out with respect to the six-dimensional paravector space $V_6$ and  
the antisymmetric matrix $Y$ has the explicit form
(To simplify the display we have omitted the lower triangle
matrix elements.)
\begin{landscape}
\begin{eqnarray}
\label{ntwofurtheradd103}
&&\hspace{-1.5cm}Y\ =\ -\ Y^T\ =\nonumber\\[1.3cm]
&&\left(
\begin{array}{*{6}{c}}
0&
- \frac{\displaystyle h^{(1,2\vert 1,2)}}{\displaystyle 2}&
- {\rm Im}\;h^{(1,2\vert 2)}&
{\rm Re}\;h^{(1,2\vert 2)}&
{\rm Im}\;h^{(1,2\vert 1)}&
- {\rm Re}\;h^{(1,2\vert 1)}\\[0.3cm]
.&
0&
{\rm Re}\left(2 h^{(1\vert 0)} + h^{(1,2\vert 2)}\right)&
{\rm Im}\left(2 h^{(1\vert 0)} + h^{(1,2\vert 2)}\right)&
{\rm Re}\left(2 h^{(2\vert 0)} - h^{(1,2\vert 1)}\right)&
{\rm Im}\left(2 h^{(2\vert 0)} - h^{(1,2\vert 1)}\right)\\[0.3cm]
.&
.&
0&
h^{(1\vert 1)}&
{\rm Im}\left( h^{(1,2\vert 0)} + h^{(1\vert 2)}\right)&
- {\rm Re}\left( h^{(1,2\vert 0)} - h^{(1\vert 2)}\right)\\[0.3cm]
.&
.&
.&
0&
- {\rm Re}\left(  h^{(1,2\vert 0)} + h^{(1\vert 2)}\right)&
- {\rm Im}\left( h^{(1,2\vert 0)} - h^{(1\vert 2)}\right)\\[0.3cm]
.&
.&
.&
.&
0&
h^{(2\vert 2)}\\[0.3cm]
.&
.&
.&
.&
.&0
\end{array}
\right)\ . \ \ \ \ 
\end{eqnarray}
\end{landscape}
\noindent
Eq.\ (\ref{ntwofurtheradd102b}) expresses the Hamiltonian 
(\ref{ntwofurtheradd100}) in terms of a biparavector 
[${\rm tr}_{\, V_6}\left( Y \hat{\hat{\hat{\cal C}\, }}\right)$; plus some
constant]. This biparavector stands in an one-to-one correspondence
to an antisymmetric matrix (exactly as this is the case for
ordinary bivectors). The further analysis of the 
Hamiltonian $H^\prime$ will be based on this correspondence.
It should be mentioned here that recently Uskov and Rau 
\citep{usko1}, Appendix B, p.\ 022331-8 [see the first equation below from
eq.\ (B.1)] have given a representation of the Hamiltonian
analogous to eq.\ (\ref{ntwofurtheradd102a}). To explicitly see
that both formulations are equivalent one must rely on our 
equations (\ref{ntwofurtheradd807a})-(\ref{ntwofurtheradd807f})
given in the Appendix \ref{appc}.\\

The matrix $Y$ can be brought to the standard  block diagonal form 
given by the matrix $Z$ 
\begin{eqnarray}
\label{ntwofurtheradd104}
Z\ =\ -\ Z^T&=&
\left(
\begin{array}{*{6}{c}}
0&
\nu_{(-1)0}&
0&
0&
0&
0\\[0.3cm]
-\nu_{(-1)0}&
0&
0&
0&
0&
0\\[0.3cm]
0&
0&
0&
\nu_{12}&
0&
0\\[0.3cm]
0&
0&
-\nu_{12}&
0&
0&
0\\[0.3cm]
0&
0&
0&
0&
0&
\nu_{34}\\[0.3cm]
0&
0&
0&
0&
-\nu_{34}&0
\end{array}
\right)
\end{eqnarray}
by means of an orthogonal transformation $L$: $Z = L\ Y\ L^T$.
Eq.\ (\ref{ntwofurtheradd102b}) then reads
\begin{eqnarray} 
\label{ntwofurtheradd105}
H^\prime&=&\left(h^{(0\vert 0)}\ +\ \frac{1}{2}\ h^{(1\vert 1)}
\ +\ \frac{1}{2}\ h^{(2\vert 2)}
\ -\ \frac{1}{4}\ h^{(1,2\vert 1,2)}\right)\ \openone_4\nonumber\\[0.3cm]
&&\ -\ \frac{i}{4}\ 
{\rm tr}_{\, V_6}\left( Z \hat{\hat{\hat{\cal D}\, }}\right)\\[0.3cm]
&=&\left(h^{(0\vert 0)}\ +\ \frac{1}{2}\ h^{(1\vert 1)}
\ +\ \frac{1}{2}\ h^{(2\vert 2)}
\ -\ \frac{1}{4}\ h^{(1,2\vert 1,2)}\right)\ \openone_4\nonumber\\[0.3cm]
&&\ +\ \frac{i}{2}\ \nu_{(-1)0}\ \hat{\hat{\hat{d}\ }}_{(-1)0}
\ +\ \frac{i}{2}\ \nu_{12}\ \hat{\hat{\hat{d}\ }}_{12}
\ +\ \frac{i}{2}\ \nu_{34}\ \hat{\hat{\hat{d}\ }}_{34}\\[0.3cm]
\label{ntwofurtheradd105b}
&=&\left(h^{(0\vert 0)}\ +\ \frac{1}{2}\ h^{(1\vert 1)}
\ +\ \frac{1}{2}\ h^{(2\vert 2)}
\ -\ \frac{1}{4}\ h^{(1,2\vert 1,2)}\right)\ \openone_4\nonumber\\[0.3cm]
&&\ +\ \frac{i}{2}\ \nu_{(-1)0}\ 
i\hat{d}_1\hat{d}_2\hat{d}_3\hat{d}_4
\ -\ \frac{i}{2}\ \nu_{12}\ \hat{d}_1\hat{d}_2
\ -\ \frac{i}{2}\ \nu_{34}\ \hat{d}_3\hat{d}_4\\[0.3cm]
\label{ntwofurtheradd105c}
&=&\left[h^{(0\vert 0)}\ +\ \frac{1}{2}\ 
\left(h^{(1\vert 1)} - \nu_{12}\right)
\ +\ \frac{1}{2}\ 
\left(h^{(2\vert 2)} - \nu_{34}\right)\right.\nonumber\\[0.3cm]
&&\left.\ -\ \frac{1}{4}\ \left(h^{(1,2\vert 1,2)} + 2 \nu_{(-1)0}\right)
\right]\ \openone_4\nonumber\\[0.3cm]
&&\ \ +\ \nu_{12}\ \hat{b}_1^\dagger\hat{b}_1\ 
\ +\ \nu_{34}\ \hat{b}_2^\dagger\hat{b}_2\ 
\ -\ 2\ \nu_{(-1)0} \ 
\hat{b}_1^\dagger\hat{b}_2^\dagger\hat{b}_1\hat{b}_2
\end{eqnarray}
with $\hat{\hat{\hat{\cal D}\ }} = L\ \hat{\hat{\hat{\cal C}\, }} L^T$
[cf.\ eq.\ (\ref{ntwofurtheradd82})]. The Hamiltonian is 
given here in terms of three (commuting) Cartan elements: 
$\hat{b}_1^\dagger\hat{b}_1$, $\hat{b}_2^\dagger\hat{b}_2$,
$\hat{b}_1^\dagger\hat{b}_2^\dagger\hat{b}_1\hat{b}_2$ [This 
is the maximal number for the group $SO(6;\mathbb{R}) \simeq SU(4)$.].
For a related consideration see ref.\ \citep{birm1}.
In a general situation,
the number of nonvanishing pairs of eigenvalues $\pm i\nu_K$ 
[$K = (-1)0,12,34$] of the (similar) matrices $Y$, $Z$
is called the {\it length} (\citep{marc1}, vol.\ 2, 
Sec.\ 4.1, Exercise 26, pp.\ 52-53), 
{\it rank} (\citep{shaw2}, Vol.\ II, Sec.\ 9.3.5, p.\ 331),
or {\it mass} (\citep{kozl2}, p. 67) of the related 
bi(para)vectors. We will use the term rank to diminish
the risk of any misunderstanding in any physics-related context.
We should point out here that
due to the one-to-one relation between antisymmetric matrices
and bivectors on one hand and the one-to-one relation 
between antisymmetric matrices and biparavectors on the other hand
results available in the literature 
concerning bivectors carry over to biparavectors (we are concerned with)
with little change. The standard orthogonal
decomposition of bivectors\footnote{\citep{scho1}, all eds., 
Chap.\ II, \S 11, pp.\ 35-36,
\citep{west1},
\citep{shaw2}, Vol.\ II, Sec.\ 9.3.5, p.\ 331,
\citep{kozl2},
\citep{mcda1}, Sec.\ 1, p.\ 184,
\citep{kozl1}, p.\ 103 (p.\ 2251 of the English transl.).}
relies on the fact that any antisymmetric matrix can be brought 
to block diagonal form [the diagonal blocks are proportional to
{\tiny $\left(
\begin{array}{*{2}{c}}
0&1\\
-1&0
\end{array}
\right)$}; see, for example, 
\citep{gant1}, Chap. XI, \S 4, various pp.\ in the 
different Russian editions (Vol.\ 2, pp.\ 12-18 of the English
transl.)]. For the general case
of three mutually different pairs of eigenvalues $\pm i\nu_{(-1)0}$,
$\pm i\nu_{12}$, $\pm i\nu_{34}$, of the (similar) matrices $Y$, $Z$
a geometric algebra formalism to calculate these eigenvalues
and the orthogonal transformation $L$ 
to transform the matrix $Y$ into the matrix $Z$ exists
(We will not review it here. Cf.\ \citep{hest1}, Chap.\ 3, Sec.\ 4, p.\ 78,
\citep{hitz1}.). This general situation, for example, will
apply for Hamiltonians of (quasi)fermions which have been obtained
from Jordan-Wigner transformations of two-spin-$\frac{1}{2}$ systems
(Those will be discussed in some detail further below.).
However, in the present subsection primarily we have in mind
the interpretation of $H^\prime$ as the Hamiltonian of 
two (equivalent) fermionic modes. Consequently,
we expect for a physical Hamiltonian that two of the eigenvalue
pairs [with an eye to eq.\ (\ref{ntwofurtheradd105c}) chosen to
be $\pm i\nu_{12}$, $\pm i\nu_{34}$] should be equal $i\nu_{12} = i\nu_{34}$
[i.e., the Hamiltonian should not change under exchange of the 
fermion indices 1 and 2 - this entails further identities for 
the coefficients in eq.\ (\ref{ntwofurtheradd100})]. Consequently, 
eq.\ (\ref{ntwofurtheradd105c}) exhibits a residual $O(2)\otimes O(2)$
symmetry related to the twofold degenerate eigenspaces of the matrices
$Y$, $Z$ related to the eigenvalues $i\nu_{12} = i\nu_{34}$ 
and $-i\nu_{12} = -i\nu_{34}$, and a certain 
arbitrariness exists in constructing the orthogonal transformation $L$
(and the related canonical transformation) \citep{loun}, Sec.\ 17.3,
p.\ 222 (In this reference, this fact is expressed in terms of 
a bivector expansion.).\\

The eigenvalues $\lambda = \pm i\nu_K$ 
[$K = (-1)0,12,34$] of the matrix $Y$ can be calculated by means of 
the characteristic equation (For the trace operation we omit here
the subscript $V_6$ because no misunderstanding may occur.)
\begin{eqnarray} 
\label{ntwofurtheradd107}
0&=&\det\left( Y - \lambda \openone_6\right)\nonumber\\[0.3cm]
&=&\lambda^6\ -\ \frac{1}{2}\left({\rm tr}\; Y^2\right)\ \lambda^4
\ +\ \frac{1}{8}\left[\left({\rm tr}\; Y^2\right)^2 \ -\
2\ {\rm tr}\; Y^4\right]\lambda^2\ +\ \det Y
\end{eqnarray}
which is a cubic equation in terms of $\lambda^2$.
The invariants of the matrices  $Y$, $Z$ entering the above
equation are 
\begin{eqnarray} 
\label{ntwofurtheradd108a}
{\rm tr}\; Y^2&=&{\rm tr}\; Z^2\ =\ 
-2\ \left(\nu^2_{(-1)0} + \nu^2_{12} + \nu^2_{34}\right),\\[0.3cm]
\label{ntwofurtheradd108b}
{\rm tr}\; Y^4&=&{\rm tr}\; Z^4\ =\ 
2\ \left(\nu^4_{(-1)0} + \nu^4_{12} + \nu^4_{34}\right),\\[0.3cm]
\label{ntwofurtheradd108c}
\det Y&=&\det Z\ =\ \left(\nu_{(-1)0}\nu_{12}\nu_{34}\right)^2\ .
\end{eqnarray}
By calculating the discriminant $\Delta$ 
of the equation (\ref{ntwofurtheradd107})
(We do not display here the explicit expression of it, see any
standard reference on cubic equations.) one can check if indeed two
of the eigenvalue pairs $\lambda = \pm i\nu_K$  agree as expected.
If this is the case the eqs.\ 
(\ref{ntwofurtheradd108a})-(\ref{ntwofurtheradd108c}) read
\begin{eqnarray} 
\label{ntwofurtheradd109a}
{\rm tr}\; Y^2&=&{\rm tr}\; Z^2\ =\ 
-2\ \left(\nu^2_{(-1)0} + 2 \nu^2\right),\\[0.3cm]
\label{ntwofurtheradd109b}
{\rm tr}\; Y^4&=&{\rm tr}\; Z^4\ =\ 
2\ \left(\nu^4_{(-1)0} + 2 \nu^4 \right),\\[0.3cm]
\label{ntwofurtheradd109c}
\det Y&=&\det Z\ =\ \left(\nu_{(-1)0}\nu^2\right)^2\ ,
\end{eqnarray}
and one can quickly find from these equations the eigenvalues
$\lambda = \pm i\nu_{(-1)0}$, 
$\lambda = \pm\nu = \pm i\nu_{12} =  \pm i\nu_{34}$
without the need to resort to the standard machinery of solving
a cubic equation in general. From the eqs.\ 
(\ref{ntwofurtheradd109a}), (\ref{ntwofurtheradd109b}) one finds 
for the squared eigenvalues $\lambda^2$
\begin{eqnarray} 
\label{ntwofurtheradd110a}
\nu^2&=&\frac{1}{6}\left(
-{\rm tr}\; Y^2\ \pm\
\sqrt{3\ {\rm tr}\; Y^4  - \frac{1}{2} \left({\rm tr}\; Y^2\right)^2}\right),
\\[0.3cm]
\label{ntwofurtheradd110b}
\nu^2_{(-1)0}&=&\frac{1}{6}\left(
-{\rm tr}\; Y^2\ \mp\ 2
\sqrt{3\ {\rm tr}\; Y^4  - \frac{1}{2} \left({\rm tr}\; Y^2\right)^2}\right)\ .
\end{eqnarray}
Note, that 
$3\; {\rm tr}\; Y^4  - \frac{1}{2} \left({\rm tr}\; Y^2\right)^2 =
4\left(\nu^2 - \nu^2_{(-1)0}\right)^2$. Consequently, in the eqs.\ 
(\ref{ntwofurtheradd110a}), (\ref{ntwofurtheradd110b}) on the r.h.s.\
the upper sign applies to weak (quasi)fermion coupling
($\nu^2  >\nu^2_{(-1)0}$) and the lower sign to strong
(quasi)fermion coupling ($\nu^2 < \nu^2_{(-1)0}$). Which sign applies
can be determined by calculating the determinant (or the Pfaffian)
of the matrix $Y$.\\

One may wonder why the analysis of the 4-level Hamiltonian $H^\prime$ 
[eq.\ (\ref{ntwofurtheradd100})] which can be 
represented by means of a $4\times 4$ matrix leads to a cubic equation
and not to a quartic one. The answer is as follows:
Writing the Hamiltonian $H^\prime$
in terms of a biparavector plus some constant [eq.\ (\ref{ntwofurtheradd102a})]
amounts to splitting the Hamiltonian $H^\prime$ into a traceless part
plus some diagonal term (proportional to $\openone_4$). Consequently, 
as this way one degree of freedom for the eigenvalues of 
the Hamiltonian $H^\prime$ has been separated out, the remaining
characteristic equation for the traceless part of $H^\prime$
is reduced in order by one degree to a cubic equation.\\

Finally, let us dwell on certain general considerations.
Qualitatively, the class of physical Hamiltonians $H^\prime$ (with 
$\nu_{12} = \nu_{34} = \nu$) can have rank 2 ($\nu_{(-1)0} = 0$)
or 3 ($\nu_{(-1)0}\neq 0$) - interpreted in terms of the biparavector(s)
related to it. For the generic case $\nu_{(-1)0}\neq 0$ (i.e., rank 3)
it is clear that it is impossible to write the Hamiltonian $H^\prime$
as the sum of two noninteracting quasiparticle fermion oscillators, i.e.,
it is impossible to define canonical transformations of the fermion 
creation ($\hat{a}_1^\dagger$, $\hat{a}_2^\dagger$) and annihilation 
($\hat{a}_1$, $\hat{a}_2$) operators such a way that the Hamiltonian
$H^\prime$ can be written in terms of noninteracting quasiparticles
[(quasi)fermions]
($\hat{b}_1$, $\hat{b}_2$). However, if 
\begin{eqnarray} 
\label{ntwofurtheradd106}
\det Y \ =\ \left({\rm Pf\; Y}\right)^2&=& 
\det Z\ =\ \left({\rm Pf\; Z}\right)^2\ =\ 
\left(\nu_{(-1)0}\ \nu^2\right)^2\ =\ 0
\end{eqnarray}
(i.e., $\nu_{(-1)0} = 0$) this is possible.\\

We conclude this subsection with some comments on the related literature.
The Hamiltonian (\ref{ntwofurtheradd105c}) with 
$\nu_{12} = \nu_{34} = \nu$ has been studied for fermion systems in
\citep{delb1,delb2,thom1,barr1,thom3,caba1,caba2,caba3}, 
while in the references \citep{gira1jmp,souz1,souz2}
the somewhat more general situation where not necessarily  
$\nu_{12} = \nu_{34}$ is being considered.
Particular versions of the general Hamiltonian (\ref{ntwofurtheradd100})
have been studied in \citep{delb1,delb2,agui1}. 
However, the case studied in \citep{agui1}
can be treated by means of linear Bogolyubov-Valatin transformations.\\

\subsection[Nonlinear spin transformations
\texorpdfstring{--}{-} 
Diagonalizing spin Hamiltonians]{\label{twospin}Nonlinear 
spin transformations\\
-- Diagonalizing spin Hamiltonians}
\subsubsection{\label{twospinone}Two-spin-\texorpdfstring{$\frac{1}{2}$}{1/2} 
systems}

It is well-known that fermion systems can be related to spin systems
by means of Jordan-Wigner transformations \citep{jord1}. This, of course,
also applies to the two-fermion system under consideration. 
From a group-theoretical point of view, each elementary 
spin $\left(\frac{1}{2}\right)$ 
corresponds to some $SU(2)$ subgroup of some larger group.
Inasmuch as spin operators (Pauli operators) of different spins 
commute among each other a system of $n$ spins corresponds to 
$n$ pairwise commuting $SU(2)$ subgroups of the larger group related to the 
system under consideration. In our setting, the transition 
from fermion operators to spin (Pauli) operators is related to 
the choice of two commuting $\mathsf{su}(2)$ Lie subalgebras within the
$\mathsf{su}(4)$ Lie algebra of the two-fermion system. By means of this
choice a two-fermion system is mapped to a two-spin-$\frac{1}{2}$
system. Such systems have been studied from various points of view 
in the past 
\citep{banw1,erdo1,erdo2,patt1,luca1,kuip1,maks1,laws1,kubl1}, 
\citep{wagn1}, Chap.\ 2, Sec.\ 13, pp.\ 52-56, Table 8, pp.\ 294-299, 
\citep{cado1}, \citep{town1},  Chap.\ 5, p.\ 120,
\citep{sire1,bowd1,rau1,vala3,efre1,zhan1,have1,rau2,bagr1,jord10,volk1,bald1,seba1}.
We leave here aside the broad range of studies performed in recent
years concerning the problem of entanglement in 
two-spin-$\frac{1}{2}$ systems -- the "harmonic oscillator of quantum
information theory". It is clear that there
are many possible Jordan-Wigner transformations (This question seems
not to have been studied systematically in the literature so far.).
Here, we will follow \citep{sire1} with some modifications arising
from certain esthetic considerations related to the equations
(\ref{ntwofurtheraddz66a})-(\ref{ntwofurtheraddz66d}). 
We want to choose a version 
of the Jordan-Wigner transformation that is fairly symmetric with
respect to the mode number indices involved. 
The authors of ref.\ \citep{sire1} define (p.\ 13835, above of
eq.\ (5); $[S_k^x,S_k^y] = i S_k^z$, $S_k^\pm = S_k^x \pm i S_k^y$)
\begin{eqnarray} 
\label{ntwofurtheradd901a}
S_1^+&=&\hat{a}_1^\dagger\ \left[\openone_4 - (1+i)\ 
\hat{a}_2^\dagger \hat{a}_2\right]\ ,\\[0.3cm]
\label{ntwofurtheradd901b}
S_1^-&=&\hat{a}_1\ \left[\openone_4 - (1-i)\ \hat{a}_2^\dagger \hat{a}_2\right]
\ ,\\[0.3cm]
\label{ntwofurtheradd901c}
S_1^z&=&\hat{a}_1^\dagger \hat{a}_1\ -\ \frac{1}{2}\ \openone_4\ ,\\[0.3cm]
\label{ntwofurtheradd901d}
S_2^+&=&\hat{a}_2^\dagger\ \left[\openone_4 - (1-i)\ 
\hat{a}_1^\dagger \hat{a}_1\right]\ ,\\[0.3cm]
\label{ntwofurtheradd901e}
S_2^-&=&\hat{a}_2\ \left[\openone_4 - (1+i)\ \hat{a}_1^\dagger \hat{a}_1\right]
\ ,\\[0.3cm]
\label{ntwofurtheradd901f}
S_2^z&=&\hat{a}_2^\dagger \hat{a}_2\ -\ \frac{1}{2}\ \openone_4\ .
\end{eqnarray}
One can convince oneself without any difficulty that the inverse 
Jordan-Wigner transformation from the spin-$\frac{1}{2}$ (Pauli) operators
to fermion operators is given by the following equations.
\begin{eqnarray} 
\label{ntwofurtheradd902a}
\hat{a}_1&=&-\ i\ S_1^-\ \left[\frac{1}{2} (1+i)\ \openone_4
+ (1-i)\ S_2^z\right]\\[0.3cm]
\label{ntwofurtheradd902b}
\hat{a}_1^\dagger&=&\ \ \ i\ S_1^+\ 
\left[\frac{1}{2} (1-i)\ \openone_4 + (1+i)\ S_2^z\right]\\[0.3cm]
\label{ntwofurtheradd902c}
\hat{a}_2&=&\ \ \ i\ S_2^-\ 
\left[\frac{1}{2} (1-i)\ \openone_4 + (1+i)\ S_1^z\right]\\[0.3cm]
\label{ntwofurtheradd902d}
\hat{a}_2^\dagger&=&
-\ i\ S_2^+\ \left[\frac{1}{2} (1+i)\ \openone_4 + (1-i)\ S_1^z\right]
\end{eqnarray}
In difference to ref.\ \citep{sire1}, we define spin-$\frac{1}{2}$ 
operators by means of the following equations [Our choice is related
to the choice for the eqs.\ 
(\ref{ntwofurtheraddz66a})-(\ref{ntwofurtheraddz66d}).].
\begin{eqnarray} 
\label{ntwofurtheradd901aa}
S_1^+&=&\ \ \ \frac{i}{\sqrt{2}}\ \hat{a}_1\ 
\left[(1 + i)\ \openone_4 - 2\ \hat{a}_2^\dagger \hat{a}_2\right]\\[0.3cm]
\label{ntwofurtheradd901bb}
S_1^-&=&- \frac{i}{\sqrt{2}}\ \hat{a}_1^\dagger\ 
\left[(1 - i)\ \openone_4 - 2\ \hat{a}_2^\dagger \hat{a}_2\right]\\[0.3cm]
\label{ntwofurtheradd901cc}
S_1^z&=&\frac{1}{2}\ \openone_4\ -\ \hat{a}_1^\dagger \hat{a}_1\\[0.3cm]
\label{ntwofurtheradd901dd}
S_2^+&=&- \frac{i}{\sqrt{2}}\ \hat{a}_2\ \left[(1 - i)\ \openone_4 - 2\ 
\hat{a}_1^\dagger \hat{a}_1\right]\\[0.3cm]
\label{ntwofurtheradd901ee}
S_2^-&=&\ \ \ \frac{i}{\sqrt{2}}\ \hat{a}^\dagger_2\ 
\left[(1 + i)\ \openone_4 - 2\ \hat{a}_1^\dagger \hat{a}_1\right]\\[0.3cm]
\label{ntwofurtheradd901ff}
S_2^z&=&\frac{1}{2}\ \openone_4\ -\ \hat{a}_2^\dagger \hat{a}_2
\end{eqnarray}
And the inverse transformation is given by
\begin{eqnarray} 
\label{ntwofurtheradd902aa}
\hat{a}_1&=&- \frac{1}{\sqrt{2}}\ S_1^+\ 
\left[\openone_4 - 2i\ S_2^z\right]\ ,\\[0.3cm]
\label{ntwofurtheradd902bb}
\hat{a}_1^\dagger&=&- \frac{1}{\sqrt{2}}\ S_1^-\ 
\left[\openone_4 + 2i\ S_2^z\right]\ ,
\\[0.3cm]
\label{ntwofurtheradd902cc}
\hat{a}_2&=&
- \frac{1}{\sqrt{2}}\ S_2^+\ 
\left[\openone_4 + 2i\ S_1^z\right]\ ,\\[0.3cm]
\label{ntwofurtheradd902dd}
\hat{a}_2^\dagger&=&- \frac{1}{\sqrt{2}}\ S_2^-\ 
\left[\openone_4 - 2i\ S_1^z\right]\ .
\end{eqnarray}
It is clear that any two-spin-$\frac{1}{2}$ 
Hamiltonian can be transformed by means
of the eqs.\ (\ref{ntwofurtheradd901a})-(\ref{ntwofurtheradd901f}) to
the general form (\ref{ntwofurtheradd100}) of the two-fermion
Hamiltonian. It is furthermore possible, as discussed above, to bring
any two-fermion Hamiltonian to the special form (\ref{ntwofurtheradd105c}).
Of course, also the quasifermion operators $\hat{b}_1$, $\hat{b}_2$ 
can be related 
to quasi-spin-$\frac{1}{2}$ operators $T_k^+$, $T_k^-$, $T_k^z$ ($k=1,2$)
by means of a Jordan-Wigner 
transformation. In analogy to the above equations we can write
\begin{eqnarray} 
\label{ntwofurtheradd903a}
\hat{b}_1&=&- \frac{1}{\sqrt{2}}\ T_1^+\ 
\left[\openone_4 - 2i\ T_2^z\right]\ ,\\[0.3cm]
\label{ntwofurtheradd903b}
\hat{b}_1^\dagger&=&- \frac{1}{\sqrt{2}}\ T_1^-\ 
\left[\openone_4 + 2i\ T_2^z\right]\ ,
\\[0.3cm]
\label{ntwofurtheradd903c}
\hat{b}_2&=&
- \frac{1}{\sqrt{2}}\ T_2^+\ 
\left[\openone_4 + 2i\ T_1^z\right]\ ,\\[0.3cm]
\label{ntwofurtheradd903d}
\hat{b}_2^\dagger&=&- \frac{1}{\sqrt{2}}\ T_2^-\ 
\left[\openone_4 - 2i\ T_1^z\right]\ .
\end{eqnarray}
Then, the quasifermion Hamiltonian $H^\prime$ 
[cf.\ eq.\ (\ref{ntwofurtheradd105c})] 
can be written as a two-quasi-spin-$\frac{1}{2}$ Hamiltonian. It reads
\begin{eqnarray} 
\label{ntwofurtheradd904}
H^\prime&=&\left(h^{(0\vert 0)}\ +\ \frac{1}{2}\ h^{(1\vert 1)}
\ +\ \frac{1}{2}\ h^{(2\vert 2)}
\ -\ \frac{1}{4}\ h^{(1,2\vert 1,2)}\right)\ \openone_4
\nonumber\\[0.3cm]
&&\ \ -\ \left(\nu_{(-1)0} + \nu_{12}\right)\ T_1^z\ 
\ -\ \left(\nu_{(-1)0} +  \nu_{34}\right)\ T_2^z\ 
\ +\ 2\ \nu_{(-1)0} \ T_1^z T_2^z\ .\ \ \ \ \ \ 
\end{eqnarray}
Consequently, any two-spin-$\frac{1}{2}$ Hamiltonian can be brought to
the above form. A related result has been found (for a somewhat restricted
class of two-spin-$\frac{1}{2}$ Hamiltonians) 
in \citep{wagn1}, Chap.\ 2, Sec.\ 13, 
pp.\ 52-56. The unitary transformations reducing any 
two-spin-$\frac{1}{2}$ Hamiltonian to the form (\ref{ntwofurtheradd904})
correspond, in general, to nonlinear spin transformations. Such 
transformations have been considered in a somewhat different context
in \citep{matt1}, see p.\ 1186, eq.\ (6a), 
\citep{lind1,lind2,lind3,gara1,safo2,safo1}.\\

\subsubsection{Single-spin-\texorpdfstring{$\frac{3}{2}$}{3/2} 
systems}

Two-fermion Hamiltonians are not only related to two-spin-$\frac{1}{2}$ 
systems but can also be related to a single-spin-$\frac{3}{2}$ system
(A concrete choice for the latter relation can be found in \citep{dobr1}.).
Consequently, any system which is given in terms of one
of these representations can also equivalently be formulated in 
terms of the other two. The study of single spin-$\frac{3}{2}$ systems
has a long history (For certain theoretical aspects see 
\citep{cree1,muha1,bain1,bain2} and references therein.) and they have found 
recent attention in the field of quantum computation 
\citep{kess1,kess2,kess3,kess4}. To make the relation of any two-fermion
system to a single-spin-$\frac{3}{2}$ system explicit we will follow here
\citep{dobr1}. We will apply the same line of reasoning as for a
two-spin-$\frac{1}{2}$ system applied above. The three (radial) 
spin-$\frac{3}{2}$ spin operators $I^+$, $I^-$, $I^z$ can be given in
terms of the fermion creation and annihilation operators the following
way (\citep{dobr1}, p.\ L506, eqs.\ (19), (20)).
\begin{eqnarray} 
\label{ntwofurtheradd905a}
I^+&=&\sqrt{3}\ \hat{a}_2\ +\ 2\ \hat{a}_2^\dagger \hat{a}_1 
\\[0.3cm]
\label{ntwofurtheradd905b}
I^-&=&\sqrt{3}\ \hat{a}_2^\dagger\ +\ 2\ \hat{a}_1^\dagger \hat{a}_2\\[0.3cm]
\label{ntwofurtheradd905c}
I^z&=&\frac{1}{2}\left[I^+,I^-\right]\ =\ 
2\ \hat{a}_1^\dagger \hat{a}_1\ +\ 
\hat{a}_2^\dagger \hat{a}_2\ -\ \frac{3}{2}\ \openone_4
\end{eqnarray}
The inverse transformation reads (\citep{dobr1}, p.\ L506, eq.\ (21)):
\begin{eqnarray} 
\label{ntwofurtheradd906a}
\hat{a}_1^\dagger&=&-\ \frac{1}{\sqrt{3}}\ I^+ I^z I^+\ ,\\[0.3cm]
\label{ntwofurtheradd906b}
\hat{a}_1&=&-\ \frac{1}{\sqrt{3}\ }I^- I^z I^-\ ,\\[0.3cm]
\label{ntwofurtheradd906c}
\hat{a}_2^\dagger&=&\frac{1}{\sqrt{3}}\ I^+ 
\left(\frac{1}{2}\openone_4  + I^z\right)^2\ ,\\[0.3cm]
\label{ntwofurtheradd906d}
\hat{a}_2&=&\frac{1}{\sqrt{3}}\ 
\left(\frac{1}{2} \openone_4 + I^z\right)^2 I^-\ .
\end{eqnarray}
Again, any single-spin-$\frac{3}{2}$ 
Hamiltonian can be transformed by means
of the eqs.\ (\ref{ntwofurtheradd905a})-(\ref{ntwofurtheradd905c}) to
the general form (\ref{ntwofurtheradd100}) of the two-fermion
Hamiltonian. It is furthermore possible, as discussed above, to bring
any two-fermion Hamiltonian to the special form (\ref{ntwofurtheradd105c}).
Then, also the quasifermion operators $\hat{b}_1$, $\hat{b}_2$ 
can be related to quasi-spin-$\frac{3}{2}$ operators $J^+$, $J^-$, $J^z$
by means of a Jordan-Wigner 
transformation. In analogy to the above equations we can write
\begin{eqnarray} 
\label{ntwofurtheradd907a}
J^+&=&\sqrt{3}\ \hat{b}_2\ +\ 2\ \hat{b}_2^\dagger \hat{b}_1\ , 
\\[0.3cm]
\label{ntwofurtheradd907b}
J^-&=&\sqrt{3}\ \hat{b}_2^\dagger\ +\ 2\ \hat{b}_1^\dagger \hat{b}_2\ ,
\\[0.3cm]
\label{ntwofurtheradd907c}
J^z&=&\frac{1}{2}\left[J^+,J^-\right]\ =\ 
2\ \hat{b}_1^\dagger \hat{b}_1\ +\ 
\hat{b}_2^\dagger \hat{b}_2\ -\ \frac{3}{2}\ \openone_4\ ,
\end{eqnarray}
and
\begin{eqnarray} 
\label{ntwofurtheradd908a}
\hat{b}_1^\dagger&=&-\ \frac{1}{\sqrt{3}}\ J^+ J^z J^+\ ,\\[0.3cm]
\label{ntwofurtheradd908b}
\hat{b}_1&=&-\ \frac{1}{\sqrt{3}}\ J^- J^z J^-\ ,\\[0.3cm]
\label{ntwofurtheradd908c}
\hat{b}_2^\dagger&=&\frac{1}{\sqrt{3}}\ J^+ 
\left(\frac{1}{2} \openone_4 + J^z\right)^2\ ,\\[0.3cm]
\label{ntwofurtheradd908d}
\hat{b}_2&=&\frac{1}{\sqrt{3}}\ 
\left(\frac{1}{2} \openone_4 + J^z\right)^2 J^-\ .
\end{eqnarray}
The quasifermion Hamiltonian $H^\prime$ 
[cf.\ eq.\ (\ref{ntwofurtheradd105c})] 
can be written as a single-quasi-spin-$\frac{3}{2}$ Hamiltonian. It then reads
\begin{eqnarray} 
\label{ntwofurtheradd909}
H^\prime&=&\left(h^{(0\vert 0)}\ +\ \frac{1}{2}\ h^{(1\vert 1)}
\ +\ \frac{1}{2}\ h^{(2\vert 2)}
\ -\ \frac{1}{4}\ h^{(1,2\vert 1,2)}
\ +\ \frac{13}{8}\ \nu_{(-1)0}\right)\ \openone_4\nonumber\\[0.3cm]
&&\ +\ \frac{1}{12}
\left(17\; \nu_{(-1)0} - 5\; \nu_{12} + 22\; \nu_{34}\right)\ J^z 
\ +\ \frac{1}{2}\ \nu_{(-1)0}\ \left(J^z\right)^2
\nonumber\\[0.3cm]
&&\ +\ \frac{1}{3}
\left(\nu_{(-1)0} -\nu_{12} + 2\; \nu_{34}\right)\ 
\left(J^z\right)^3\ .\ \ \ \ \ \ 
\end{eqnarray}

Finally, we would like to mention that on the basis of the two above 
(generalized) Jordan-Wigner transformations for two-spin-$\frac{1}{2}$
and single-spin-$\frac{3}{2}$ systems relations can be established between
spin-$\frac{1}{2}$ and spin-$\frac{3}{2}$ operators\footnote{There seems
to have been done some related research in the past \citep{grub1}.}. 
Combined with the 
full range of nonlinear Bogolyubov-Valatin transformations discussed we thus
obtain a large manifold of expressing single-spin-$\frac{3}{2}$ 
operators in terms of two-spin-$\frac{1}{2}$ operators and vice versa.\\

\section{\label{param}\texorpdfstring{$SU(4)$}{SU(4)}, 
\texorpdfstring{$SO(6;\mathbb{R})$}{SO(6;R)}, 
\texorpdfstring{$SO(6;\mathbb{R})/\mathbb{Z}_2$}{SO(6;R)/Z\_2} 
transformations and their parametric relations}

In the previous sections unitary $SU(4)$ transformations ${\sf U}$,
orthogonal $SO(6;\mathbb{R})$ transformations $L$, and 
$SO(6;\mathbb{R})/\mathbb{Z}_2$ transformations $\chi$ have played
an important role. So far, we have not discussed (except for the 
expression of $\chi$ in terms of $L$) their concrete mutual relationship
which is of importance for any explicit calculation. In this section 
we finally will consider this technical problem. To set the frame 
for this discussion we will first specify which sort of parametrization
for the matrices ${\sf U}$ and $L$ we are going to use.\\

Let us write for the unitary $4\times 4$ matrix
${\sf U} = {\sf U}\left(\{\lambda\};\{\hat{c}\}\right)$ implementing
a given nonlinear Bogolyubov-Valatin transformation
($T_0$ is a complex number while $T^{m_1 m_2}$ are the matrix elements 
of a complex antisymmetric $6\times 6$ matrix $T$.)
\begin{eqnarray}
\label{ntwofurtheradd316}
{\sf U}&=&T_0\ \openone_4\ +\ T^M\ \hat{\hat{\hat{c}\,}}_M
\ =\ T_0\ \openone_4\ +\ 
\frac{1}{2}\ T^{m_1 m_2}\ \hat{\hat{\hat{c}\,}}_{m_1 m_2}\ .
\end{eqnarray}
From the unitarity condition ${\sf U} {\sf U}^\dagger = \openone_4$ follow
the equations
\begin{eqnarray}
\label{ntwofurtheradd317a}
\vert T_0\vert^2\ +\ T^M\ \overline{T}_M&=&1\ ,\\[0.3cm]
\label{ntwofurtheradd317b}
-\ T_0\; \overline{T}_P\ +\ \overline{T}_0\; T_P
\ -\ T_{p_1}^{\ \ m}\; \overline{T}_{m p_2}  
\ +\ \overline{T}_{p_1}^{\ \ m}\; T_{m p_2}
\ +\ i\ T^M\ \overline{T}^N\ \epsilon_{MNP}&=&0\ .\ \ \ \ \ \ \ \ 
\end{eqnarray}
These equations have been given earlier (in some less 
general notation) in \citep{zaik1}, p.\ 243, eq.\ (35).
For another version of these equations see eq.\ (\ref{ntwofurtheradd317l}).
For any given unitary matrix ${\sf U}$ the coefficients 
$T_0$, $T_M$ can be calculated by means of the equations
[cf.\ eq.\ (\ref{ntwofurtheradd63za})]
\begin{eqnarray}
\label{ntwofurtheradd316b}
T_0&=&\frac{1}{4}\ {\rm tr}\; {\sf U}\ ,\\[0.3cm]
\label{ntwofurtheradd316c}
T_M&=&- \frac{1}{4}\ {\rm tr}\left(\hat{\hat{\hat{c}\,}}_M\ {\sf U}\right)\ .
\end{eqnarray}

For the orthogonal $6\times 6$ matrix $L$ we rely on the Cayley representation
\begin{eqnarray}
\label{ntwofurtheradd401}
L&=&\frac{\openone_6 + A}{\openone_6- A}\ =\ 
\openone_6\ +\ \frac{2 A}{\openone_6 - A}\ =\ 
- \openone_6\ +\ \frac{2}{\openone_6 - A}
\end{eqnarray}
in terms of a real antisymmetric matrix $A$ (We disregard here all
problems of the Cayley representation related to any eigenvalues = -1
of the orthogonal matrix $L$.). The matrix $L$
can be expressed as a sum over a finite number of powers of the 
matrix A. For an explicit expression see eq.\ (\ref{ntwofurtheradd415})
further down. In turn, the antisymmetric
matrix $A$ can be expressed in terms of $L$ as
\begin{eqnarray}
\label{ntwofurtheradd411b}
A&=&\frac{L - \openone_6}{L + \openone_6} 
\ =\ \frac{L^{(-)}}{\openone_6 + L^{(+)}}\ ,\\[0.3cm]
\label{ntwofurtheradd411ba}
L^{(+)}&=&\frac{1}{2}\; \left(L + L^T\right)\ ,\\[0.3cm]
\label{ntwofurtheradd411bb}
L^{(-)}&=&\frac{1}{2}\; \left(L - L^T\right)\ . 
\end{eqnarray}
The antisymmetric matrix $A$ can also be expressed as a sum 
over a finite number of powers of the matrix $L$, however, already
for the present case these expressions are quite involved and,
therefore, we will not display them here (for details
see \citep{bogu3}).\\

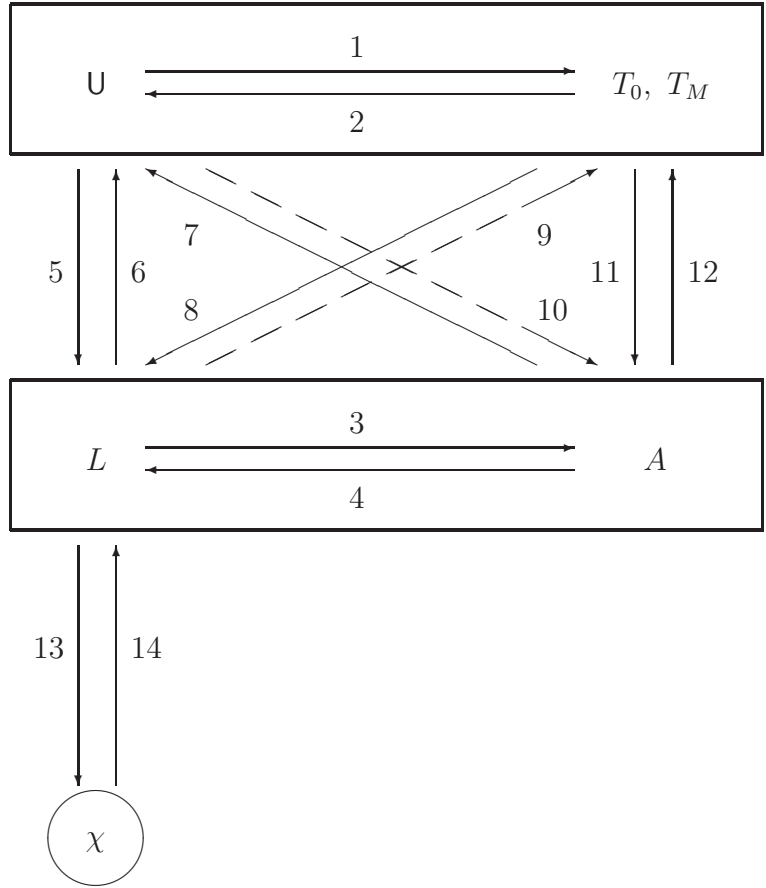
\begin{figure}[ht]
\unitlength1.mm
\begin{picture}(150,130)
\put(15,-3){
\begin{picture}(130,130)
\linethickness{0.3mm}
\put(10,122){\line(1,0){100}}
\put(10,102){\line(1,0){100}}
\put(10,72){\line(1,0){100}}
\put(10,52){\line(1,0){100}}
\put(10,102){\line(0,1){20}}
\put(110,102){\line(0,1){20}}
\put(10,52){\line(0,1){20}}
\put(110,52){\line(0,1){20}}
\put(21.3,11){\circle{12.5}}
\linethickness{0.15mm}
\put(24,74){\vector(0,1){26}}
\put(19,100){\vector(0,-1){26}}
\put(98,74){\vector(0,1){26}}
\put(93,100){\vector(0,-1){26}}
\put(24,18){\vector(0,1){32}}
\put(19,50){\vector(0,-1){32}}
\put(28,113){\vector(1,0){57}}
\put(85,110){\vector(-1,0){57}}
\put(28,63){\vector(1,0){57}}
\put(85,60){\vector(-1,0){57}}
\put(78,79){\vector(2,-1){10}}
\put(72,82){\line(2,-1){4}}
\put(66,85){\line(2,-1){4}}
\put(60,88){\line(2,-1){4}}
\put(54,91){\line(2,-1){4}}
\put(48,94){\line(2,-1){4}}
\put(42,97){\line(2,-1){4}}
\put(36,100){\line(2,-1){4}}
\put(80,74){\vector(-2,1){52}}
\put(78,95){\vector(2,1){10}}
\put(72,92){\line(2,1){4}}
\put(66,89){\line(2,1){4}}
\put(60,86){\line(2,1){4}}
\put(54,83){\line(2,1){4}}
\put(48,80){\line(2,1){4}}
\put(42,77){\line(2,1){4}}
\put(36,74){\line(2,1){4}}
\put(80,100){\vector(-2,-1){52}}
\put(20,110){\parbox[b]{10mm}{${\sf U}$}}
\put(90,110){\parbox[b]{15mm}{$T_0,\ T_M$}}
\put(20,60){\parbox[b]{10mm}{$L$}}
\put(90,60){\parbox[b]{10mm}{$\ \ \ A$}}
\put(20,10){\parbox[b]{10mm}{$\chi$}}
\put(55,115){\parbox[b]{10mm}{1}}
\put(55,105){\parbox[b]{10mm}{2}}
\put(55,65){\parbox[b]{10mm}{3}}
\put(55,55){\parbox[b]{10mm}{4}}
\put(15,85){\parbox[b]{10mm}{5}}
\put(26,85){\parbox[b]{10mm}{6}}
\put(33,90){\parbox[b]{10mm}{7}}
\put(33,80){\parbox[b]{10mm}{8}}
\put(80,90){\parbox[b]{10mm}{9}}
\put(80,80){\parbox[b]{10mm}{10}}
\put(87,85){\parbox[b]{10mm}{11}}
\put(100,85){\parbox[b]{10mm}{12}}
\put(13,35){\parbox[b]{10mm}{13}}
\put(26,35){\parbox[b]{10mm}{14}}
\end{picture}  }
\end{picture}
\caption{\label{overview}Schematic overview over the content of
Sec.\ \ref{param}.}
\end{figure}

To simplify navigation through the present section we now
give in Figure \ref{overview} a schematic overview of its content.
Each arrow in Figure \ref{overview} stands for an equation expressing the 
quantity at the end point of the arrow by another one at its 
starting point.\\

\pagebreak

\begin{itemize}
\item[]
 {\bf List of reference points for Figure \ref{overview}}:
\begin{itemize}
\item[Arrow 1:] 
Eqs.\ (\ref{ntwofurtheradd316b}), (\ref{ntwofurtheradd316c}).

\item[Arrow 2:]
Eq.\ (\ref{ntwofurtheradd316}).

\item[Arrow 3:]
Eq.\ (\ref{ntwofurtheradd411b}) and  ref.\ \citep{bogu3}.

\item[Arrow 4:]
Eq.\ (\ref{ntwofurtheradd401}), 
Subsec.\ \ref{fedorov}, eq.\ (\ref{ntwofurtheradd415}).

\item[Arrow 5:]
Subsec.\ \ref{double}, 
eqs.\ (\ref{ntwofurtheradd318}), (\ref{ntwofurtheradd318b}).

\item[Arrow 6:] 
Subsec.\ \ref{klotz}, eq.\ (\ref{ntwofurtheradd421}).

\item[Arrow 7:] 
Subsec.\ \ref{fedorov}, eq.\ (\ref{ntwofurtheradd411a}).

\item[Arrow 8:] 
Subsec.\ \ref{double}, eq.\ (\ref{ntwofurtheradd319b}).

\item[Arrow 9:]
No equation displayed, can be composed of the equations 
corresponding to the arrows 6 and 1 (or the arrows 3 and 12).

\item[Arrow 10:]
No equation displayed, can be composed of the equations 
corresponding to the arrows 5 and 3 (or the arrows 1 and 11).

\item[Arrow 11:]
Subsec.\ \ref{fedorov}, eq.\ (\ref{ntwofurtheradd412}).

\item[Arrow 12:]
Subsec.\ \ref{fedorov}, eqs.\ (\ref{ntwofurtheradd411c}), 
(\ref{ntwofurtheradd411d}).

\item[Arrow 13:]
Subsec.\ \ref{global}, paragraph III, eq.\ (\ref{ntwofurtheradd78c}).

\item[Arrow 14:]
Subsec.\ \ref{chiinverse}, eq.\ (\ref{ntwofurtheradd211}).

\end{itemize}
\end{itemize}

\subsection{\label{double}From 
\texorpdfstring{${\sf U}\in SU(4)$}{U in SU(4)} 
to \texorpdfstring{$L \in SO(6;\mathbb{R})$}{L in SO(6;R)}} 

In this and the following subsection, we will study the 
explicit relation between the $SO(6;\mathbb{R})$ 
transformations $L$ and the corresponding unitary 
$SU(4)$ transformations ${\sf U}\left(\{\lambda\};\{\hat{c}\}\right)$. 
In the present subsection, we will assume that  
${\sf U} = {\sf U}\left(\{\lambda\};\{\hat{c}\}\right)$ is known and
derive from it an expression for $L$. We will perform this task by relying on
the discussion presented in \citep{step1,step2} (for follow-up work
see \citep{lyub1}).\\

To begin with, let
us start by repeating some elements of the very lucid account 
given in \citep{step2} (We will follow here also the notation used
in this article with slight modifications.). 
Consider a set of six linearly independent 
antisymmetric (complex) $4\times 4$ 
matrices\footnote{For an early 
discussion of the role of antisymmetric matrices in the present 
context see \citep{lema1}, related discussions
can be found in \citep{scho4,vebl1}, \citep{stru1}, Chap.\ IV, Secs.\
5-9, pp.\ 36-40, \citep{scho2} (There are many other papers by
Schouten and collaborators containing related but less focused material.), 
\citep{haan1,lee1} (very clear mathematical accounts), 
\citep{nord1}, \S\ 11, p.\ 49-52, 
\citep{este1},
\citep{buch1} (also note \citep{buch2}), 
\citep{klot1}, 
\citep{post1}, 
{\cyrrm Lekts.} [Lekts.]\ 13, pp.\ 258-299 
(pages 247-285 of the English transl.), and in 
\citep{andr4,andr2,andr3,andr1}. 
A somewhat related discussion for the
case $SL(2;\mathbb{C})$, $SO(1,3;\mathbb{R})$ is given in 
\citep{elli1}. A different approach of dealing with the homomorphism
between the groups $SU(4)$ and $SO(6;\mathbb{R})$ 
related to the physical problem
of the 3-particle problem is discussed in \citep{louc1}, Secs.\ VI.A,
VI.B, pp.\ 557-559.}:
$\Gamma_k^{\hspace{-0.5cm}+\atop\ } = -\
\Gamma_k^{\hspace{-0.25cm}{+\atop\ }\ T}$,
$k = -1,\ldots ,4$. These matrices represent spin (space) bivectors
($\in\bigwedge^2 (\mathbb{C}_4)$, where $\mathbb{C}_4$ is the spin space)
and can be chosen to
obey ($\epsilon^{abcd}$, $\epsilon_{abcd}$,
$\epsilon^{1234} = \epsilon_{1234} = 1$ is the completely 
antisymmetric tensor operating in spin space $\mathbb{C}_4$)
\begin{eqnarray}
\label{ntwofurtheradd301}
\frac{1}{8}\ \epsilon^{abcd}\
\left(\Gamma_k^{\hspace{-0.5cm}+\atop\ }\right)_{ab}\ 
\left(\Gamma_l^{\hspace{-0.5cm}+\atop\ }\right)_{cd}&=&\delta_{kl}
\end{eqnarray}
(\citep{step2}, p.\ 10, eq.\ (1), an example for these matrices 
can be found in \citep{step1}, p.\ 814, eq.\ (12); 
also see our Appendix \ref{appc}). One can then 
define antisymmetric $4\times 4$ matrices 
$\Gamma_k^{\hspace{-0.5cm}-\atop\ } = -\
\Gamma_k^{\hspace{-0.25cm}{-\atop\ }\ T}$ (the spin space
Hodge duals of $\Gamma_k^{\hspace{-0.5cm}+\atop\ }$)\footnote{Incidentally, 
the minus sign on top of the $\Gamma$ symbol should not be confused with the
sign for complex conjugation -- a longer bar.}
by writing
\begin{eqnarray}
\label{ntwofurtheradd302}
\left(\Gamma_k^{\hspace{-0.5cm}-\atop\ }\right)^{ab}&=&
-\ \frac{1}{2}\ \epsilon^{abcd}\
\left(\Gamma_k^{\hspace{-0.5cm}+\atop\ }\right)_{cd}\ .
\end{eqnarray}
Then, eq.\ (\ref{ntwofurtheradd301}) can be written compactly as
\begin{eqnarray}
\label{ntwofurtheradd301b}
\frac{1}{4}\ {\rm tr}\left(\Gamma_k^{\hspace{-0.5cm}+\atop\ }
\Gamma_l^{\hspace{-0.5cm}-\atop\ }\right)&=&\delta_{kl}\ .
\end{eqnarray}
From this equation one 
recognizes\footnote{Cf.\
\citep{scho4}, p.\ 410, eq.\ (2.14),
\citep{stru1}, p.\ 36 (p.\ 216 of the whole volume), 
bottom of the page,\\[-0.6cm]
\begin{tabbing}
\citep{scho2}, p.\ 184, eq.\ (60),\hskip1.6cm \=
\citep{haan1}, p.\ 48, eq.\ (2),\hskip1.2cm \=
\citep{lee1}, p.\ 139, eq.\ (1.5),\\
\citep{buch1}, p.\ 368, eq.\ (13.8), 
\>\citep{step2}, pp.\ 10-11,
\>\citep{andr3}, Subsec.\ 2.1.1, p.\ 19, eq.\ (49),\\
\citep{andr1}, p.\ 11 (p.\ 9 of the English transl.), eq.\ (2),
\>
\>\citep{cheu1}, p.\ 24, Appendix A, eqs.\ (A2).
\end{tabbing} }
that the matrices 
$\Gamma_k^{\hspace{-0.5cm}+\atop\ }$,
$\Gamma_k^{\hspace{-0.5cm}-\atop\ }$ obey the 
equation\footnote{Incidentally, such a relation has also emerged in 
\citep{ingr1}, p.\ 1066, eq.\ (6). For a somewhat related 
discussion also see \citep{miln1}, in particular, p.\ 3, eq.ß (5).\\
\ }
\begin{eqnarray}
\label{ntwofurtheradd303}
\Gamma_k^{\hspace{-0.5cm}+\atop\ } 
\Gamma_l^{\hspace{-0.5cm}-\atop\ } 
\ +\ \Gamma_l^{\hspace{-0.5cm}+\atop\ } 
\Gamma_k^{\hspace{-0.5cm}-\atop\ }\ =\ 
\Gamma_k^{\hspace{-0.5cm}-\atop\ } 
\Gamma_l^{\hspace{-0.5cm}+\atop\ } 
\ +\ \Gamma_l^{\hspace{-0.5cm}-\atop\ } 
\Gamma_k^{\hspace{-0.5cm}+\atop\ }&=&2\delta_{kl}\ \openone_4\ .
\end{eqnarray}
One can also derive the following useful 
relations\footnote{Cf.\ \citep{vebl1}, p.\ 510, eq.\ (4.10),
\citep{scho2}, p.\ 177, eq.\ (7),
\citep{haan1}, p.\ 48, eq.\ (6),
\citep{lee1}, p.\ 138, eqs.\ (1.2), 
\citep{nord1}, p.\ 50, eq.\ (33) (incidentally, in the 
preceding eq.\ (32)
$\eta_4^{25}$ and $\eta^4_{25}$ should correctly read
$\eta_4^{23}$ and $\eta^4_{23}$, respectively),\\[-1.15cm]
\begin{tabbing}
\hskip5cm \=
\citep{nord2}, p.\ 159, eq.\ (21),\hskip0.9cm \=
\citep{buch1}, p.\ 367, eq.\ (13.6),\\
\citep{lord4}, p.\ 93, eq.\ (2.10),
\>\citep{step2}, p.\ 11, eqs.\ (11), (12),
\>\citep{andr2}, p.\ 44, eq.\ (2),\\
\citep{andr3}, Subsec.\ 2.1.1, pp.\ 17-20,
\>\citep{andr1}, p.\ 11 (p.\ 9 of the English transl.), eq.\ (1),
\>\\
\citep{cheu1}, p.\ 24, Appendix A, eqs.\ (A4)-(A6).
\end{tabbing}}
\begin{eqnarray}
\label{ntwofurtheradd303b}
\left(\Gamma_k^{\hspace{-0.5cm}+\atop\ }\right)_{ab} 
\left(\Gamma^{\hspace{-0.5cm} ^-\atop\ }
\ ^{\hspace{-3.5mm}k}\right)^{cd}
&=&2\ \delta_a^{\ d}\delta_b^{\ c}\ -\ 2\ \delta_a^{\ c}\delta_b^{\ d}\ ,\\[0.3cm]
\label{ntwofurtheradd303c}
\left(\Gamma_k^{\hspace{-0.5cm}-\atop\ }\right)^{ab} 
\left(\Gamma^{\hspace{-0.5cm} ^-\atop\ }
\ ^{\hspace{-3.5mm}k}\right)^{cd}
&=&2\ \epsilon^{abcd}\ ,\\[0.3cm]
\label{ntwofurtheradd303d}
\left(\Gamma_k^{\hspace{-0.5cm}+\atop\ }\right)_{ab} 
\left(\Gamma^{\hspace{-0.5cm} ^+\atop\ }
\ ^{\hspace{-3.5mm}k}\right)_{cd}
&=&2\ \epsilon_{abcd}\ .
\end{eqnarray}
It is now possible to represent the paravectors 
$\hat{\hat{\hat{c}\,}}_k$ in terms of 
the antisymmetric matrices $\Gamma_k^{\hspace{-0.5cm}+\atop\ }$,
$\Gamma_k^{\hspace{-0.5cm}-\atop\ }$. One 
finds\footnote{\label{bibi}Also cf.\ 
\citep{scho4}, p.\ 410, eq.\ (2.13), 
\citep{stru1}, p.\ 37 (p.\ 217 of the whole volume), 
\citep{haan1}, p.\ 49, eq.\ (7).
The analogous equation for an arbitrary biparavector
$\hat{\hat{\hat{c}\,}}_M$ [see eq.\ (\ref{ntwofurtheradd307b}),
Subsec.\ \ref{bibidis}] 
is given in\\[-0.7cm]
\begin{tabbing}
\citep{buch1}, p.\ 367, eq.\ (13.7),\hskip0.5cm \=
\citep{klot1}, p.\ 2244, eq.\ (9),\hskip1.9cm \=
\citep{step2}, p.\ 11, eq.\ (18),\\
\citep{andr2}, p.\ 44, eq.\ (5a),
\>\citep{lauf1}, Sec.\ 5.3, p.\ 167, eq.\ (5.21),
\>\citep{andr3}, Subsec.\ 6.1.1, p.\ 97, eq.\ (472),\\
\citep{andr1}, p.\ 12 (p.\ 10 of the English transl.), 
the first eq.\ on the page.
\end{tabbing} }
\begin{eqnarray}
\label{ntwofurtheradd304}
\hat{\hat{\hat{c}\,}}_k&=&-\ \Gamma_0^{\hspace{-0.5cm}+\atop\ }
\Gamma_k^{\hspace{-0.5cm}-\atop\ }\ .
\end{eqnarray}
The choice of $\Gamma_0^{\hspace{-0.5cm}+\atop\ }$ singles out a
certain spin bivector and in the spin bivector space 
$\bigwedge^2 (\mathbb{C}_4)$ a certain direction (vector, line).
Haantjes calls this direction (line) the {\it axis} 
of the pentade $\hat{\hat{\hat{c}\,}}_k$, $k = -1,1\ldots,4$ 
(\citep{haan1}, p.\ 51, stelling 4 [proposition 4]).
Furthermore, it is possible to choose the matrices 
$\Gamma_k^{\hspace{-0.5cm}+\atop\ }$,
$\Gamma_k^{\hspace{-0.5cm}-\atop\ }$ such a way that 
\begin{eqnarray}
\label{ntwofurtheradd305}
\Gamma_k^{{\hspace{-0.5cm}+\atop\ }\dagger}&=&
\Gamma_k^{\hspace{-0.5cm}-\atop\ }
\end{eqnarray}
applies. The condition (\ref{ntwofurtheradd305}) determines in the 
(complex six-dimensional) spin bivector space 
$\bigwedge^2 (\mathbb{C}_4)$ a (real six-dimensional) subspace
which we denote in view of eq.\ (\ref{ntwofurtheradd301b})
by $\bigwedge^2_{\mathbb{R}_6} (\mathbb{C}_4) \simeq\mathbb{R}_6$
[also see the comment below from eq.\ (\ref{ntwofurtheradd308b})].
As a technical comment, we would like to mention here that
$\Gamma_k^{{\hspace{-0.5cm}+\atop\ }\dagger}$ on the l.h.s.\ of eq.\
(\ref{ntwofurtheradd305}) is an object with 
upper (spin) indices related by hermitian conjugation to the object 
$\Gamma_k^{\hspace{-0.5cm}+\atop\ }$ with lower (spin) indices. 
To cut a long story short, this is related to the fact that
for bivectors in $\bigwedge^2_{\mathbb{R}_6} (\mathbb{C}_4) \simeq\mathbb{R}_6$, 
we can raise and lower indices 
in spin space by means of $s_{ab^\prime} = \delta_{ab^\prime}$, 
$s^{ab^\prime} = \delta^{ab^\prime}$
supplemented by an operation of hermitian conjugation
[cf.\ \cite{andr3}, Subsec.\ 2.2.3,
p.\ 40, table 1, and Subsec.\ 2.2.2, p.\ 31/32, 
{\cyrrm Teorema} [Teorema]/[Theorem] 2, eq.\ (121);  
\cite{andr1}, p.\ 15 (p.\ 12 of the English
transl.), table, and p.\ 13 (p.\ 11 of the English transl.), 
{\cyrrm Teorema} [Teorema]/[Theorem] 2, eq.\ (10)]\footnote{We are indebted to 
K.\ V.\ Andreev for pointing out to us these subtleties and 
refer the interested reader
to his publications \cite{andr2,andr3,andr1} for further details.}.\\

By virtue of eq.\ (\ref{ntwofurtheradd302}), eq.\  (\ref{ntwofurtheradd305})
leads to the condition (--- denotes complex conjugation)
\begin{eqnarray}
\label{ntwofurtheradd306}
\overline{\left(\Gamma_k^{\hspace{-0.5cm}+\atop\ }\right)_{ba}}&=&
\left(\Gamma_k^{\hspace{-0.5cm}-\atop\ }\right)^{ab}\ =\ 
- \frac{1}{2}\ \epsilon^{abcd}\
\left(\Gamma_k^{\hspace{-0.5cm}+\atop\ }\right)_{cd}
\end{eqnarray}
which entails that the real part of the matrix 
$\Gamma_k^{\hspace{-0.5cm}+\atop\ }$ should be a selfdual matrix
while the imaginary part should be an anti-selfdual matrix.  
We give in Appendix \ref{appc} an explicit example for the set of matrices 
$\Gamma_k^{\hspace{-0.5cm}+\atop\ }$ that obey 
eq.\ (\ref{ntwofurtheradd305}) [Up to the choice of signs
this set of matrices agrees with that given in 
\citep{step1}, p.\ 814, eq.\ (12).].
Taking into account eq.\ (\ref{ntwofurtheradd305}) we find
\begin{eqnarray}
\label{ntwofurtheradd307}
\hat{\hat{\hat{c}\hspace{0.3mm}}}_k^\dagger
&=&-\ \Gamma_k^{\hspace{-0.5cm}+\atop\ }
\Gamma_0^{\hspace{-0.5cm}-\atop\ }\ .
\end{eqnarray}
Equations analogous to the eqs.\ (\ref{ntwofurtheradd304}),
(\ref{ntwofurtheradd307}) also hold for $\hat{\hat{\hat{d}\;}}_k$,
$\hat{\hat{\hat{d}\;}}_k^\dagger$ with the matrices 
$\Gamma_k^{\hspace{-0.5cm}+\atop\ }$ replaced by some other matrices
$\Gamma_k^{\hspace{-0.25cm}{+\atop\ }\,\prime}$. These two sets of matrices
are related by the equation\footnote{\citep{step1}, p.\ 815, eq.\ (10),  
p.\ 821, eq.\ (47), \citep{step2},  p.\ 12, below from eq.\ (23), 
p.\ 13, eq.\ (30). Note that Stepanovski\u\i\
has studied the more general case 
$SL(4;\mathbb{C})/\mathbb{Z}_2 \simeq SO(6;\mathbb{C})$ and that
we have interchanged ${\sf U}$ and 
${\sf U}^\dagger$ compared with the use in \citep{step1,step2}.} 
\begin{eqnarray}
\label{ntwofurtheradd308a}
\Gamma_k^{\hspace{-0.25cm}{+\atop\ }\,\prime}&=&L_{kl}\ 
\Gamma_l^{\hspace{-0.5cm}+\atop\ }\ =\ 
{\sf U}\; 
\Gamma_k^{\hspace{-0.5cm}+\atop\ }{\sf U}^T\ ,\\[0.3cm]
\label{ntwofurtheradd308b}
\Gamma_k^{\hspace{-0.25cm}{-\atop\ }\,\prime}&=&L_{kl}\ 
\Gamma_l^{\hspace{-0.5cm}-\atop\ }\ =\ 
\overline{\sf U}\; \Gamma_k^{\hspace{-0.5cm}-\atop\ }{\sf U}^\dagger\ .
\end{eqnarray}
In general, here the matrix $L$ may  be an element of the group 
$SO(6;\mathbb{C})$, however,
by virtue of the condition (\ref{ntwofurtheradd305}) the matrix $L$
must be real in our case and, therefore, belongs 
to the group $SO(6;\mathbb{R})$. 
It seems worth mentioning that eq.\  (\ref{ntwofurtheradd306})
is invariant under the transformations (\ref{ntwofurtheradd308a}),
(\ref{ntwofurtheradd308b}) (To see this one has to rely on the Laplace
expansion of the determinant of the $SU(4)$ matrix ${\sf U}$.).
The eqs.\ (\ref{ntwofurtheradd308a}),
(\ref{ntwofurtheradd308b}) provide us with additional motivation to
denote the subspace of the
spin bivector space $\bigwedge^2 (\mathbb{C}_4)$ we are using
by the symbol $\mathbb{R}_6$ because its invariance group is
$SO(6;\mathbb{R})$.
The antisymmetric matrices $\Gamma_k^{\hspace{-0.5cm}+\atop\ }$
can be understood as the orthonormal basis in the
spin bivector space $\bigwedge^2 (\mathbb{C}_4)$
[cf.\ eq.\ (\ref{ntwofurtheradd301b})].\\ 

From eq.\ (\ref{ntwofurtheradd308b}) one finds the 
relation\footnote{Cf.\ 
\citep{step1}, p.\ 822, eq.\ (52), also see 
\citep{lord1}, p.\ 766, the equation below from eq.\ (2.4), 
\citep{baru2}, p.\ 13, eq.\ (4.9), 
\citep{klot1}, p.\ 2244, eq.\ (6), 
\citep{andr3}, Subsec.\ 2.2.1, p.\ 22, eq.\ (62), and 
\citep{andr1}, p.\ 12 (p.\ 10 of the English transl.), eq.\ (4). 
A related result can be found in
\citep{macf1}, p.\ 139, eq.\ (3.13).}
\begin{eqnarray}
\label{ntwofurtheradd318}
L_{kl}&=&\frac{1}{4}\ {\rm tr} \left( 
\Gamma_k^{\hspace{-0.25cm}{-\atop\ }\,\prime}\; 
\Gamma_l^{\hspace{-0.5cm}+\atop\ }\right)\ =\ 
\frac{1}{4}\ {\rm tr}\left(\Gamma_k^{\hspace{-0.5cm}-\atop\ }{\sf U}^\dagger\; 
\Gamma_l^{\hspace{-0.5cm}+\atop\ }\overline{\sf U}\right)\\[0.3cm]
\label{ntwofurtheradd318b}
&=&\frac{1}{4}\ {\rm tr} \left( 
\Gamma_k^{\hspace{-0.25cm}{+\atop\ }\,\prime}\; 
\Gamma_l^{\hspace{-0.5cm}-\atop\ }\right)\ =\ 
\frac{1}{4}\ {\rm tr}\left(\Gamma_k^{\hspace{-0.5cm}+\atop\ }{\sf U}^T\; 
\Gamma_l^{\hspace{-0.5cm}-\atop\ }{\sf U}\right)\ .
\end{eqnarray}
Inserting now eq.\ (\ref{ntwofurtheradd316}) into it and calculating the 
occurring traces by means of the eqs.\ 
(\ref{ntwofurtheradd309})-(\ref{ntwofurtheradd314}) we obtain
[cf.\ \citep{step1}, p.\ 822, eq.\ (53)]
\begin{eqnarray}
\label{ntwofurtheradd319a}
L&=&\left(\overline{T}_0^2 + 
\overline{T}^M\; \overline{T}_M\right)\ \openone_6\ -\
2\ \overline{T}_0\ \overline{T}
\ +\ 2\ \overline{T}^2\ -\ 2i\ Pc^{\; 2}(\overline{T})
\nonumber\\[0.3cm]
\label{ntwofurtheradd319b}
&=&\left(T_0^2 + T^M\; T_M\right)\ \openone_6\ -\
2\ T_0\ T\ +\ 2\ T^2\ +\ 2i\ Pc^{\; 2}(T)\ .
\end{eqnarray}
Here, we have introduced the matrix $Pc^{\; 2}(T)$ for the 
antisymmetric matrix $T$ by defining 
\begin{eqnarray}
\label{ntwofurtheradd411e}
Pc^{\; 2}(T)_{kl}&=&\frac{\partial}{\partial T_{lk}}\ {\rm Pf}\; T\ =\
-\frac{1}{2}\ T^M T^N \epsilon_{MNkl}\ .
\end{eqnarray}
We call the matrix $Pc^{\; 2}(T)$ the (second) {\it supplementary
Pfaffian compound matrix}. We have defined it 
in analogy to the definition of supplementary compound matrices 
from a determinant [cf.\ Appendix \ref{appcomp}, eq.\ (\ref{A1f})]. 
The matrix $Pc^{\; 2}(T)$ obeys the 
following equations.
\begin{eqnarray}
\label{ntwofurtheradd411f}
Pc^{\; 2}(T)\ T&=&T\ Pc^{\; 2}(T)\ =\ {\rm Pf}\; T\ \openone_6\\[0.3cm]
\label{ntwofurtheradd411g}
\left[Pc^{\; 2}(T)\right]^2&=&-\frac{1}{4}
\left[\frac{1}{2}\left({\rm tr}\; T^2\right)^2
- {\rm tr}\; T^4\right]\ \openone_6\ +\
\frac{1}{2}\ T^2\ {\rm tr}\; T^2\ -\ T^4
\end{eqnarray}
Under the assumption ${\rm Pf}\; T \neq 0$ we can derive from the
above two equations the relation [by multiplying eq.\ 
(\ref{ntwofurtheradd411g}) by $T$]
\begin{eqnarray}
\label{ntwofurtheradd411h}
Pc^{\; 2}(T)&=&-\ \frac{T}{{\rm Pf}\; T}\left\{
\frac{1}{4}
\left[\frac{1}{2}\left({\rm tr}\; T^2\right)^2
- {\rm tr}\; T^4\right]\ \openone_6\ -\
\frac{1}{2}\ T^2\ {\rm tr}\; T^2\ +\ T^4\right\}\ .\ \ \ \ \ 
\end{eqnarray}
Eq.\ (\ref{ntwofurtheradd411f}) can be viewed as a special form 
of the Pfaffian analogue of the Laplace expansion of a 
determinant\footnote{Cf.\ \citep{caia1}, Part 1, Chap.\ 1, 
Sec.\ 3, p.\ 6, eq.\ (1.5).
This expansion is originally due to Tanner \citep{tann1}, has also been 
noted later by Baker \citep{bake1}, and proved in \citep{zaja1,zaja2,bril1}. 
For a recent discussion see \citep{knut1}.}.\\

It is clear that the imaginary part of the expressions on the
r.h.s.\ of eq.\ (\ref{ntwofurtheradd319b}) must vanish, 
consequently the following relations apply.
\begin{eqnarray}
\label{ntwofurtheradd320}
{\rm Im}\left(T_0^2 + T^M\; T_M\right)\ \openone_6\
+\ 2\ {\rm Im}\left(T^2\right)&=&0\\[0.3cm]
\label{ntwofurtheradd321}
{\rm Im}\left(T_0\ T\right)
\ -\ {\rm Re}\left(Pc^{\; 2}(T)\right)&=&0
\end{eqnarray}
Eq.\ (\ref{ntwofurtheradd320}) originates from the symmetric 
part of eq.\ (\ref{ntwofurtheradd319b}) while 
eq.\ (\ref{ntwofurtheradd321}) is derived from its antisymmetric part. 
From eq.\ (\ref{ntwofurtheradd320}) we can obtain the equation
\begin{eqnarray}
\label{ntwofurtheradd326}
{\rm Im}\left(T_0^2\right)&=&\ {\rm Im}\left({\rm tr}\; T^2\right)\ .
\end{eqnarray}

\subsection{From 
\texorpdfstring{$SO(6;\mathbb{R})$}{SO(6;R)} 
back to its double cover 
\texorpdfstring{$SU(4)$}{SU(4)}} 

Having obtained eq.\ (\ref{ntwofurtheradd319a}), 
we will now reverse reasoning and 
study the relation between the $SO(6;\mathbb{R})$ transformations $L$ 
and the corresponding unitary 
$SU(4)$ transformations ${\sf U}\left(\{\lambda\};\{\hat{c}\}\right)$
in the opposite direction. We will now assume that $L$ is known and
derive from it an expression for 
${\sf U} = {\sf U}\left(\{\lambda\};\{\hat{c}\}\right)$. In the literature,
one finds two approaches to this task. One is due to Fedorov and collaborators
\citep{bogu2,bogu1} and another one has been given by Klotz \citep{klot1}
(relying on earlier work by Macfarlane \citep{macf2,macf1}).
In the first part of the following discussion (Subsec.\ \ref{fedorov})
we will rely  to a large extent on the former while in a second
part (Subsec.\ \ref{klotz}) we will describe the approach by Klotz.

\subsubsection{\label{fedorov}Plane orthogonal transformations 
\texorpdfstring{--}{-} 
The approach by Fedorov and collaborators}

To begin with, it turns out to be useful to recall that any orthogonal
transformation $SO(6;\mathbb{R})$ can be represented in terms of
3 commuting orthogonal transformations within mutually orthogonal
planes \citep{scho11,vita1}, \citep{cart1}, Act.\ Sci.\ Ind.\ 643, 
\S 45, pp.\ 46/47 (p.\ 36 of the English transl.). 
In the Cayley representation of an orthogonal
transformation $L$ [eq.\ (\ref{ntwofurtheradd401})]
each of these plane orthogonal transformations can be given in 
terms of a decomposable (real antisymmetric) matrix $A$. For such a
plane orthogonal transformation [obeying 
$A^3 = \frac{1}{2}\ \left({\rm tr}\; A^2\right)\ A$] the orthogonal matrix $L$
can be written as ($a\neq 1$; \citep{bogu2}, p.\ 1034, 
p.\ 1350 of the English transl.)
\begin{eqnarray}
\label{ntwofurtheradd402}
L&=&\openone_6\ +\ \frac{2 A}{1 - a}\ \left(\openone_6 + A\right)\ ,\ \ \ 
a\ =\ \frac{1}{2}\ {\rm tr}\; A^2\ .
\end{eqnarray}
Now, comparing the antisymmetric parts of 
the equations (\ref{ntwofurtheradd319b}) and
(\ref{ntwofurtheradd402}) we immediately find the relation
[The last term on the r.h.s.\ of eq.\ (\ref{ntwofurtheradd319b}) vanishes for
a plane orthogonal transformation.] 
\begin{eqnarray}
\label{ntwofurtheradd403}
\frac{A}{1 - a}&=&-\ T_0\ T
\end{eqnarray}
where 
\begin{eqnarray}
\label{ntwofurtheradd404}
T_0&=&\pm \left(1 + \frac{1}{2}\ {\rm tr}\; T^2\right)^{\frac{1}{2}}
\ =\ \pm \left(1 - a\right)^{-\frac{1}{2}}
\ =\ \pm \left(1 - \frac{1}{2}\ {\rm tr}\; A^2\right)^{-\frac{1}{2}}
\end{eqnarray}
in view of eq.\ (\ref{ntwofurtheradd317a}) [$T_0$ and $T$
are real for a plane orthogonal transformation.]. For a plane orthogonal 
transformation parametrized by the antisymmetric matrix $A$ it holds
\begin{eqnarray}
\label{ntwofurtheradd405}
1 - \frac{1}{2}\ {\rm tr}\; A^2&=&\det\left(\openone_6 - A\right)\ .
\end{eqnarray}
Consequently, eq.\ (\ref{ntwofurtheradd403}) can be written as
\begin{eqnarray}
\label{ntwofurtheradd406a}
T&=&\mp \left(1 - \frac{1}{2}\ {\rm tr}\; A^2\right)^{-\frac{1}{2}}\ A
\ =\ \mp\ \frac{A}{\sqrt{\det\left(\openone_6 - A\right)}}\ ,\ \\[0.3cm]
\label{ntwofurtheradd406b}
A&=&-\ \frac{T}{T_0}\ =\ -\ {\rm sign}\; T_0\ 
\left(1 + \frac{1}{2}\ {\rm tr}\; T^2\right)^{-\frac{1}{2}}\ T\ .
\end{eqnarray}
Finally, any unitary transformation ${\sf U}$ standing in correspondence to a
plane orthogonal transformation $L = L(A)$ can be written as
\begin{eqnarray}
\label{ntwofurtheradd407}
{\sf U}\ =\ {\sf U}(A)&=&\pm 
\left[\det\left(\openone_6 - A\right)\right]^{-\frac{1}{2}}
\left(\openone_4 - A^N \hat{\hat{\hat{c}\,}}_N\right)\ .
\end{eqnarray}

Having obtained this representation we can now go over 
to the general case. Any real antisymmetric matrix $A$ representing
a general orthogonal transformation $SO(6;\mathbb{R})$
can be written as the sum of three decomposable real antisymmetric
matrices $A_k$ that represent three (commuting) plane orthogonal 
transformations in mutually orthogonal planes: 
\begin{eqnarray}
\label{ntwofurtheradd408}
A&=&A_1\ +\ A_2\ +\ A_3\ ,\ \ \ A_k\ A_l\ =\ 0\ \  {\rm for}\ k\neq l\ . 
\end{eqnarray}
We have 
\begin{eqnarray}
\label{ntwofurtheradd409}
L\ =\ L(A)&=&L(A_1)\;L(A_2)\;L(A_3)
\end{eqnarray}
and, consequently,
\begin{eqnarray}
\label{ntwofurtheradd410}
{\sf U}\ =\ {\sf U}(A)&=&{\sf U}(A_1)\;{\sf U}(A_2)\;{\sf U}(A_3)\ .
\end{eqnarray}
Inserting eq.\ (\ref{ntwofurtheradd407}) into 
eq.\  (\ref{ntwofurtheradd410}) and performing the algebra
[by means of eq.\ (\ref{ntwofurtheradd63za}) and taking into
account the decomposability of the matrices 
$A_k$ (no summation over $k$ here): 
$A_k^M A_k^N \epsilon_{MNP} = 0$ ] we find
\begin{eqnarray}
\label{ntwofurtheradd411a}
{\sf U}\ =\ {\sf U}(A)&=&T_0\ \openone_4\ +\ T^P\ \hat{\hat{\hat{c}\,}}_P
\nonumber\\[0.3cm]
&=&\pm \left[\det\left(\openone_6 - A\right)\right]^{-\frac{1}{2}}
\nonumber\\[0.3cm]
&&\ \ \ \ \ \ \times\ \left[\left(1 - i\ {\rm Pf} A\right)\openone_4 
- \left(A^P - i\ Pc^{\; 2}(A)^P\right) 
\hat{\hat{\hat{c}\,}}_P\right]
\ ,\ \ \ \ \ \ \ \\[0.3cm]
\label{ntwofurtheradd411c}
T_0&=&\pm \left[\det\left(\openone_6 - A\right)\right]^{-\frac{1}{2}}
\left(1 - i\ {\rm Pf} A\right)\ ,\ \ \ \ \ \\[0.3cm]
\label{ntwofurtheradd411d}
T_P&=&\mp \left[\det\left(\openone_6 - A\right)\right]^{-\frac{1}{2}}
\left(A_P - i\ Pc^{\; 2}(A)_P\right)\ .
\end{eqnarray}
Eq.\ (\ref{ntwofurtheradd411a}) agrees with eq.\ (27) of
\citep{bogu1} (p.\ 988, p.\ 258 of the English transl.). 
For a related result see \citep{klot1}, p.\ 2245, eq.\ (13)
[or its discussion in our Subsec.\ \ref{klotz}, 
eq.\ (\ref{ntwofurtheradd421})]. It also yields an alternative to 
eq.\ (\ref{ntwofurtheradd319b}) by relying on the Cayley representation
[eq.\ (\ref{ntwofurtheradd401})] of the matrix $L$ 
[cf.\ eq.\ (26) of \citep{bogu1} (p.\ 988, p.\ 258 of the English transl.)]
\begin{eqnarray}
\label{ntwofurtheradd412}
A&=&-\ \frac{{\rm Re}\; T}{{\rm Re}\;  T_0}\ .
\end{eqnarray}

The eqs.\ (\ref{ntwofurtheradd411c}), (\ref{ntwofurtheradd411d}) 
can now be inserted into eq.\ (\ref{ntwofurtheradd319b}) to obtain 
the following further version of eq.\ (\ref{ntwofurtheradd401}).
\begin{eqnarray}
\label{ntwofurtheradd413a}
L&=&\openone_6\ +\ 
\frac{2 A (\openone_6 + A)}{\det\left(\openone_6 - A\right)}
\left[\left(1 - \frac{1}{2}\ {\rm tr}\; A^2\right) \openone_6 + A^2 -
\left[Pc^{\; 2}(A)\right]^2\right]
\end{eqnarray}
Besides more elementary ones we have made use of the following result
in deriving eq.\ (\ref{ntwofurtheradd413a}).
\begin{eqnarray}
\label{ntwofurtheradd414}
\epsilon_{KMN}\ {\rm Re}\; T^M\ {\rm Im}\; T^N
&=&\frac{1}{\det\left(\openone_6 - A\right)}
\left[- \frac{1}{2}\ A_K\ {\rm tr}\; A^2\ 
+ \left(A^3\right)_K\right]
\end{eqnarray}
Eq.\ (\ref{ntwofurtheradd413a}) can be further transformed to read 
\begin{eqnarray}
\label{ntwofurtheradd415}
L&=&\openone_6\ +\ 2 A\nonumber\\[0.3cm]
&&-\ \frac{2 (\openone_6 + A)}{\det\left(\openone_6 - A\right)}
\left[\left(\det A\right) \openone_6 - A^2\left(
\left(1 - \frac{1}{2}\ {\rm tr}\; A^2\right) \openone_6 + A^2\right)
\right]\ .
\end{eqnarray}
This form agrees with the general eq.\ (3) in ref.\ \citep{bogu2}.\\

\subsubsection{\label{klotz}The approach by Klotz}

The approach by Klotz 
\citep{klot1}\footnote{Note, that 
an analogous approach has been used earlier by Macfarlane (leaving
aside here the problem of the signature in the spaces used)
in the cases of $SO(4;\mathbb{R})$ \citep{macf2} and 
$SO(5;\mathbb{R})$ \citep{macf1}, p.\ 145.}  
(adapted to our case) for relating
$SO(6;\mathbb{R})$ transformations $L$ 
and the corresponding unitary 
$SU(4)$ transformations ${\sf U}\left(\{\lambda\};\{\hat{c}\}\right)$
starts by considering the relation
\begin{eqnarray}
\label{ntwofurtheradd416}
\left(\hat{\hat{\hat{c}\hspace{0.3mm}}}_N\right)_a^{\ b}
\left(\hat{\hat{\hat{c}\hspace{0.3mm}}}\hspace{-1mm}\ ^N\right)_c^{\ d}&=&
\delta_a^{\ b}\delta_c^{\ d}\ -\ 4\ \delta_a^{\ d}\delta_c^{\ b}
\end{eqnarray}
which goes back to Pauli\footnote{\citep{paul1}, p.\ 32 
(p.\ 725 of Vol.\ 2 of the reprint \citep{kron1}), eq.\ (I), 

\noindent
\citep{paul2}, p.\ 118
(p.\ 762 of Vol.\ 2 of the reprint \citep{kron1}), eq.\ (II)
Also cf.\ e.g., \\[-0.75cm]
\begin{tabbing}
\citep{good1}, p.\ 210, eq.\ (233),\hskip1.6cm \=
\citep{hris4}, p.\ 35, eq.\ (11),\hskip1cm \=
\citep{macf1}, p.\ 137, eq.\ (2.22),\\
\citep{buch1}, p.\ 361, eq.\ (4.2),
\>\citep{klot1}, p.\ 2245, Sec.\ IV,
\>\citep{nash1}, p.\ 208, eq.\ (50),\\
\citep{lauf1}, Sec.\ 5.3, p.\ 167, eq.\ (5.22),
\>\citep{andr2}, p.\ 44, eq.\ (5a),
\>\citep{andr3}, Subsec.\ 2.1.1, p.\ 18, eq.\ (47),\\
\citep{andr1}, p.\ 12 (p.\ 10 of the English transl.), eq.\ ($2^\prime$).
\end{tabbing}}. 
Here, $\left(\hat{\hat{\hat{c}\,}}_N\right)_a^{\ b}$ denotes
the matrix elements of the operator ($4\times 4$) matrix
$\hat{\hat{\hat{c}\,}}_N$. Eq.\ (\ref{ntwofurtheradd416})
can easily be derived by means of the eqs.\ 
(\ref{ntwofurtheradd65aa}),
(\ref{ntwofurtheradd304}), (\ref{ntwofurtheradd307}),
(\ref{ntwofurtheradd303b})-(\ref{ntwofurtheradd303d}).
Multiplying eq.\ (\ref{ntwofurtheradd416})
by ${\sf U}_{a^\prime}^{\ \ a}$ and $\overline{\sf U}^{\; c}_{\;\ b}$ 
[i.e., certain matrix elements of the unitary operator
${\sf U}\left(\{\lambda\};\{\hat{c}\}\right) \in SU(4)$] and performing
the sum over the indices $a$, $b$, $c$, one can derive the following
equation (in operator notation). 
\begin{eqnarray}
\label{ntwofurtheradd417}
{\sf U}\ {\rm tr}\;\overline{\sf U}&=&
\frac{1}{4}\left[\openone_4 - 
{\sf U}\; \hat{\hat{\hat{c}\,}}_N\; {\sf U}^\dagger\;
\hat{\hat{\hat{c}\,}}\hspace{-1mm}\ ^N\right]
\end{eqnarray}
Recognizing, that ${\sf U}\; \hat{\hat{\hat{c}\,}}_N\; {\sf U}^\dagger$
can be written as $\chi_N^{\ \ \ M}\hat{\hat{\hat{c}\,}}_M$
and taking into account the relation 
${\rm tr}\;\overline{\sf U} = 4\ \overline{T}_0$
[cf.\ eq.\ (\ref{ntwofurtheradd316b})] we find the expression
\begin{eqnarray}
\label{ntwofurtheradd418}
{\sf U}&=&
\frac{T_0}{16\ \vert T_0\vert^2} \left[\openone_4\ -\ 
\chi_{NM}\; \hat{\hat{\hat{c}\,}}\hspace{-1mm}\ ^M 
\hat{\hat{\hat{c}\,}}\hspace{-1mm}\ ^N\right]\ .
\end{eqnarray}
By taking the trace of eq.\ (\ref{ntwofurtheradd418}) one can
derive the relation (${\rm tr}\;\chi = \chi_M^{\ \ \ M}$) 
\begin{eqnarray}
\label{ntwofurtheradd419}
\vert T_0\vert^2&=&\frac{1}{16}\left[\;
1 + {\rm tr}\;\chi\; \right]\ =\ \frac{1}{16}\left[
1 + \frac{1}{2}\; \left({\rm tr}\; L\right)^2
- \frac{1}{2}\; {\rm tr}\; L^2\right]
\end{eqnarray}
which, however, still leaves the (complex) phase of $T_0$ undetermined.
One can convince oneself that this result agrees with the corresponding
expression derived from eq.\ (\ref{ntwofurtheradd411c}). 
To see this explicitly one has to rely on the relation
\begin{eqnarray}
\label{ntwofurtheradd420}
\frac{1}{6}\; \left({\rm tr}\; L\right)^4
\ -\ \left({\rm tr}\; L\right)^2\; {\rm tr}\; L^2
\ +\ \frac{4}{3}\; {\rm tr}\; L\; {\rm tr}\; L^3&&\nonumber\\[0.3cm]
\ +\ \frac{1}{2}\; \left({\rm tr}\; L^2\right)^2
\ -\ {\rm tr}\; L^4
&=&2\ \left({\rm tr}\; L\right)^2\ -\ 2\ {\rm tr}\; L^2
\end{eqnarray}
which can be derived from eq.\ (\ref{ntwofurtheradd78de})
by means of taking appropriate trace operations on both sides.  
Further progress in discussing eq.\ (\ref{ntwofurtheradd418})
can now be made by relying on eq.\ (\ref{ntwofurtheradd63za}).
Finally, eq.\ (\ref{ntwofurtheradd418}) is found to read
[${\rm tr}\; L = {\rm tr}\; L^{(+)}$,
$Pc^2(L) = Pc^2(L^{(-)})$]
\begin{eqnarray}
\label{ntwofurtheradd421b}
{\sf U}&=&T_0 \left[ \openone_4\ - \frac{1}{16\ \vert T_0\vert^2}
\left(\chi_{nk\ \ l}^{\ \ \ n}\ -\ 
i\ \chi^{PQ} \epsilon_{PQkl}\right)\; 
\hat{\hat{\hat{c}\,}}\hspace{-1mm}\ ^{kl}\;\right]\\[0.3cm]
\label{ntwofurtheradd421}
&=&T_0 \left[ \openone_4\ -\ 
\frac{1}{8\ \vert T_0\vert^2} \left\{ L^{(-)}\left[ 
\left({\rm tr}\; L\right) \openone_6 - 2\; L^{(+)}\right]
- 2 i\ Pc^2(L)\right\}_M \hat{\hat{\hat{c}\,}}\hspace{-1mm}\ ^M 
\right]\ .\ \ \ \
\end{eqnarray}
This entails 
\begin{eqnarray}
\label{ntwofurtheradd422}
T_M&=&-\ \frac{2\ T_0}{1 + 
\frac{1}{2}\; \left({\rm tr}\; L\right)^2
- \frac{1}{2}\; {\rm tr}\; L^2}\nonumber\\[0.3cm]
&&\ \ \ \times\ \ 
\left\{ L^{(-)}\left[ 
\left({\rm tr}\; L\right) \openone_6 - 2\; L^{(+)}\right]
- 2 i\ Pc^2(L)\right\}_M
\end{eqnarray}
which stands in correspondence to eq.\ 
(\ref{ntwofurtheradd411d})\footnote{Infinitesimally 
$L^{(-)}\simeq 2 A $, and for this domain 
one immediately recognizes that eq.\ (\ref{ntwofurtheradd422})
agrees with eq.\ (\ref{ntwofurtheradd411d}).
However, we have not attempted to explicitly check the equivalence
of the eqs.\ (\ref{ntwofurtheradd411d}) and (\ref{ntwofurtheradd422})
in general.}.\\

It now remains to calculate the (complex) phase of $T_0$. 
Klotz \citep{klot1} incorrectly states on p.\ 2245
[above of eq.\ (13)] that the (complex) phase of $T_0$ can assume the (fixed!)
values $0,\frac{\pi}{2},\pi,\frac{3\pi}{2}$ only. This statement does 
not follow (as Klotz suggests) from the condition $\det{\sf U} = 1$
and is also in contradiction to eq.\ (\ref{ntwofurtheradd411c}).
Following the reasoning of Klotz \citep{klot1},
Laufer \citep{lauf1} has also discussed (Sec.\ 5.3, p.\ 162-179)
the homomorphism between $SU(2,2)$ and $SO(2,4;\mathbb{R})$ (The difference in 
signature to our case $SU(4)$, $SO(6;\mathbb{R})$, 
can be disregarded for the present purpose.) and makes the same 
erroneous statement\footnote{Bottom of p.\ 169; incidentally, 
the problem of the complex phase does not occur for smaller groups 
and, therefore, does not show up in the work of Macfarlane \citep{macf1}.}. 
As a matter of fact, $T_0$
can be calculated from the condition $\det{\sf U} = 1$ as some
fourth (complex!) root and any root will differ from the other three
by one of the phases $\frac{\pi}{2},\pi,\frac{3\pi}{2}$ (mod $2\pi$)
but can be complex itself, in principle. The determinant can be 
calculated by means of the 
formula\footnote{Cf.\ \citep{eddi5}, Sec.\ 3.6, p.\ 42, eq.\ (3.62), 
\citep{mccr2},
\citep{hari1}, p.\ 37 (reprint \# 1: p.\ 55, reprint \# 2: p.\ 948), 
eq.\ (44), 
\citep{eddi6}, Sec.\ 65, p.\ 129, eq.\ (65.1),
\citep{take1}, Part II, \S 11, pp.\ 160/161,
\citep{step1}, Sec.\ 3, p.\ 819, eq.\ (40). Be aware of misprints and
errors in the various versions of the equation as displayed 
in the references cited.}
\begin{eqnarray}
\label{ntwofurtheradd423}
\det{\sf U}&=&T_0^4\ -\ T_0^2\ {\rm tr}\; T^2
\ -\ \frac{1}{4}\left({\rm tr}\; T^2\right)^2
\ +\ {\rm tr}\; T^4
\ +\ 8i\ T_0\ {\rm Pf}\; T\ .
\end{eqnarray}
The sign of the last term is determined by our choice of the 
paravector space (specifically, our choice for the sign of the
component $\hat{\hat{\hat{c}\,}}_0 = - \openone_4$). 
The explicit calculation of $T_0$ in terms of the matrix
$L$ by means of eq.\ (\ref{ntwofurtheradd423}) turns out to be quite 
tedious. In view of the fact, that eq.\ (\ref{ntwofurtheradd411c})
provides us with an explicit expression of $T_0$ in terms of the 
matrix $A$ we will not further pursue this subject here. The 
algebraic complexity of the calculation of $T_0$ in terms of the matrix
$L$ seems to be related to the fact mentioned below from 
eq.\ (\ref{ntwofurtheradd411bb}) that the expression relating the 
antisymmetric matrix $A$ to a sum of a finite number of powers of 
the matrix $L$ is algebraically involved.\\

The alert reader will have noticed that the method of Klotz \citep{klot1}
delivers twice as many results for the matrix ${\sf U}$
than the method of Fedorov and collaborators \citep{bogu2,bogu1} 
(Of course, the difference lies in the number of allowed values for the 
total phase of ${\sf U}$ only.) 
and than is required for the group $SU(4)$ as the 
\underline{double} cover of $SO(6;\mathbb{R})$. The explanation lies in
the respective starting points of the two methods. In our version 
of the approach, the starting point
of the method of Fedorov and collaborators, eqs.\ (\ref{ntwofurtheradd318}), 
(\ref{ntwofurtheradd318b}), is 
$\mathbb{Z}_2$-invariant for the elements of $SU(4)$. On the other
hand, eq.\ (\ref{ntwofurtheradd417}), the starting point of Klotz
is invariant under the center $\mathbb{Z}_4$ of $SU(4)$.\\

\subsection{\label{chiinverse}From 
\texorpdfstring{$SO(6;\mathbb{R})/\mathbb{Z}_2$}{SO(6;R)/Z\_2} 
back to its double cover 
\texorpdfstring{$SO(6;\mathbb{R})$}{SO(6;R)}} 

We will now invert eq.\ (\ref{ntwofurtheradd78c}), i.e., we will express
the elements of the matrix $L$ in terms of the elements 
of the (compound) matrix $\chi\ = \ C_2(L)$. This task occurs if
one wants to find from the coefficients $\chi_{0k}^{\ \ M}$ 
of a given Bogolyubov-Valatin
transformation the $SO(6;\mathbb{R})$ matrix $L$ generating it.
We would like to discuss two possible lines of attack to this
problem which, however, both have their deficiencies.\\

The first method relies on eq.\ (\ref{ntwofurtheradd418}) 
[or eq.\ (\ref{ntwofurtheradd421b})] which can
be inserted into eq.\ (\ref{ntwofurtheradd318}) to obtain the matrix
$L$ directly. As a first step, this method requires the
(tedious) calculation of the complete 
matrix $\chi_{MN}$ from the coefficients 
$\chi_{0k}^{\ \ M}$ [eq.\ (\ref{ntwofurther44h}) and its generalizations]. 
Furthermore, $T_0$ is known from eq.\ (\ref{ntwofurtheradd419}) 
up to its phase only and its calculation entails  
the difficulties discussed at the end of Subsec.\ \ref{klotz}.\\ 
 
The second method to calculate the matrix $L$ from the coefficients
$\chi_{0k}^{\ \ M}$ takes a different route. Consider the 
following matrix elements (We also allow $k,l = 0$ for which 
$\chi_{0k\ 0l}$ vanishes individually.)
\begin{eqnarray}
\label{ntwofurtheradd201}
\chi_{0k\ 0l}&=&\ \beta_I^{\ m}\ \beta_I^{\ n}\ \chi_{mk\ nl}\nonumber\\[0.3cm]
&=&\left(\beta_I L \beta^T_I\right) L_{kl}\ -\ 
\left(L \beta^T_I\right)_k \Big(\beta_I L\Big)_l
\ = \ L_{00} L_{kl}\ -\ L_{k0} L_{0l}\ .
\end{eqnarray}
Of course, the choice of $\beta = \beta_I$ [eq.\ (\ref{ntwofurther60c})]
involves a certain element
of arbitrariness in expressing the matrix $L$ in terms of $\chi$ and
is dictated here by physics considerations and convenience.
We can now formally write (assuming $L_{00}\neq 0$)
\begin{eqnarray}
\label{ntwofurtheradd202}
L_{kl}&=&
\frac{\chi_{0k\ 0l}\ +\ \left(L \beta^T_I\right)_k \Big(\beta_I L\Big)_l
}{\beta_I L \beta^T_I}\\[0.3cm]
\label{ntwofurtheradd202a}
&=&\frac{\chi_{0k\ 0l}\ +\ L_{k0} L_{0l}}{L_{00}}\ .
\end{eqnarray}
Note that $\chi_{0(-1)\ 0l}$ is given by
\begin{eqnarray}
\label{ntwofurtheradd202b}
- \chi_{0(-1)\ p_1 p_2}&=&\chi_{(-1)0\ p_1 p_2}\nonumber\\[0.3cm]
&=&\frac{1}{4!}\ \epsilon_{(-1)0 k_1 k_2 l_1 l_2}\
\epsilon_{p_1 p_2 q_1 q_2 s_1 s_2}\nonumber\\[0.3cm]
&&\ \ \ \ \ \ \ 
\chi_{0 k_1\ m q_1}\ \chi_{0 k_2\ m q_2}\
\chi_{0 l_1\ n s_1}\ \chi_{0 l_2\ n s_2}
\end{eqnarray}
which is obtained by relying on the eqs.\ (\ref{ntwofurtheradd78b}), 
(\ref{ntwofurtheradd65a}) and 
$\hat{\hat{d}\ }_{\hspace{-1mm}(-1)}
\ =\ \hat{\hat{\hat{d}\ }}_{\hspace{-1mm}(-1)0}\ = 
i\hat{d}_1\hat{d}_2\hat{d}_3\hat{d}_4$. The other components of 
$\chi_{0k\ 0l}$ can be read off from eq.\ (\ref{ntwofurther63b}).
We now have to express $L_{k0}$, $L_{0l}$ in terms of $\chi_{0k\ 0l}$ yet.
Let us start with $L_{00}$. Taking the determinant on both sides of
eq.\ (\ref{ntwofurtheradd202}) we find 
\begin{eqnarray}
\label{ntwofurtheradd203}
\det\; L&=&\frac{\det\; \chi_{00}}{L_{00}^4}
\end{eqnarray}
where $\chi_{00}$ denotes the $5\times 5$ matrix with the matrix elements 
$\chi_{0k\ 0l}$ ($k,l\neq 0$). Taking into account the orthogonality of the 
matrix $L$ ($\det\; L = 1$) we finally obtain the relation
\begin{eqnarray}
\label{ntwofurtheradd204}
\vert L_{00}\vert &=&\left(\det\; \chi_{00}\right)^\frac{1}{4}\ .
\end{eqnarray}

The non-diagonal elements $L_{k0}$, $L_{0l}$ can be found
by means of the following consideration.
Taking into account the orthogonality relation $L L^T = \openone_6$
we find the following equation.
\begin{eqnarray}
\label{ntwofurtheradd205}
\chi_{0k}^{\ \ 0m}\; \chi_{0l\ 0m}
&=&\left(L_{k0} - L_{00}\ \delta_{k0}\right)
\left(L_{l0} - L_{00}\ \delta_{l0}\right)\ +\ 
L_{00}^2\ \left(\delta_{kl} - \delta_{k0}\delta_{l0}\right)
\end{eqnarray}
This equation can be transformed to read
\begin{eqnarray}
\label{ntwofurtheradd206}
\left(L_{k0} - L_{00}\ \delta_{k0}\right)
\left(L_{l0} - L_{00}\ \delta_{l0}\right)&=&
\chi_{0k}^{\ \ 0m}\; \chi_{0l\ 0m}\ -\
L_{00}^2 \left(\delta_{kl} - \delta_{k0}\delta_{l0}\right)\ .
\end{eqnarray}
Setting $k=l\ (\neq 0)$ we immediately find
\begin{eqnarray}
\label{ntwofurtheradd207}
\vert L_{k0}\vert &=&\sqrt{\chi_{0k}^{\ \ 0m}\; \chi_{0k\ 0m} - L_{00}^2}
\ =\ \sqrt{\chi_{0k}^{\ \ 0m}\; \chi_{0k\ 0m} - 
\left(\det\; \chi_{00}\right)^\frac{1}{2}}\ .
\end{eqnarray}
Considering $\chi^{0m}_{\ \ \ 0k}\; \chi_{0m\ 0l}$ we 
obtain in an analogous way
\begin{eqnarray}
\label{ntwofurtheradd208}
\vert L_{0k}\vert &=&\sqrt{\chi^{0m}_{\ \ \ 0k}\; \chi_{0m\ 0k} - L_{00}^2}
\ =\ \sqrt{\chi^{0m}_{\ \ \ 0k}\; \chi_{0m\ 0k} - 
\left(\det\; \chi_{00}\right)^\frac{1}{2}}\ .
\end{eqnarray}
As an aside, we also find the following expressions for $k\neq l$ ($\neq 0$)
\begin{eqnarray}
\label{ntwofurtheradd209a}
L_{k0} L_{l0}&=&\chi_{0k}^{\ \ 0m}\; \chi_{0l\ 0m}\ ,
\\[0.3cm]
\label{ntwofurtheradd209b}
L_{0k} L_{0l}&=&\chi^{0m}_{\ \ \ 0k}\; \chi_{0m\ 0l}\ ,
\end{eqnarray}
which entail the following relations for 
the matrix $\chi_{00}$ (no summation over $k,l$ on the r.h.s.)
\begin{eqnarray}
\label{ntwofurtheradd210a}
\chi_{0k}^{\ \ 0m}\; \chi_{0l\ 0m}\ -\ \delta_{kl}\
\left(\det\; \chi_{00}\right)^\frac{1}{2}\hspace{-4cm}&&\nonumber\\[0.3cm]
&=&
\sqrt{\left(\chi_{0k}^{\ \ 0m}\; \chi_{0k\ 0m} - 
\left(\det\; \chi_{00}\right)^\frac{1}{2}\right)
\left(\chi_{0l}^{\ \ 0m}\; \chi_{0l\ 0m} - 
\left(\det\; \chi_{00}\right)^\frac{1}{2}\right)}\ ,\ \ \ \ \ \ 
\\[0.3cm]
\label{ntwofurtheradd210b}
\chi^{0m}_{\ \ \ 0k}\; \chi_{0m\ 0l}\ -\ \delta_{kl}\
\left(\det\; \chi_{00}\right)^\frac{1}{2}\hspace{-4cm}&&\nonumber\\[0.3cm]
&=&
\sqrt{\left(\chi^{0m}_{\ \ \ 0k}\; \chi_{0m\ 0k} - 
\left(\det\; \chi_{00}\right)^\frac{1}{2}\right)
\left(\chi^{0m}_{\ \ \ 0l}\; \chi_{0m\ 0l} - 
\left(\det\; \chi_{00}\right)^\frac{1}{2}\right)}\ .\ \ \ \ \ \ 
\end{eqnarray}

Collecting all results and inserting them into 
eq.\ (\ref{ntwofurtheradd202a}), our final expression for the matrix
$L$ in terms of the matrix $\chi$ (more specifically, in terms of 
its submatrix $\chi_{00}$) reads
\begin{eqnarray}
\label{ntwofurtheradd211}
\vert L_{kl}\vert &=&\left(\det\; \chi_{00}\right)^{-\frac{1}{4}}
\Bigg\vert\chi_{0k\ 0l}\nonumber\\[0.3cm]
&&\ \ \ \ \ \left.\pm \ \sqrt{\left(\chi_{0k}^{\ \ 0m}\; \chi_{0k\ 0m} - 
\left(\det\; \chi_{00}\right)^\frac{1}{2}\right)
\left(\chi^{0m}_{\ \ \ 0l}\; \chi_{0m\ 0l} - 
\left(\det\; \chi_{00}\right)^\frac{1}{2}\right)}\ \ \right\vert\ .
\nonumber\\
&& 
\end{eqnarray}
Already from eq.\  (\ref{ntwofurtheradd78c}) it is clear that the matrix 
elements of the matrix $L$ can be determined from the 
matrix $\chi$ up to a minus sign only. The $\pm$-sign in 
(\ref{ntwofurtheradd211}) reflects an
uncertainty and, therefore, weakness of the applied method
to reconstruct the matrix $L$ from its second compound matrix 
$\chi = C_2(L)$.\\

\section{\label{disc}Discussion}

In this section we collect the discussion of various aspects 
of the study performed in the preceeding parts of the article.
We have postponed these considerations until now in order to streamline
the presentation in the other sections.\\

\subsection{Bogolyubov-Valatin transformations}

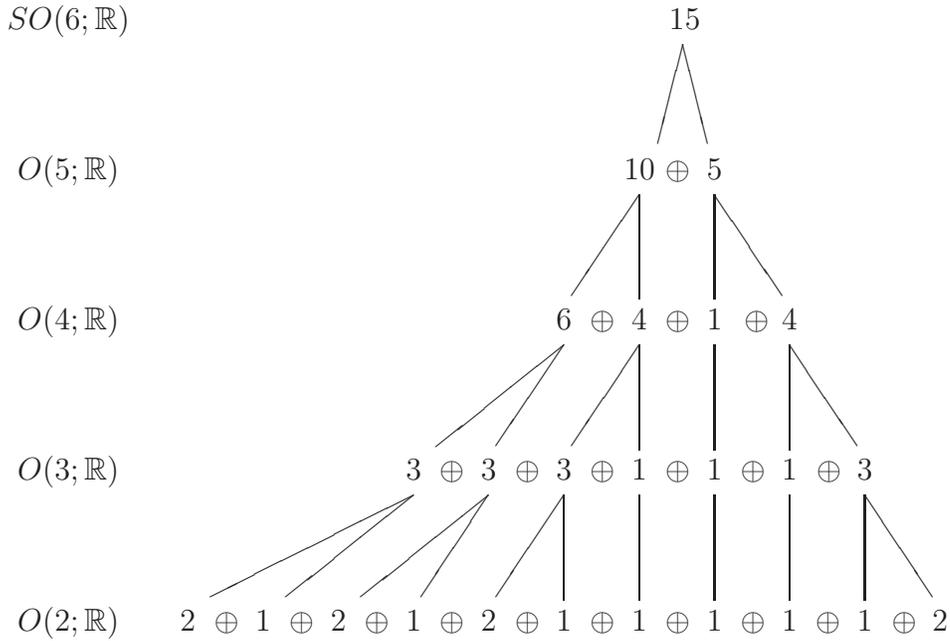
\begin{figure}[ht]
\unitlength1.mm
\begin{picture}(150,100)
\put(2,90){$SO(6;\mathbb{R})$}
\put(90,90){15}
\put(91.8,88){\line(-1,-4){3.3}}
\put(91.8,88){\line(1,-4){3.3}}
\put(2,70){$\ O(5;\mathbb{R})$}
\put(84,70){10}
\put(89.5,70){$\oplus$}
\put(95,70){5}
\put(86,68){\line(0,-1){14}}
\put(86,68){\line(-2,-3){9}}
\put(96,68){\line(0,-1){14}}
\put(96,68){\line(2,-3){9}}
\put(2,50){$\ O(4;\mathbb{R})$}
\put(75,50){6}
\put(79.5,50){$\oplus$}
\put(85,50){4}
\put(89.5,50){$\oplus$}
\put(95,50){1}
\put(100,50){$\oplus$}
\put(105,50){4}
\put(76,48){\line(-5,-4){17}}
\put(76,48){\line(-2,-3){9}}
\put(86,48){\line(0,-1){14}}
\put(86,48){\line(-2,-3){9}}
\put(96,48){\line(0,-1){14}}
\put(106,48){\line(0,-1){14}}
\put(106,48){\line(2,-3){9}}
\put(2,30){$\ O(3;\mathbb{R})$}
\put(55,30){3}
\put(59.5,30){$\oplus$}
\put(65,30){3}
\put(69.5,30){$\oplus$}
\put(75,30){3}
\put(79.5,30){$\oplus$}
\put(85,30){1}
\put(89.5,30){$\oplus$}
\put(95,30){1}
\put(99.5,30){$\oplus$}
\put(105,30){1}
\put(109.6,30){$\oplus$}
\put(115,30){3}
\put(56,28){\line(-2,-1){27}}
\put(56,28){\line(-5,-4){17}}
\put(66,28){\line(-5,-4){17}}
\put(66,28){\line(-2,-3){9}}
\put(76,28){\line(0,-1){14}}
\put(76,28){\line(-2,-3){9}}
\put(86,28){\line(0,-1){14}}
\put(96,28){\line(0,-1){14}}
\put(106,28){\line(0,-1){14}}
\put(116,28){\line(0,-1){14}}
\put(116,28){\line(2,-3){9}}
\put(2,10){$\ O(2;\mathbb{R})$}
\put(25,10){2}
\put(29.5,10){$\oplus$}
\put(35,10){1}
\put(39.5,10){$\oplus$}
\put(45,10){2}
\put(49.5,10){$\oplus$}
\put(55,10){1}
\put(59.5,10){$\oplus$}
\put(65,10){2}
\put(69.5,10){$\oplus$}
\put(75,10){1}
\put(79.5,10){$\oplus$}
\put(85,10){1}
\put(89.5,10){$\oplus$}
\put(95,10){1}
\put(99.5,10){$\oplus$}
\put(105,10){1}
\put(109.5,10){$\oplus$}
\put(115,10){1}
\put(119.6,10){$\oplus$}
\put(125,10){2}
\end{picture}
\caption{\label{groupchain}Decomposition of the 
15-dimensional biparavector space $\bigwedge^2 (V_6)$
under the subgroup chain of $SO(6;\mathbb{R})$.}
\end{figure}

To prepare ourselves for the following discussion we first
have a look at the problem how the 15-dimensional
irreducible representation of the Bogolyubov-Valatin 
group $SO(6;\mathbb{R})/\mathbb{Z}_2$ operating in the
biparavector space $\bigwedge^2 (V_6)$ decomposes into
representations of smaller dimension under the subgroup chain
$SO(6;\mathbb{R})\supset O(5;\mathbb{R}) \supset 
O(4;\mathbb{R}) \supset O(3;\mathbb{R}) \supset  O(2;\mathbb{R})$.
This decomposition can be described by the graphical
representation given in Figure \ref{groupchain}.

\subsubsection{Linear Bogolyubov-Valatin transformations}

Now we will examine the well-known class of 
\underline{linear} [$O(2n = 4;\mathbb{R})$ $\subset$ 
$SO(6;\mathbb{R})$]\footnote{Cf.\ 
e.g., \citep{blai1}, Sec.\ 2.2, p.\ 38. For further references 
concerning linear Bogolyubov-Valatin transformations
see \citep{vanh1}, p.\ 10247, below from eq.\ (16).}
Bogolyubov-Valatin transformations
(Then, the only nonvanishing coefficients in the 
eq.\ (\ref{ntwo3a}) are
$\lambda^{(1\vert 0)}_k$, $\lambda^{(0\vert 1)}_k$, 
$\lambda^{(2\vert 0)}_k$, $\lambda^{(0\vert 2)}_k$.) and see
how the decomposition of the bipara\-vector space 
$\bigwedge^2 (V_6)$ is structured under this class of 
canonical transformations. 
The $6\times 6$ matrix $L\in SO(6;\mathbb{R})$ then has the form
(somewhat symbolically written)
\begin{eqnarray}
\label{ntwofurtheradd80}
L&=&\left(
\begin{array}{*{3}{c}}
\det A&
0&
0\\
0&
1&
0\\
0&
0&
A\end{array}
\right)
\end{eqnarray}
where the $4\times 4$ matrix $A\in O(4;\mathbb{R})$.
The matrix $A = A\left(\{\lambda\}\right)$ reads
[cf.\ the eqs.\ (\ref{ntwoadd13a}), (\ref{ntwoadd13b}),
and (\ref{ntwoadd33})]:
\begin{eqnarray}
\label{ntwofurther3}
A\ =\ A\left(\{\lambda\}\right)&=&\left(
\begin{array}{*{4}{c}}
{\rm Re}\;\kappa_1^{(1\vert 0)}&
{\rm Re}\;\kappa_1^{(0\vert 1)}&
{\rm Re}\;\kappa_1^{(2\vert 0)}&
{\rm Re}\;\kappa_1^{(0\vert 2)}\\
{\rm Im}\;\kappa_1^{(1\vert 0)}&
{\rm Im}\;\kappa_1^{(0\vert 1)}&
{\rm Im}\;\kappa_1^{(2\vert 0)}&
{\rm Im}\;\kappa_1^{(0\vert 2)}\\
{\rm Re}\;\kappa_2^{(1\vert 0)}&
{\rm Re}\;\kappa_2^{(0\vert 1)}&
{\rm Re}\;\kappa_2^{(2\vert 0)}&
{\rm Re}\;\kappa_2^{(0\vert 2)}\\
{\rm Im}\;\kappa_2^{(1\vert 0)}&
{\rm Im}\;\kappa_2^{(0\vert 1)}&
{\rm Im}\;\kappa_2^{(2\vert 0)}&
{\rm Im}\;\kappa_2^{(0\vert 2)}\end{array}
\right)\ =\ \chi^{[1]}\ .\ \ \ 
\end{eqnarray}
Here, $\chi^{[1]}$ denotes the matrix with row number $k$ and 
column number $m$ matrix elements $\chi^{[1]\ (m)}_k$. Eq.\ 
(\ref{ntwofurtheradd80}) can be found by relying on Subsec.\
\ref{chiinverse} and on eq.\ (\ref{ntwofurtheradd78}).\\

Consider now the linear space $W$ (of operators in Fock space) 
which is the direct sum of the space spanned by the identity operator
in spin space $\mathbb{C}_4$ (by the way, an invariant
subspace) and the space of 
biparavectors $\bigwedge^2 (V_6)$. It is 
spanned by the basis\footnote{To achieve a more compact graphical
display, we give here the
Hermitian conjugate $a^\dagger$. Be aware of the minus signs and note
that in the last line for simplicity we have used a somewhat imprecise
notation [cf.\ eq.\ (\ref{ntwofurtheradd79})].}
\begin{eqnarray}
\label{ntwofurther1a}
a^\dagger&=&\left(\openone_4,
-i\hat{a}_1^{[1]}\wedge\hat{a}_1^{[2]}\wedge\hat{a}_2^{[1]}\wedge\hat{a}_2^{[2]},\right.\nonumber\\[0.3cm]
&&\ \ 
i\hat{a}_1^{[2]}\wedge\hat{a}_2^{[1]}\wedge\hat{a}_2^{[2]},
-i\hat{a}_1^{[1]}\wedge\hat{a}_2^{[1]}\wedge\hat{a}_2^{[2]},
i\hat{a}_1^{[1]}\wedge\hat{a}_1^{[2]}\wedge\hat{a}_2^{[2]},
-i\hat{a}_1^{[1]}\wedge\hat{a}_1^{[2]}\wedge\hat{a}_2^{[1]},
\nonumber\\[0.3cm]
&&\ \ 
-\hat{a}_1^{[1]},
-\hat{a}_2^{[1]},
-\hat{a}_1^{[2]},
-\hat{a}_2^{[2]},\nonumber\\[0.3cm]
&&\ \ 
-\hat{a}_1^{[1]}\wedge\hat{a}_1^{[2]},
-\hat{a}_1^{[1]}\wedge\hat{a}_2^{[1]},
-\hat{a}_1^{[1]}\wedge\hat{a}_2^{[2]},\nonumber\\[0.3cm]
&&\ \ \left.
-\hat{a}_1^{[2]}\wedge\hat{a}_2^{[1]},
-\hat{a}_1^{[2]}\wedge\hat{a}_2^{[2]},
-\hat{a}_2^{[2]}\wedge\hat{a}_2^{[2]}\right)\\[0.3cm]
\label{ntwofurther1aa}
&=&\left(\openone_4,-i\hat{c}_1\wedge\hat{c}_2\wedge\hat{c}_3\wedge\hat{c}_4,
\right.\nonumber\\[0.3cm]
&&\ \ 
i\hat{c}_2\wedge\hat{c}_3\wedge\hat{c}_4,
-i\hat{c}_1\wedge\hat{c}_3\wedge\hat{c}_4,
i\hat{c}_1\wedge\hat{c}_2\wedge\hat{c}_4,
-i\hat{c}_1\wedge\hat{c}_2\wedge\hat{c}_3,\nonumber\\[0.3cm]
&&\ \ 
-\hat{c}_1,
-\hat{c}_2,
-\hat{c}_3,
-\hat{c}_4,\nonumber\\[0.3cm]
&&\ \ \left.
-\hat{c}_1\wedge\hat{c}_2,
-\hat{c}_1\wedge\hat{c}_3,
-\hat{c}_1\wedge\hat{c}_4,
-\hat{c}_2\wedge\hat{c}_3,
-\hat{c}_2\wedge\hat{c}_4,
-\hat{c}_3\wedge\hat{c}_4,\right)\nonumber\\[0.3cm]
&=&\left(\openone_4,\hat{\hat{\hat{c}\hspace{0.4mm}}}^\dagger_{(-1)0},
\hat{\hat{\hat{c}\hspace{0.4mm}}}^\dagger_{(-1)1},
\hat{\hat{\hat{c}\hspace{0.4mm}}}^\dagger_{(-1)2},
\hat{\hat{\hat{c}\hspace{0.45mm}}}^\dagger_{(-1)3},
\hat{\hat{\hat{c}\hspace{0.3mm}}}^\dagger_{(-1)4},
\right.\nonumber\\[0.3cm]
&&\ \ \ \ \left.
\hat{\hat{\hat{c}\hspace{0.4mm}}}^\dagger_{01},
\hat{\hat{\hat{c}\hspace{0.3mm}}}^\dagger_{02},
\hat{\hat{\hat{c}\hspace{0.3mm}}}^\dagger_{03},
\hat{\hat{\hat{c}\hspace{0.4mm}}}^\dagger_{04},
\hat{\hat{\hat{c}\hspace{0.3mm}}}^\dagger_{12},
\hat{\hat{\hat{c}\hspace{0.3mm}}}^\dagger_{13},
\hat{\hat{\hat{c}\hspace{0.3mm}}}^\dagger_{14},
\hat{\hat{\hat{c}\hspace{0.3mm}}}^\dagger_{23},
\hat{\hat{\hat{c}\hspace{0.4mm}}}^\dagger_{24},
\hat{\hat{\hat{c}\hspace{0.3mm}}}^\dagger_{34}
\right)\nonumber\\[0.3cm]
&=&\left(\openone_4,\hat{\hat{\hat{C}\hspace{0.8mm}}}^\dagger\right)\ .
\end{eqnarray}
Also defining\footnote{Below we have not indicated that the wedge product
relates to a different vector space than in eq.\ (\ref{ntwofurther1a}), i.e.: 
$\wedge\ =\ \stackrel{b}{\displaystyle \wedge}\ =\ 
\stackrel{d}{\displaystyle \wedge}$.}
\begin{eqnarray}
\label{ntwofurther1b}
b^\dagger&=&\left(\openone_4,
-i\hat{b}_1^{[1]}\wedge\hat{b}_1^{[2]}\wedge\hat{b}_2^{[1]}\wedge\hat{b}_2^{[2]},\right.\nonumber\\[0.3cm]
&&\ \ 
i\hat{b}_1^{[2]}\wedge\hat{b}_2^{[1]}\wedge\hat{b}_2^{[2]},
-i\hat{b}_1^{[1]}\wedge\hat{b}_2^{[1]}\wedge\hat{b}_2^{[2]},
i\hat{b}_1^{[1]}\wedge\hat{b}_1^{[2]}\wedge\hat{b}_2^{[2]},
-i\hat{b}_1^{[1]}\wedge\hat{b}_1^{[2]}\wedge\hat{b}_2^{[1]},
\nonumber\\[0.3cm]
&&\ \ 
-\hat{b}_1^{[1]},
-\hat{b}_2^{[1]},
-\hat{b}_1^{[2]},
-\hat{b}_2^{[2]},\nonumber\\[0.3cm]
&&\ \ 
-\hat{b}_1^{[1]}\wedge\hat{b}_1^{[2]},
-\hat{b}_1^{[1]}\wedge\hat{b}_2^{[1]},
-\hat{b}_1^{[1]}\wedge\hat{b}_2^{[2]},\nonumber\\[0.3cm]
&&\ \ \left.
-\hat{b}_1^{[2]}\wedge\hat{b}_2^{[1]},
-\hat{b}_1^{[2]}\wedge\hat{b}_2^{[2]},
-\hat{b}_2^{[2]}\wedge\hat{b}_2^{[2]}\right)\\[0.3cm]
\label{ntwofurther1bb}
&=&\left(\openone_4,-i\hat{d}_1\wedge\hat{d}_2\wedge\hat{d}_3\wedge\hat{d}_4,
\right.\nonumber\\[0.3cm]
&&\ \ 
i\hat{d}_2\wedge\hat{d}_3\wedge\hat{d}_4,
-i\hat{d}_1\wedge\hat{d}_3\wedge\hat{d}_4,
i\hat{d}_1\wedge\hat{d}_2\wedge\hat{d}_4,
-i\hat{d}_1\wedge\hat{d}_2\wedge\hat{d}_3,\nonumber\\[0.3cm]
&&\ \ 
-\hat{d}_1,
-\hat{d}_2,
-\hat{d}_3,
-\hat{d}_4,\nonumber\\[0.3cm]
&&\ \ \left.
-\hat{d}_1\wedge\hat{d}_2,
-\hat{d}_1\wedge\hat{d}_3,
-\hat{d}_1\wedge\hat{d}_4,
-\hat{d}_2\wedge\hat{d}_3,
-\hat{d}_2\wedge\hat{d}_4,
-\hat{d}_3\wedge\hat{d}_4,\right)\nonumber\\[0.3cm]
&=&\left(\openone_4,\hat{\hat{\hat{d}\hspace{1.3mm}}}^\dagger_{(-1)0},
\hat{\hat{\hat{d}\hspace{1.3mm}}}^\dagger_{(-1)1},
\hat{\hat{\hat{d}\hspace{1.3mm}}}^\dagger_{(-1)2},
\hat{\hat{\hat{d}\hspace{1.3mm}}}^\dagger_{(-1)3},
\hat{\hat{\hat{d}\hspace{1.5mm}}}^\dagger_{(-1)4},
\right.\nonumber\\[0.3cm]
&&\ \ \ \ \left.
\hat{\hat{\hat{d}\hspace{1.3mm}}}^\dagger_{01},
\hat{\hat{\hat{d}\hspace{1.3mm}}}^\dagger_{02},
\hat{\hat{\hat{d}\hspace{1.5mm}}}^\dagger_{03},
\hat{\hat{\hat{d}\hspace{1.3mm}}}^\dagger_{04},
\hat{\hat{\hat{d}\hspace{1.3mm}}}^\dagger_{12},
\hat{\hat{\hat{d}\hspace{1.3mm}}}^\dagger_{13},
\hat{\hat{\hat{d}\hspace{1.5mm}}}^\dagger_{14},
\hat{\hat{\hat{d}\hspace{1.3mm}}}^\dagger_{23},
\hat{\hat{\hat{d}\hspace{1.3mm}}}^\dagger_{24},
\hat{\hat{\hat{d}\hspace{1.3mm}}}^\dagger_{34}
\right)\nonumber\\[0.3cm]
&=&\left(\openone_4,\hat{\hat{\hat{D}\hspace{0.5mm}}}^\dagger\right)
\end{eqnarray}
we can write any two-mode Bogolyubov-Valatin transformation
as a basis transformation in the linear space $W$.
\begin{eqnarray}
\label{ntwofurther2}
b&=& {\cal A}\left(\{\lambda\}\right)\ a
\end{eqnarray}
For a linear Bogolyubov-Valatin transformation,
the matrix $\cal A$ assumes a block diagonal form. It reads
[cf.\ eq.\ (\ref{ntwofurtheradd78})]:
\begin{eqnarray}
\label{ntwofurther4a}
{\cal A}\left(\{\lambda\}\right)&=&
{\rm diag}\left[C_0(A),C_4(A),C_3(A^T)^\star,C_1(A),C_2(A)\right]
\end{eqnarray}
Here, $C_l(A)$ denotes compound matrices of the matrix $A$
(cf.\ Appendix \ref{appcomp}).
Let us denote the subspace
in which the compound matrix $C_l(A)$ operates by $W_l$
($W = W_0\oplus W_4\oplus W_3\oplus W_1\oplus W_2)$.
With the usual relations for compound matrices and the 
orthogonality condition for the matrix $A$ we can write
eq.\ (\ref{ntwofurther4a}) as:
\begin{eqnarray}
\label{ntwofurther4b}
{\cal A}\left(\{\lambda\}\right)&=&
{\rm diag}\left[1,\det A,(\det A) A,A,C_2(A)\right]
\end{eqnarray}
For the indices of the compound matrices we have applied here the
usual convention of the lexicographical order of the indices.
Note that the above representation of ${\cal A}$ in terms of 
compound matrices requires (and, therefore, induces) a certain
order in the sequence 
of the basis elements of the linear subspaces $W_2$,
$W_3$ invariant under the $O(4;\mathbb{R})$ transformation
[cf.\ eqs.\ 
(\ref{ntwofurther1a}), (\ref{ntwofurther1b})] once the sequence
in the subspace $W_1$  (spanned by $\hat{a}_1^{[1]}=\hat{c}_1$, 
$\hat{a}_1^{[2]}=\hat{c}_2$, $\hat{a}_2^{[1]}=\hat{c}_3$,
$\hat{a}_2^{[2]}=\hat{c}_4$) has been chosen.\\

\subsubsection{Nonlinear Bogolyubov-Valatin transformations 
for one fer\-mi\-on mode}

To make further contact with previously known results,
we consider now the case of a full nonlinear 
$SO(3;\mathbb{R})$ Bogolyubov-Valatin transformation of \underline{one} 
fermion mode (say, with mode number 1) considered in \citep{vanh1}.
\begin{eqnarray}
\label{ntwofurther8add1}
\hat{b}_1&=&{\sf B}_1\left(\{\lambda\};\{\hat{a}\}\right)
\ =\ \lambda_1^{(0\vert 1)}\ \hat{a}_1\ +\
\lambda_1^{(1\vert 0)}\ \hat{a}_1^\dagger\ +\
\lambda_1^{(1\vert 1)}\ 
\left(\hat{a}_1^\dagger\hat{a}_1 -\frac{1}{2}\ \openone_4\right)
\end{eqnarray}
The coefficients $\lambda_1^{(0\vert 1)}$, $\lambda_1^{(1\vert 0)}$, 
$\lambda_1^{(1\vert 1)}$ obey the equations (cf.\ \citep{vanh1}, p.\
10247)
\begin{eqnarray}
\label{ntwofurther8add2a}
\vert\lambda_1^{(1\vert 0)}\vert^2\ +\
\vert\lambda_1^{(0\vert 1)}\vert^2\ +\
\frac{1}{2}\ \vert\lambda_1^{(1\vert 1)}\vert^2&=&1\ ,\\[0.3cm]
\label{ntwofurther8add2b}
4\ \lambda_1^{(1\vert 0)}\lambda_1^{(0\vert 1)}\ +\ 
\left(\lambda_1^{(1\vert 1)}\right)^2&=&0\ .
\end{eqnarray}
For this case, the $6\times 6$ matrix $L\in SO(6;\mathbb{R})$ has the form
[cf.\ eq.\ (\ref{ntwofurtheradd78})]
\begin{eqnarray}
\label{ntwofurther8add3}
L&=&\left(
\begin{array}{*{6}{c}}
P_{(-1)(-1)}&
0&
0&
0&
P_{(-1)3}&
P_{(-1)4}\\
0&
Q_{00}&
Q_{01}&
Q_{02}&
0&
0\\
0&
Q_{10}&
Q_{11}&
Q_{12}&
0&
0\\
0&
Q_{20}&
Q_{21}&
Q_{22}&
0&
0\\
P_{3(-1)}&
0&
0&
0&
P_{33}&
P_{34}\\
P_{4(-1)}&
0&
0&
0&
P_{43}&
P_{44}
\end{array}
\right)
\end{eqnarray}
where the $3\times 3$ matrices $P,\ Q\in SO(3;\mathbb{R})$.
The orthogonal matrix $Q$ is related to the 
orthogonal matrix $A\left(\{\lambda\}\right)$
used in eq.\ (23), p.\ 10248 in \citep{vanh1} the following way
\begin{eqnarray}
\label{ntwofurther8add4a}
A\left(\{\lambda\}\right)&=&
\left(
\begin{array}{*{3}{c}}
{\rm Re}\;\kappa^{(1\vert 0)}&
{\rm Re}\;\kappa^{(0\vert 1)}&
{\rm Re}\;\kappa^{(1\vert 1)}\\
{\rm Im}\;\kappa^{(1\vert 0)}&
{\rm Im}\;\kappa^{(0\vert 1)}&
{\rm Im}\;\kappa^{(1\vert 1)}\\
{\rm Im}\left(\overline{\kappa^{(0\vert 1)}}\kappa^{(1\vert 1)}\right)&
{\rm Im}\left(\overline{\kappa^{(1\vert 1)}}\kappa^{(1\vert 0)}\right)&
{\rm Im}\left(\overline{\kappa^{(1\vert 0)}}\kappa^{(0\vert 1)}\right)
\end{array}
\right)\nonumber\\[0.3cm]
&=&C_2\left(Q\right)\ .
\end{eqnarray}
Here,
\begin{eqnarray}
\label{ntwofurther8add4b}
\kappa^{(1\vert 0)}&=&\lambda^{(0\vert 1)}\ +\ \lambda^{(1\vert 0)}\ ,\\[0.3cm]
\label{ntwofurther8add4c}
\kappa^{(0\vert 1)}&=&
i\left(\lambda^{(0\vert 1)}\ -\ \lambda^{(1\vert 0)}\right)\ ,\\[0.3cm]
\label{ntwofurther8add4d}
\kappa^{(1\vert 1)}&=&\lambda^{(1\vert 1)}\ .
\end{eqnarray}
Relying on the equation (for the notation applied 
cf.\ Appendix \ref{appcomp}) 
\begin{eqnarray}
\label{ntwofurther8add5a}
A^{-1}&=&A^T\ =\ C_1\left(Q\right)^\star\ =\ Q^\star\ =\ 
\varepsilon\ Q^T\ \varepsilon 
\end{eqnarray}
where
\begin{eqnarray}
\label{ntwofurther8add5b}
\varepsilon &=&\left(
\begin{array}{*{3}{c}}
0&0&1\\
0&-1&0\\
1&0&0
\end{array}
\right)
\end{eqnarray}
we quickly find 
\begin{eqnarray}
\label{ntwofurther8add5c}
Q&=&\varepsilon\ A\left(\{\lambda\}\right)\ \varepsilon\nonumber\\[0.3cm]
&=&\left(
\begin{array}{*{3}{c}}
{\rm Im}\left(\overline{\kappa^{(1\vert 0)}}\kappa^{(0\vert 1)}\right)&
-{\rm Im}\left(\overline{\kappa^{(1\vert 1)}}\kappa^{(1\vert 0)}\right)&
{\rm Im}\left(\overline{\kappa^{(0\vert 1)}}\kappa^{(1\vert 0)}\right)\\
-{\rm Im}\;\kappa^{(1\vert 1)}&
\ \ {\rm Im}\;\kappa^{(0\vert 1)}&
-{\rm Im}\;\kappa^{(1\vert 0)}\\
\ \ {\rm Re}\;\kappa^{(1\vert 1)}&
-{\rm Re}\;\kappa^{(0\vert 1)}&
\ \ {\rm Re}\;\kappa^{(1\vert 0)}
\end{array}
\right)\ .
\end{eqnarray}
The orthogonal $3\times 3$ matrix $P$ must be a representation
of the special orthogonal transformation given by the matrix $Q$
(in view of the group properties of general Bogolyubov-Valatin 
transformations).
The simplest choice consists in setting $P = \openone_3$, however,
also putting $P = Q$ (or setting it equal to some equivalent 
representation matrix) is possible. At this point we are mainly 
interested in the question of whether
a full nonlinear Bogolyubov-Valatin
transformation of the fermion mode with mode number 1 necessarily 
induces a related (nonidentical) transformation of the fermion 
mode with mode number 2 in such a way that the combined transformation
is canonical. One can now convince oneself of the fact that no
admissible choice of the matrix $P$ leaves the fermion mode with
mode number 2 unchanged. For the "minimal" choice $P = \openone_3$
we find
\begin{eqnarray}
\label{ntwofurther8add6}
\hat{b}_2&=&{\sf B}_2\left(\{\lambda\};\{\hat{a}\}\right)\nonumber\\[0.3cm]
&=&\left[\left(\vert\lambda_1^{(0\vert 1)}\vert^2 - 
\vert\lambda_1^{(1\vert 0)}\vert^2\right)\ \openone_4
\ +\ \left(\lambda_1^{(1\vert 1)}\overline{\lambda_1^{(0\vert 1)}}
- \overline{\lambda_1^{(1\vert 1)}}\lambda_1^{(1\vert 0)}\right)\ 
\hat{a}_1\right.\nonumber\\[0.3cm]
&&\left.\ +\ \left(\lambda_1^{(1\vert 1)}\overline{\lambda_1^{(1\vert 0)}}
- \overline{\lambda_1^{(1\vert 1)}}\lambda_1^{(0\vert 1)}\right)\ 
\hat{a}_1^\dagger\right]\ \hat{a}_2\ .
\end{eqnarray}
Of course, for a linear Bogolyubov-Valatin transformation of mode 1 (i.e.,
$\lambda_1^{(1\vert 1)} = 0$) the two fermion modes decouple.\\

This consideration demonstrates the truly nonlinear character of 
general Bogo\-lyu\-bov-Valatin transformations. A nonlinear 
Bogolyubov-Valatin transformation for one mode necessarily requires us
to perform corresponding changes for the other mode(s) in order 
to maintain the correct CAR. The picture of independent quasi-particles
characteristic for linear Bogolyubov-Valatin transformations 
no longer applies in general.\\

\subsubsection{Nonlinear Bogolyubov-Valatin transformations 
\texorpdfstring{--}{-} 
Some comment on the ansatz of Fukutome and collaborators}

One of the earlier studies of nonlinear Bogolyubov-Valatin transformations
is contained in an article be Fukutome and collaborators \citep{fuku2}.
Our study of these transformations allows us to see the ansatz used
in ref.\ \citep{fuku2} from a different perspective. For the interested reader
(who, for easy reference, is advised to have a copy of the article 
\citep{fuku2} close at hand),
we would like to mention that our eq.\ (\ref{ntwofurtheradd65a})
bears some resemblance to eq.\ (2.11) on p.\ 1557 in 
\citep{fuku2} specialized to the case of two fermion modes. 
However, while in part it corresponds to our 
more general equation (\ref{ntwofurtheradd65a})
the eq.\ (2.11) in \citep{fuku2} does not adequately
reflect in its display the general group theoretical structure
of nonlinear Bogolyubov-Valatin transformations. To see this note that
the case considered in ref.\ \citep{fuku2} corresponds to the choice 
$\chi^{[4]\ast}_k = 0$, $\chi^{[3]\ast}_k = 0$ in our eq.\ 
(\ref{ntwofurther63b}). Consequently, under the reduction
$SO(5;\mathbb{R})\subset SO(6;\mathbb{R})$ 
considered in \citep{fuku2} the 15-dimensional
irreducible representation of the nonlinear Bogolyubov-Valatin
transformation [equivalent to $SO(6;\mathbb{R})/\mathbb{Z}_2$]
operating in biparavector space $\bigwedge^2 (V_6)$
splits into two irreducible 
representations of $SO(5;\mathbb{R})$ - a 5-dimensional representation
[operating in the space
of biparavectors with indices $k = (-1)$ and $l\neq (-1)$]
and a 10-dimensional representation [operating in the space
of biparavectors with indices $k,l \neq (-1)$ ]; cf.\ Figure \ref{groupchain}. 
The 5-dimensional linear space (of operators in Fock space) 
is spanned by products of three or four
different fermion creation and annihilation operators while the 
10-dimensional space (of operators in Fock space)
is spanned by the fermion creation and annihilation operators
themselves and all possible products of two (different) of them. 
The unit operator in Fock space does not transform under 
nonlinear Bogolyubov-Valatin transformations and it spans an 
invariant subspace of the 16-dimensional space of linear operators
in Fock space. Consequently, eq.\ (2.11) on p.\ 1557 in 
\citep{fuku2} must be the above characterized 5-dimensional 
representation in disguise and not a transformation in paravector space
[associated with the Clifford algebra $C(0,4)$] as its display suggests.\\

\subsubsection{The group of nonlinear Bogolyubov-Valatin transformations 
as the adjoint representation of the group 
\texorpdfstring{$SU(2^n = 4)$}{SU(2\textasciicircum n = 4)} 
}

Recalling eq.\ (\ref{ntwofurtheradd78})
\begin{eqnarray}
\label{ntwofurtheradd78dissa}
{\sf U}\; \hat{\hat{\hat{c}\,}}_M\; {\sf U}^\dagger &=& 
\chi_M^{\ \ N}\ \hat{\hat{\hat{c}\,}}_N
\end{eqnarray}
we recognize (\citep{corn1}, Vol.\ II, Chap.\ 11,
Sec.\ 5, Theorem III, p.\ 417)
that the matrix $\chi^T$ (i.e., the transposed of $\chi$)
is the representation matrix
of the group element ${\sf U}$ of the Lie group $SU(2^n = 4)$ taken in
the adjoint representation, i.e., we can write
\begin{eqnarray}
\label{ntwofurtheradd78dissb}
\chi^T&=& \left[\chi\left({\sf U}\right)\right]^T
\ =\  {\sf Ad\left(U\right)}\ .
\end{eqnarray}
Consequently, we find [cf.\ eq.\  (\ref{ntwofurtheradd78})]
\begin{eqnarray}
\label{ntwofurtheradd78dissc}
{\sf Ad\left(U\right)}&=&C_2(L^T)
\end{eqnarray}
where the matrix $L = L\left({\sf U}\right)\in SO(6;\mathbb{R})$
is given by [cf.\ eq.\  (\ref{ntwofurtheradd318b})]
\begin{eqnarray}
\label{ntwofurtheradd78dissd}
L_{kl}&=\frac{1}{4}\  
{\rm tr}\left(\Gamma_k^{\hspace{-0.5cm}+\atop\ }{\sf U}^T\; 
\Gamma_l^{\hspace{-0.5cm}-\atop\ }{\sf U}\right)\ .
\end{eqnarray}
Clearly, the adjoint representation of the Lie group $SU(2^n = 4)$
is not faithful because $SU(2^n = 4)$ posesses a non-trivial center 
(i.e., $\mathbb{Z}_4$). Let us finally note that the adjoint 
representation of the group $SU(3)$ has been discussed in 
\citep{macf3}.

\subsection[Biparavectors and parametrizations of 
\texorpdfstring{$SU(4)$}{SU(4)} 
and \texorpdfstring{$SO(6;\mathbb{R})$}{SO(6;R)} 
transformations]{Biparavectors and parametrizations 
of \texorpdfstring{$SU(4)$}{SU(4)} 
and\\
\texorpdfstring{$SO(6;\mathbb{R})$}{SO(6;R)} 
transformations}

Low-dimensional unitary and orthogonal matrices play an essential role in many
investigations in theoretical physics and beyond. The choice of 
representation for them can significantly facilitate or inhibit
their use. In the mathematical
and physical literature many different parametrizations can be found.
Concerning the unitary matrix ${\sf U}\in SU(4)$
we are of course interested in representations in terms of 
the biparavectors $\hat{\hat{\hat{c}\,}}_M$, 
i.e., in terms of fermion creation and annihilation operators
(For a related discussion see \citep{bogu4,redk1}.).
We have discussed in Sec.\ \ref{param}
such a representation of $SU(4)$ matrices in terms of biparavectors 
[cf.\ eq.\ (\ref{ntwofurtheradd316})]. In this subsection, we would like to 
supplement this discussion by some further comments.\\

\subsubsection{\label{ostlund}The representation of 
\texorpdfstring{$SU(4)$}{SU(4)} 
transformations by \"Ostlund and Mele}

An interesting representation of the unitary operator ${\sf U}$
related to eq.\ (\ref{ntwofurtheradd316}) (i.e., in terms of 
biparavectors)
has been considered by \"Ostlund and Mele \citep{ostl1}.
Define the following states (We have changed somewhat the numbering in
comparison with ref.\ \citep{ostl1}.):
\begin{eqnarray}
\label{ntwofurtheradd317c}
\Psi_{(-1)}&=&\vert 2\rangle_{(2,1)}\ =\ 
\hat{a}_2^\dagger \hat{a}_1^\dagger\vert 0\rangle\ ,\\[0.3cm]
\label{ntwofurtheradd317d}
\Psi_0&=&\vert 0\rangle\ ,\\[0.3cm]
\label{ntwofurtheradd317e}
\Psi_1&=&\vert 1\rangle_1\ =\ \hat{a}_1^\dagger\vert 0\rangle\ ,\\[0.3cm]
\label{ntwofurtheradd317f}
\Psi_2&=&\vert 1\rangle_2\ =\ \hat{a}_2^\dagger\vert 0\rangle\ .
\end{eqnarray}
Then, one defines the elements of the matrix 
$\hat{\hat{\hat{m}\hspace{0.1mm}}}$  
($\equiv m$, in the notation used in \citep{ostl1})
by means of the relation 
\begin{eqnarray}
\label{ntwofurtheradd317g}
\hat{\hat{\hat{m}\hspace{0.1mm}}}_{ij}&=&\Psi_i \Psi_j^\dagger\ .
\end{eqnarray}
It is then useful to recall the relation
\begin{eqnarray}
\label{ntwofurtheradd317ga}
\vert 0\rangle \langle 0\vert&=&\Psi_0 \Psi_0^\dagger\ =\ 
\left(\openone_4 - \hat{a}_1^\dagger \hat{a}_1\right)
\left(\openone_4 - \hat{a}_2^\dagger \hat{a}_2\right)
\end{eqnarray}
to explicitly construct the matrix $\hat{\hat{\hat{m}\hspace{0.1mm}}}\equiv m$
in terms of the creation and annihilation operators
$\hat{a}_k^\dagger$, $\hat{a}_k$ (cf.\ eq.\ (1) of ref.\ \citep{ostl1}).
Consequently, the matrix elements 
$\hat{\hat{\hat{m}\hspace{0.1mm}}}_{ij}$ of the matrix
$\hat{\hat{\hat{m}\hspace{-0.1mm}}}$ are linear combinations of the operators 
$\hat{\hat{\hat{c}\,}}_M$ (and transform under $SO(6;\mathbb{R})/\mathbb{Z}_2$ 
transformations correspondingly). For the explicit relations see
Appendix \ref{appe}.\\

One can now write ($X_{ij}\in \mathbb{C}$)
\begin{eqnarray}
\label{ntwofurtheradd317h}
{\sf U}&=&{\sf U}(X)\ =\ \sum_{i,j = -1}^{2} X_{ji}\  
\hat{\hat{\hat{m}\hspace{-0.1mm}}}_{ij}\ .
\end{eqnarray}
A remarkable property of the representation of the unitary operator
${\sf U}$ in terms of the matrix $\hat{\hat{\hat{m}\hspace{-0.1mm}}}$
is represented by the following equation where $XY$ denotes the 
matrix multiplication of the two $4\times 4$ matrices $X$ and $Y$.
\begin{eqnarray}
\label{ntwofurtheradd317k}
{\sf U}(X) {\sf U}(Y) &=&{\sf U}(XY)
\end{eqnarray}
Taking into account the relation
$\left[{\sf U}(X)\right]^\dagger = {\sf U}(X^\dagger)$ (eq.\ (3)
of \citep{ostl1}), the unitarity condition for the unitary operator 
${\sf U}$ is then found to read [This is another version of
the eqs.\  (\ref{ntwofurtheradd317a}), (\ref{ntwofurtheradd317b}).]
\begin{eqnarray}
\label{ntwofurtheradd317l}
X X^\dagger&=&\openone_4\ .
\end{eqnarray}
Consequently, eq.\ (\ref{ntwofurtheradd317h}) provides us with a 
representation of the unitary matrix ${\sf U}$ by means of some, 
in general other, unitary matrix $X$.\\

\subsubsection{Exponential representation}

Next we want to have a short look at some other version of the
parametrizations of the unitary [${\sf U}\in SU(4)$] (and orthogonal 
[$L \in SO(6;\mathbb{R})$]) transformations we have applied. 
Besides the representation we are using [see eq.\ (\ref{ntwofurtheradd316})]
of most interest appears to be an exponential representation of the unitary
matrix ${\sf U}$
\begin{eqnarray}
\label{ntwofurtheradd501}
{\sf U}&=&T_0\ \openone_4\ +\ T^M\ \hat{\hat{\hat{c}\,}}_M
\ =\ T_0\ \openone_4\ +\ 
\frac{1}{2}\ T^{m_1 m_2}\ \hat{\hat{\hat{c}\,}}_{m_1 m_2}\nonumber\\[0.3cm]
&=&\exp\left(V^M\ \hat{\hat{\hat{c}\,}}_M\right)
\ = \ 
\exp\left(\frac{1}{2}\ V^{m_1 m_2}\ \hat{\hat{\hat{c}\,}}_{m_1 m_2}\right)
\end{eqnarray}
in terms of a real antisymmetric matrix $V$ (remember: 
$\hat{\hat{\hat{c}\,}}^\dagger_M = -\hat{\hat{\hat{c}\,}}_M$).
Formally, 
\begin{eqnarray}
\label{ntwofurtheradd502}
V^M\ \hat{\hat{\hat{c}\,}}_M&=&
\frac{1}{2}\ V^{m_1 m_2}\ \hat{\hat{\hat{c}\,}}_{m_1 m_2}\ =\ \ln {\sf U}\ .
\end{eqnarray}
There is a comprehensive mathematical literature on the subject of the 
logarithm of a matrix and its subtleties. Restricting our attention
to the main sheet of the logarithmic function, a calculation of the
matrix $V$ in terms of $T_0$ and the matrix $T$ could rely on the
approach presented in \citep{unti1}, in particular Sec.\ 2.C.3, p.\ 021108-5
(also note \citep{simi1}). However, we will not pursue this 
calculation here.\\

Concerning some of the subtleties of the
matrix logarithm it is useful to recall that any $n\times n$ 
unitary matrix operating in a complex space of $n$ dimensions 
can be represented as a $2n\times 2n$ orthogonal matrix 
operating in a real space of dimension $2n$. Consequently, the
problem of finding the exponential representation of the unitary matrix
${\sf U}$ is closely related to the problem of the exponential 
representation of the orthogonal matrix $L$ in terms of an antisymmetric
matrix $B$
\begin{eqnarray}
\label{ntwofurtheradd503}
L&=&{\rm e}^{\displaystyle\; B}\ .
\end{eqnarray}
For our case of the $6\times 6$ matrix $L$,
by virtue of the Cayley-Hamilton theorem the antisymmetric
($6\times 6$) matrix $B$ must be a
linear combination of the matrices $A^k$, $k = 1,3,5$,
in the vicinity of the unity transformation $L = \openone_6$.
For the subject of the logarithm of orthogonal matrices we
refer the reader to \citep{bach1,shaw1,kurt1}. Finally, we want to mention
that for bi(para)vectors two different notions of the exponential 
function exist (see \citep{loun2}, \citep{loun}, Sec.\ 17.3, p.\ 221, both
editions). But, we will not further discuss this subject here.\\

\subsubsection{On Cayley-Klein parameters for 
\texorpdfstring{$SO(6;\mathbb{R})$}{SO(6;R)} 
transformations}

In the light of the equations
(\ref{ntwofurtheradd308a}), (\ref{ntwofurtheradd308b})
(Subsec.\ \ref{double})
reflecting the relation between $SO(6;\mathbb{R})$ and $SU(4)$
transformations it seems to be interesting to
add a comment on the introduction of generalized Cayley-Klein 
parameters for the orthogonal transformations $L$ in the Euclidean 
space $\mathbb{R}_6$. For a good introduction to 
Cayley-Klein parameters in $\mathbb{R}_3$ we refer the reader to
\citep{gold1}, \S 4.5, 1.\ ed.\ p.\ 109,
2.\ ed.\ p.\ 148 (Note, that in the third edition \citep{gold2} this
material is no longer included.). Leaving aside the problem of the signature of
the (pseudo-) Euclidean space on which the (pseudo-) orthogonal transformations
operate, Cayley-Klein parameters for (pseudo$-$) 
orthogonal transformations
in a 4-dimensional space have been studied by Cayley already 
\citep{cayl1}. They have later been discussed in 
\citep{silb1,rose2,niko4,niko3,kwal2}, 
\citep{somm1}, also see \citep{somm2}, Mathemati\-sche Zus\"atze 
und Erg\"anzungen, Sec.\ 17, pp.\ 806-813, 
\citep{fisc1,esta1,greu1,rast1,syng1,zadd1}.
Corresponding studies of Cayley-Klein parameters
for (pseudo-) orthogonal transformations in five dimensions can
be found in \citep{erik1,koci2,koci3}. A five-dimensional space
seems to be the space of largest dimension to which the standard
Cayley-Klein parameters can be generalized. From our perspective,
the reason is that for orthogonal transformations $L\in SO(5;\mathbb{R})$ 
the 15-dimensional biparavector space $\bigwedge^2 (V_6)$
decomposes into two invariant
subspaces. One subspace is a 5-dimensional space in which 
the five-dimensional transformation $L$ acts linearly. 
The second invariant subspace is a 10-dimensional space in which
the 5-dimensional transformation $L$ acts via its second compound matrix
exactly in the same manner as an orthogonal transformation  
$L\in SO(6;\mathbb{R})$ is acting in the full 15-dimensional 
biparavector space $\bigwedge^2 (V_6)$
[cf.\ eq.\ (\ref{ntwofurtheradd78}) and Fig.\ \ref{groupchain}]. 
Consequently, generalized 
Cayley-Klein parameters for orthogonal transformations in $\mathbb{R}_6$
must be based on the spin bivector space 
$\bigwedge^2_{\mathbb{R}_6} (\mathbb{C}_4)$ 
according to the eqs.\ (\ref{ntwofurtheradd308a}), 
(\ref{ntwofurtheradd308b}), rather than on eq.\ (\ref{ntwofurtheradd78})
related to biparavector space $\bigwedge^2 (V_6)$.\\

\subsection{\label{bibidis}Biparavectors as
bi(para)vectors of the spin bivector space 
\texorpdfstring{$\bigwedge^2 (\mathbb{C}_4)$}{
/\textbackslash\textasciicircum 2 (C\_ 4)}} 

Here, we would like to extend somewhat the discussion performed in paragraph IV
of Subsec.\ \ref{global}.
By means of the eqs.\ (\ref{ntwofurtheradd304}),
(\ref{ntwofurtheradd307}) (Subsec.\ \ref{double})
one finds for the decomposable/simple biparavectors 
[cf.\ eq.\ (\ref{ntwofurtheradd65aa}); $m_1\not= m_2$]\footnote{For the related
references see footnote \ref{bibi} to eq.\ (\ref{ntwofurtheradd304}) 
in Subsec.\ \ref{double}.}
\begin{eqnarray}
\label{ntwofurtheradd307b}
\hat{\hat{\hat{c}\hspace{0.3mm}}}_{m_1 m_2}&=&\frac{1}{2}
\left(\hat{\hat{\hat{c}\hspace{0.3mm}}}^\dagger_{m_1}
\hat{\hat{\hat{c}\,}}_{m_2}
- \hat{\hat{\hat{c}\hspace{0.3mm}}}^\dagger_{m_2}\hat{\hat{\hat{c}\,}}_{m_1}
\right)\ =\ \hat{\hat{\hat{c}\hspace{0.3mm}}}^\dagger_{m_1}
\hat{\hat{\hat{c}\,}}_{m_2}\nonumber\\[0.3cm]
&=&\frac{1}{2}\left(
\Gamma_{m_1}^{\hspace{-0.5cm}+\atop\ }
\Gamma_{m_2}^{\hspace{-0.5cm}-\atop\ }
-
\Gamma_{m_2}^{\hspace{-0.5cm}+\atop\ }
\Gamma_{m_1}^{\hspace{-0.5cm}-\atop\ }\right)
\ =\ \Gamma_{m_1}^{\hspace{-0.5cm}+\atop\ }
\Gamma_{m_2}^{\hspace{-0.5cm}-\atop\ } \nonumber\\[0.3cm]
&=&\frac{1}{2}\left(
\Gamma_{m_1}^{{\hspace{-0.5cm}-\atop\ }\dagger}
\Gamma_{m_2}^{\hspace{-0.5cm}-\atop\ }
-
\Gamma_{m_2}^{{\hspace{-0.5cm}-\atop\ }\dagger}
\Gamma_{m_1}^{\hspace{-0.5cm}-\atop\ }\right)\ =\
\Gamma_{m_1}^{{\hspace{-0.5cm}-\atop\ }\dagger}
\Gamma_{m_2}^{\hspace{-0.5cm}-\atop\ }\nonumber\\[0.3cm]
&=& 
\frac{1}{2}\left(
\Gamma_{m_1}^{\hspace{-0.5cm}+\atop\ }
\Gamma_{m_2}^{{\hspace{-0.5cm}+\atop\ }\dagger}
-
\Gamma_{m_2}^{\hspace{-0.5cm}+\atop\ }
\Gamma_{m_1}^{{\hspace{-0.5cm}+\atop\ }\dagger}\right)
\ =\ \Gamma_{m_1}^{\hspace{-0.5cm}+\atop\ }
\Gamma_{m_2}^{{\hspace{-0.5cm}+\atop\ }\dagger}\ .
\end{eqnarray}
In view of eq.\ (\ref{ntwofurtheradd307b}),
we can write
\begin{eqnarray}
\label{ntwofurtheradd307c}
\bigwedge\hspace{-1mm}^2 (V_6) &=& 
\bigwedge\hspace{-1mm}^2 \left(
\bigwedge\hspace{-1mm}^2_{\mathbb{R}_6} (\mathbb{C}_4)\right)
\ =\ \bigwedge\hspace{-1mm}^2 (\mathbb{R}_6)\ , 
\end{eqnarray}
but clearly $V_6\not= \mathbb{R}_6$
(The vectors of the spin bivector space $\mathbb{R}_6$ 
are antisymmetric matrices while one direction
in the paravector space $V_6$ is given by the unit matrix $\openone_4$.).
Eq.\ (\ref{ntwofurtheradd307b})
tells us that the decomposable/simple biparavectors 
$\hat{\hat{\hat{c}\hspace{0.3mm}}}_M$ of the paravector space $V_6$
can also be understood as decomposable/simple
bi(para)vectors of the spin bivector space 
$\mathbb{R}_6$ \footnote{We tend to use the term 'biparavectors' 
of spin bivector space here because their definition involves a 
Hermitian conjugation in one of the two factors in analogy to
the biparavectors defined in eq.\ (\ref{ntwofurtheradd65aa}).}.
In other and more plain words, any ($4\times 4$) Dirac matrix and
any of their (multiple) products can be given as the product of two
antisymmetric ($4\times 4$) matrices that exhibit a certain generalized 
anticommutation relation. This insight into
the structure of decomposable/simple biparavectors 
$\hat{\hat{\hat{c}\hspace{0.3mm}}}_M$ has in part
been spelled out by
Haantjes, \citep{haan1}, p.\ 51, stelling 5 [proposition 5], and 
a corresponding comment can also be found in ref.\ \citep{buch1}, \S 14,
p.\ 269, above of eq.\ (14.8).
While we have concentrated our study on the structure of
$SO(6;\mathbb{R})$ transformations in bipa\-ra\-vector 
space $\bigwedge^2 (V_6)$ 
Haantjes \citep{haan1} and Buchdahl \citep{buch1}
reflect the decomposability property from out the 
construction of the paravectors/biparavectors in terms of 
the product of spin bivectors [cf.\ eq.\ (\ref{ntwofurtheradd304})], i.e.,
their representation in terms of products of antisymmetric matrices.
For the special choice discussed in Appendix \ref{appc} this product
representation of the biparavectors corresponds to a product 
representation in terms of two commuting sets of quaternions $\mathbb{H}$. 
Another product representation (in terms of matrices called $S_\alpha$ 
and $D_\alpha$) has been discussed by Eddington \citep{eddi1}, 
\citep{eddi5}, Chap.\ III, Secs.\ 3.1, 3.2, pp.\ 34-36. Nikol'ski\u\i\
\citep{niko1} has shown that this product representation is related
to transformations of tetrahedral coordinates (For a related 
discussion concerning the Dirac equation also see \citep{niko2}.).\\

\subsection{\label{misc}Miscellaneous}

Finally, we would like to add some comments related to the choice of
the metric (in particular its signature) in spin bivector space. 
According to the eqs.\ (\ref{ntwofurtheradd308a}), (\ref{ntwofurtheradd308b})
(Subsec.\ \ref{double}),
unitary transformations [$\in SU(4)$] in the spin space $\mathbb{C}_4$
correspond to orthogonal transformations [$\in SO(6;\mathbb{R})$] in
the spin bivector space $\mathbb{R}_6$. On the other hand, it has 
been known for a long time that quadriquaternions (or, equivalently, 
the biparavectors $\hat{\hat{\hat{c}\hspace{0.3mm}}}_M$) are related 
to conformal transformations \citep{comb1}. To arrive at these
one must change from the spin bivector space $\mathbb{R}_6$ to 
the related spin bivector space $\mathbb{R}_{5,1}$. This change in 
metric can be achieved, for example, by means of the transition
$\Gamma_{-1}^{\hspace{-0.5cm}+\atop\ }\longrightarrow i 
\ \Gamma_{-1}^{\hspace{-0.5cm}+\atop\ }$ [entailing, e.g., 
$\left(\Gamma_{-1}^{\hspace{-0.5cm}+\atop\ }\right)^2 = - \openone_4$].
The $SO(5,1;\mathbb{R})$ transformations in $\mathbb{R}_{5,1}$ then stand in 
correspondence to $SU^{\displaystyle\ast}(4)$ 
transformations in spin space\footnote{For the 
notation see, e.g., \citep{corn1}, Vol.\ 2, p.\ 392, 
Table 10.1. $SU^{\displaystyle\ast}(4)$
is the noncompact version of $SU(4)$ related
to it by the "unitary trick" of Weyl \citep{gilm1}.}.
It is then possible to represent the pseudo-orthogonal $SO(5,1;\mathbb{R})$
transformations in $\mathbb{R}_{5,1}$ by means of conformal 
transformations in a subspace $\mathbb{R}_4\subset \mathbb{R}_{5,1}$.
However, the corresponding transformations in spin space then are
no longer unitary and do not correspond to canonical transformations
in which we are interested. From a particle physics perspective,
transformations not in Euclidean space $\mathbb{R}_4$
but in Minkowski space $\mathbb{R}_{3,1}$ are interesting. 
The Minkowski space is then to be embedded in a 
six-dimensional pseudo-Euclidean space $\mathbb{R}_{4,2}$ and 
pseudo-orthogonal $SO(4,2;\mathbb{R})$ transformations in 
$\mathbb{R}_{4,2}$ then correspond to pseudo-unitary $SU(2,2)$
transformations in spin space. 
We will not further dwell on this subject here but confine
ourselves to some comments on the related literature:
For a good, general mathematical account of conformal transformations
see \citep{roze1}, Chap.\ 11, 
p.\ 480. Ref.\ \citep{kast1} covers the subject from a physical
perspective (including many references, also see \citep{kast2}). 
For articles dealing with various aspects of the above
relationship that are interesting within the present context see  
\citep{hepn1,hris5,hris6,hris9,hris10,siga1}. 
This subject has also been discussed, in part recently, 
from a twistor perspective (see, e.g., 
\citep{klot1,roch1,arca1}). Leaving aside special conformal 
transformations and dilatations in the 
pseudo-Euclidean space $\mathbb{R}_{3,1}$, the relevance of
the restricted class of $SO(4,1;\mathbb{R})$ transformations 
in $\mathbb{R}_{4,2}$ for massive (4-component) fermions
(Dirac particles) has been discussed in the physics literature
(Dirac equation in $4+1$ dimensions) 
\citep{erik1,erik2,erik3,brac1,koci1}.
Also, (generalized) Foldy-Wouthuysen transformations in spin space 
fit into the latter framework \citep{devr1,loid1}.
A discussion of the algebra of Dirac matrices related to the group
$SO(3,2;\mathbb{R})$ can be found in \citep{dira1,buch3,lee2}.
Finally, for the sake of completeness, we also want to mention 
the papers \citep{seng1} and \citep{linh1,linh2,linh3,linh4,linh5}
which have attempted to address certain problems related to the general 
spin space transformations we have studied.\\

\section{\label{concl}Conclusions}

Having reached the end of the present study of nonlinear 
Bogolyubov-Valatin transformations for two fermion modes, it seems
fair to say that even this apparently simple and basic 
case has revealed its complexity and broad significance.
Among the more general aspects, we would like to point out
that unlike linear Bogolyubov-Valatin transformations which can
be performed for each mode of quasifermions independently and
separately, nonlinear Bogolyubov-Valatin transformations exhibit
the truly nonlinear nature of the canonical fermion anticommutation 
relations because any nonlinear Bogolyubov-Valatin transformation
-- even within one fermion mode -- immediately impacts the other mode(s) 
and, in general, the (linear) picture of independent quasifermions 
no longer applies.\\

From a mathematical point of view, nonlinear Bogolyubov-Valatin
transformations are accompanied by transformations of multivectors
of different grades and, therefore, can be viewed as concrete
realizations of supersymmetric transformations. The preservation
of the canonical anticommutation relations under nonlinear 
Bogolyubov-Valatin transformation is closely connected
to the decomposability of the biparavectors generating the Clifford
algebra $C(0,4)$. This is the Clifford algebra 
that is associated with the creation and annihilation 
operators of the two fermion modes. From a somewhat different 
perspective, the preservation
of the canonical anticommutation relations under nonlinear 
Bogolyubov-Valatin transformation can be described in other words as
being closely connected to the
fact that Dirac matrices can always be written as the product of
two antisymmetric matrices. This is a fact that was first recognized
in the 1930's already by Dutch mathematicians 
(Schouten, Struik, Haantjes).\\

Looking to future generalizations 
of the present work to more than two fermion modes it should be
said that one should expect to meet mathematical structures which
have not let their presence in the study of one and two fermion
modes be known. The present investigation 
of nonlinear Bogolyubov-Valatin transformations
for $n=2$ fermion modes exhibits a number of features that are related
to fact that the group $SU(2^n = 4)$ is the double cover of the 
group $SO(6;\mathbb{R})$ [For one fermion mode $SU(2)$ is related 
to $SO(3;\mathbb{R})$.]. For $n>2$ fermion modes no such relation of the group 
$SU(2^n)$ to any of the orthogonal groups exists. This fact necessarily 
will raise its head in any future study of nonlinear 
Bogolyubov-Valatin transformation for more than two fermion modes.
This will lead to a further increase in the complexity of the problem
to be studied but certainly also to further interesting mathematical 
and physical insight into the nature of fermionic systems.\\

\vfill\ 

\subsection*{Acknowledgements}
\addcontentsline{toc}{section}{Acknowledgements}

K.\ S.\ would like to thank K.\ V.\ Andreev, G.\ J.\ T.\ Grube, and
J.\ A.\ de Wet for correspondence and comments on an earlier version
of the manuscript. For sending
copies of articles and theses, we are indebted to: 
K.\ V.\ Andreev for \citep{andr4,andr2,andr3,andr1},
A.\ Bulgac for \citep{bulg1},
Chueng-Ryong Ji for \citep{grub1}, V.\ L.\ Safonov for \cite{safo2},
J.\ A.\ de Wet for \citep{dewe1},
and D.\ Sh.\ Sabirov of {\it Vestn.\ Bashkir.\ Univ.}\ for \citep{andr2}. 
K.\ S.\ gratefully acknowledges kind hospitality at the Theoretical 
Subatomic Physics group of the Department
of Physics and Astronomy of the Vrije Universiteit Amsterdam.\\

\pagebreak
\newcounter{asection}
\setcounter{asection}{1}
\renewcommand{\thesection}{\Alph{asection}}
\renewcommand{\theequation}{\mbox{\Alph{asection}.\arabic{equation}}}
\refstepcounter{section}
\label{appa}
\section*{Appendix \Alph{asection}}
\setcounter{equation}{0}
\addcontentsline{toc}{section}{Appendix \Alph{asection}}

We define the coefficients [cf.\ Sec.\ \ref{prelim},
eqs.\ (\ref{ntwo3a}), (\ref{ntwo3b})]
\begin{eqnarray}
\label{ntwoadd11}
\kappa^{(0\vert 0)}_k&=&\lambda^{(0\vert 0)}_k\ ,\\[0.3cm]
\kappa^{(1\vert 0)}_k&=&\lambda^{(1\vert 0)}_k +
\lambda^{(0\vert 1)}_k + \frac{1}{2} \lambda^{(1,2\vert 2)}_k -
\frac{1}{2} \lambda^{(2\vert 1,2)}_k\ ,\\[0.3cm]
\kappa^{(0\vert 1)}_k&=&- i \left(\lambda^{(1\vert 0)}_k -
\lambda^{(0\vert 1)}_k + \frac{1}{2} \lambda^{(1,2\vert 2)}_k +
\frac{1}{2} \lambda^{(2\vert 1,2)}_k\right)\ ,\\[0.3cm]
\kappa^{(2\vert 0)}_k&=&\lambda^{(2\vert 0)}_k +
\lambda^{(0\vert 2)}_k - \frac{1}{2} \lambda^{(1,2\vert 1)}_k +
\frac{1}{2} \lambda^{(1\vert 1,2)}_k\ ,\\[0.3cm]
\kappa^{(0\vert 2)}_k&=&- i \left(\lambda^{(2\vert 0)}_k -
\lambda^{(0\vert 2)}_k - \frac{1}{2} \lambda^{(1,2\vert 1)}_k -
\frac{1}{2} \lambda^{(1\vert 1,2)}_k\right)\ ,\\[0.3cm]
\kappa^{(1\vert 1)}_k&=&\lambda^{(1\vert 1)}_k\ ,\\[0.3cm]
\kappa^{(1,2\vert 0)}_k&=&-\ \frac{i}{2}\ \left( \lambda^{(1,2\vert 0)}_k +
\lambda^{(0\vert 1,2)}_k + \lambda^{(1\vert 2)}_k -
\lambda^{(2\vert 1)}_k\right)\ ,\\[0.3cm]
\kappa^{(1\vert 2)}_k&=&\frac{1}{2}\ \left( - \lambda^{(1,2\vert 0)}_k +
\lambda^{(0\vert 1,2)}_k + \lambda^{(1\vert 2)}_k +
\lambda^{(2\vert 1)}_k\right)\ ,\\[0.3cm]
\kappa^{(2\vert 1)}_k&=&\frac{1}{2}\ \left( - \lambda^{(1,2\vert 0)}_k +
\lambda^{(0\vert 1,2)}_k - \lambda^{(1\vert 2)}_k -
\lambda^{(2\vert 1)}_k\right)\ ,\\[0.3cm]
\kappa^{(0\vert 1,2)}_k&=&\frac{i}{2}\ \left( \lambda^{(1,2\vert 0)}_k +
\lambda^{(0\vert 1,2)}_k - \lambda^{(1\vert 2)}_k +
\lambda^{(2\vert 1)}_k\right)\ ,\\[0.3cm]
\kappa^{(2\vert 2)}_k&=&\lambda^{(2\vert 2)}_k\ ,\\[0.3cm]
\kappa^{(1,2\vert 1)}_k&=&\ \ \frac{1}{2}\ \left( - \lambda^{(1\vert 1,2)}_k +
\lambda^{(1,2\vert 1)}_k\right)\ ,\\[0.3cm]
\kappa^{(1\vert 1,2)}_k&=&-\ \frac{i}{2}\ \left(\lambda^{(1\vert 1,2)}_k +
\lambda^{(1,2\vert 1)}_k\right)\ ,\\[0.3cm]
\kappa^{(1,2\vert 2)}_k&=&\frac{1}{2}\ \left(\lambda^{(2\vert 1,2)}_k -
\lambda^{(1,2\vert 2)}_k\right)\ ,\\[0.3cm]
\kappa^{(2\vert 1,2)}_k&=&\frac{i}{2}\ \left(\lambda^{(2\vert 1,2)}_k +
\lambda^{(1,2\vert 2)}_k\right)\ ,\\[0.3cm]
\kappa^{(1,2\vert 1,2)}_k&=&\frac{1}{2}\ \lambda^{(1,2\vert 1,2)}_k\ .
\end{eqnarray}
This in turn implies:
\begin{eqnarray}
\label{ntwoadd12}
\lambda^{(1\vert 0)}_k&=&\frac{1}{2}\ \left( \kappa^{(1\vert 0)}_k +
i \kappa^{(0\vert 1)}_k + \kappa^{(1,2\vert 2)}_k +
i \kappa^{(2\vert 1,2)}_k\right)\ ,\\[0.3cm]
\lambda^{(0\vert 1)}_k&=&\frac{1}{2}\ \left( \kappa^{(1\vert 0)}_k -
i \kappa^{(0\vert 1)}_k + \kappa^{(1,2\vert 2)}_k -
i \kappa^{(2\vert 1,2)}_k\right)\ ,\\[0.3cm]
\lambda^{(2\vert 0)}_k&=&\frac{1}{2}\ \left( \kappa^{(2\vert 0)}_k +
i \kappa^{(0\vert 2)}_k + \kappa^{(1,2\vert 1)}_k +
i \kappa^{(1\vert 1,2)}_k\right)\ ,\\[0.3cm]
\lambda^{(0\vert 2)}_k&=&\frac{1}{2}\ \left( \kappa^{(2\vert 0)}_k -
i \kappa^{(0\vert 2)}_k + \kappa^{(1,2\vert 1)}_k -
i \kappa^{(1\vert 1,2)}_k\right)\ ,\\[0.3cm]
\lambda^{(1,2\vert 0)}_k&=&\frac{1}{2}\ \left( - \kappa^{(2\vert 1)}_k -
\kappa^{(1\vert 2)}_k - i \kappa^{(1,2\vert 0)}_k +
i \kappa^{(0\vert 1,2)}_k\right)\ ,\\[0.3cm]
\lambda^{(1\vert 2)}_k&=&\frac{1}{2}\ \left( - \kappa^{(2\vert 1)}_k +
\kappa^{(1\vert 2)}_k - i \kappa^{(1,2\vert 0)}_k -
i \kappa^{(0\vert 1,2)}_k\right)\ ,\\[0.3cm]
\lambda^{(2\vert 1)}_k&=&\frac{1}{2}\ \left( - \kappa^{(2\vert 1)}_k +
\kappa^{(1\vert 2)}_k + i \kappa^{(1,2\vert 0)}_k +
i \kappa^{(0\vert 1,2)}_k\right)\ ,\\[0.3cm]
\lambda^{(0\vert 1,2)}_k&=&\frac{1}{2}\ \left( \kappa^{(2\vert 1)}_k +
\kappa^{(1\vert 2)}_k - i \kappa^{(1,2\vert 0)}_k +
i \kappa^{(0\vert 1,2)}_k\right)\ ,\\[0.3cm]
\lambda^{(1,2\vert 1)}_k&=&\kappa^{(1,2\vert 1)}_k +
i \kappa^{(1\vert 1,2)}_k\ ,\\[0.3cm]
\lambda^{(1\vert 1,2)}_k&=&-\ \kappa^{(1,2\vert 1)}_k +
i \kappa^{(1\vert 1,2)}_k\ ,\\[0.3cm]
\lambda^{(1,2\vert 2)}_k&=&-\ \kappa^{(1,2\vert 2)}_k -
i \kappa^{(2\vert 1,2)}_k\ ,\\[0.3cm]
\lambda^{(2\vert 1,2)}_k&=&\kappa^{(1,2\vert 2)}_k -
i \kappa^{(2\vert 1,2)}_k\ ,\\[0.3cm]
\lambda^{(1,2\vert 1,2)}_k&=&2 \ \kappa^{(1,2\vert 1,2)}_k\ .
\end{eqnarray}

Furthermore, the following relations apply [cf.\ eqs.\
(\ref{ntwoadd13a}), (\ref{ntwoadd13b}) and (\ref{ntwoadd32c}); $k=1,2$]
\begin{eqnarray}
\label{ntwoadd33}
\chi^{[1]\ (1)}_{2k-1}&=&{\rm Re}\;\kappa^{(1\vert 0)}_k\ ,
\ \ \ \ \ \ \ \ \ \ \ \
\chi^{[1]\ (1)}_{2k}\ =\ {\rm Im}\;\kappa^{(1\vert 0)}_k\ ,\nonumber\\[0.3cm]
\chi^{[1]\ (2)}_{2k-1}&=&{\rm Re}\;\kappa^{(0\vert 1)}_k\ ,
\ \ \ \ \ \ \ \ \ \ \ \
\chi^{[1]\ (2)}_{2k}\ =\ {\rm Im}\;\kappa^{(0\vert 1)}_k\ ,\nonumber\\[0.3cm]
\chi^{[1]\ (3)}_{2k-1}&=&{\rm Re}\;\kappa^{(2\vert 0)}_k\ ,
\ \ \ \ \ \ \ \ \ \ \ \
\chi^{[1]\ (3)}_{2k}\ =\ {\rm Im}\;\kappa^{(2\vert 0)}_k\ ,\nonumber\\[0.3cm]
\chi^{[1]\ (4)}_{2k-1}&=&{\rm Re}\;\kappa^{(0\vert 2)}_k\ ,
\ \ \ \ \ \ \ \ \ \ \ \ 
\chi^{[1]\ (4)}_{2k}\ =\ {\rm Im}\;\kappa^{(0\vert 2)}_k\ ,\nonumber\\[0.3cm]
\chi^{[2]\ (1,2)}_{2k-1}&=&{\rm Re}\;\kappa^{(1\vert 1)}_k\ ,
\ \ \ \ \ \ \ \ \ \ \,  
\chi^{[2]\ (1,2)}_{2k}\ =\ {\rm Im}\;\kappa^{(1\vert 1)}_k\ ,\nonumber\\[0.3cm]
\chi^{[2]\ (1,3)}_{2k-1}&=&{\rm Re}\;\kappa^{(1,2\vert 0)}_k\ ,
\ \ \ \ \ \ \ \ \
\chi^{[2]\ (1,3)}_{2k}\ =\ {\rm Im}\;\kappa^{(1,2\vert 0)}_k\ ,
\nonumber\\[0.3cm]
\chi^{[2]\ (1,4)}_{2k-1}&=&{\rm Re}\;\kappa^{(1\vert 2)}_k\ ,
\ \ \ \ \ \ \ \ \ \ \
\chi^{[2]\ (1,4)}_{2k}\ =\ {\rm Im}\;\kappa^{(1\vert 2)}_k\ ,\nonumber\\[0.3cm]
\chi^{[2]\ (2,3)}_{2k-1}&=&{\rm Re}\;\kappa^{(2\vert 1)}_k\ ,
\ \ \ \ \ \ \ \ \ \ \
\chi^{[2]\ (2,3)}_{2k}\ =\ {\rm Im}\;\kappa^{(2\vert 1)}_k\ ,\nonumber\\[0.3cm]
\chi^{[2]\ (2,4)}_{2k-1}&=&{\rm Re}\;\kappa^{(0\vert 1,2)}_k\ ,
\ \ \ \ \ \ \ \ \ \
\chi^{[2]\ (2,4)}_{2k}\ =\ {\rm Im}\;\kappa^{(0\vert 1,2)}_k\ ,
\nonumber\\[0.3cm]
\chi^{[2]\ (3,4)}_{2k-1}&=&{\rm Re}\;\kappa^{(2\vert 2)}_k\ ,
\ \ \ \ \ \ \ \ \ \ \ \,
\chi^{[2]\ (3,4)}_{2k}\ =\ {\rm Im}\;\kappa^{(2\vert 2)}_k\ ,\nonumber\\[0.3cm]
\chi^{[3]\ (1,2,3)}_{2k-1}&=&{\rm Re}\;\kappa^{(1,2\vert 1)}_k\ ,
\ \ \ \ \ \ \ \ 
\chi^{[3]\ (1,2,3)}_{2k}\ =\ 
{\rm Im}\;\kappa^{(1,2\vert 1)}_k\ ,\nonumber\\[0.3cm]
\chi^{[3]\ (1,2,4)}_{2k-1}&=&{\rm Re}\;\kappa^{(1\vert 1,2)}_k\ ,
\ \ \ \ \ \ \ \
\chi^{[3]\ (1,2,4)}_{2k}\ =\ 
{\rm Im}\;\kappa^{(1\vert 1,2)}_k\ ,\nonumber\\[0.3cm]
\chi^{[3]\ (1,3,4)}_{2k-1}&=&{\rm Re}\;\kappa^{(1,2\vert 2)}_k\ ,
\ \ \ \ \ \ \ \
\chi^{[3]\ (1,3,4)}_{2k}\ =\ 
{\rm Im}\;\kappa^{(1,2\vert 2)}_k\ ,\nonumber\\[0.3cm]
\chi^{[3]\ (2,3,4)}_{2k-1}&=&{\rm Re}\;\kappa^{(2\vert 1,2)}_k\ ,
\ \ \ \ \ \ \ \
\chi^{[3]\ (2,3,4)}_{2k}\ =\ 
{\rm Im}\;\kappa^{(2\vert 1,2)}_k\ ,\nonumber\\[0.3cm]
\chi^{[4]\ (1,2,3,4)}_{2k-1}&=&{\rm Re}\;\kappa^{(1,2\vert 1,2)}_k\ ,
\ \ \ \ \
\chi^{[4]\ (1,2,3,4)}_{2k}\ =\ {\rm Im}\;\kappa^{(1,2\vert 1,2)}_k\ .
\end{eqnarray}

\stepcounter{asection}
\refstepcounter{section}
\label{appb}
\section*{Appendix \Alph{asection}}
\setcounter{equation}{0}
\addcontentsline{toc}{section}{Appendix \Alph{asection}}

To work out the anticommutator (and commutator) of the operators
$\hat{d}_k$ (see Subsec.\ \ref{antico})
it turns out to be useful to calculate the following 
products first (On the r.h.s.\ terms with more than four factors
in the wedge product have been omitted as they vanish in the case
under consideration.).
\begin{eqnarray}
\label{ntwofurther40a}
\hat{c}_q\hat{d}_l&=&
-\ \chi^{[1]}_{l\ (q)}\ \openone_4
\ -\ \chi^{[2]}_{l\ (q,s)}\ \hat{c}^{\ s}
\ + \ \chi^{[1]\ (r)}_l\ \hat{c}_q\wedge\hat{c}_r
\ -\ \frac{1}{2!}\ \chi^{[3]}_{l\ (q,s,t)}\ i\
\hat{c}^s\wedge\hat{c}^{\ t}\nonumber\\[0.3cm]
&&+\ \frac{1}{2!}\ \chi^{[2]\ (r,s)}_l\
\hat{c}_q\wedge\hat{c}_r\wedge\hat{c}_s
\ -\ \frac{1}{3!}\ \chi^{[4]}_{l\ (q,s,t,u)}\
\ i\ \hat{c}^{\ s}\wedge\hat{c}^{\ t}\wedge\hat{c}^{\ u}
\nonumber\\[0.3cm]
&&+\ \frac{1}{3!}\ \chi^{[3]\ (r,s,t)}_l\ i\
\hat{c}_q\wedge\hat{c}_r\wedge\hat{c}_s\wedge\hat{c}_t
\end{eqnarray}
$p\neq q$, the r.h.s.\ is antisymmetric in $p$ and $q$:
\begin{eqnarray}
\label{ntwofurther40b}
\hat{c}_p\hat{c}_q\hat{d}_l&=&
-\ \chi^{[2]}_{l\ (p,q)}\ \openone_4
\ -\ \chi^{[1]}_{l\ (q)}\ \hat{c}_p
\ +\ \chi^{[1]}_{l\ (p)}\ \hat{c}_q
\ -\ \chi^{[3]}_{l\ (p,q,t)}\ i\ \hat{c}^{\ t}\nonumber\\[0.3cm]
&&-\ \chi^{[2]}_{l\ (q,s)}\ \hat{c}_p\wedge\hat{c}^{\ s}
\ +\ \chi^{[2]}_{l\ (p,s)}\ \hat{c}_q\wedge\hat{c}^{\ s}\nonumber\\[0.3cm]
&&-\ \frac{1}{2!}\ \chi^{[4]}_{l\ (p,q,t,u)}
\ i\ \hat{c}^{\ t}\wedge\hat{c}^{\ u}
\ +\ \chi^{[1]\ (r)}_l\ \hat{c}_p\wedge\hat{c}_q\wedge\hat{c}_r
\nonumber\\[0.3cm]
&&-\ \frac{1}{2!}\ \chi^{[3]}_{l\ (q,s,t)}\ 
i\ \hat{c}_p\wedge\hat{c}^{\ s}\wedge\hat{c}^{\ t}
\ +\ \frac{1}{2!}\ \chi^{[3]}_{l\ (p,s,t)}\ 
i\ \hat{c}_q\wedge\hat{c}^{\ s}\wedge\hat{c}^{\ t}\nonumber\\[0.3cm]
&&+\ \frac{1}{2!}\ \chi^{[2]\ (r,s)}_l\ 
\hat{c}_p\wedge\hat{c}_q\wedge\hat{c}_r\wedge\hat{c}_s\ .
\end{eqnarray}
$n\neq p\neq q$, the r.h.s.\ is completely antisymmetric in $n$, $p$ and $q$:
\begin{eqnarray}
\label{ntwofurther40c}
\hat{c}_n\hat{c}_p\hat{c}_q\hat{d}_l&=&
\chi^{[3]}_{l\ (n,p,q)}\ i\ \openone_4
\ -\ \chi^{[2]}_{l\ (p,q)}\ \hat{c}_n
\ +\ \chi^{[2]}_{l\ (n,q)}\ \hat{c}_p
\ -\ \chi^{[2]}_{l\ (n,p)}\ \hat{c}_q\nonumber\\[0.3cm]
&&+\ \chi^{[4]}_{l\ (n,p,q,u)}\ i\ \hat{c}^{\ u}
\ -\ \chi^{[1]}_{l\ (q)}\ \hat{c}_n\wedge\hat{c}_p
\ +\ \chi^{[1]}_{l\ (p)}\ \hat{c}_n\wedge\hat{c}_q\nonumber\\[0.3cm]
&&-\ \chi^{[1]}_{l\ (n)}\ \hat{c}_p\wedge\hat{c}_q
\ -\ \chi^{[3]}_{l\ (p,q,t)}\ i\ \hat{c}_n\wedge\hat{c}^{\ t}
\nonumber\\[0.3cm]
&&+\ \chi^{[3]}_{l\ (n,q,t)}\ i\ \hat{c}_p\wedge\hat{c}^{\ t}
\ -\ \chi^{[3]}_{l\ (n,p,t)}\ i\ \hat{c}_q\wedge\hat{c}^{\ t}
\nonumber\\[0.3cm]
&&-\ \chi^{[2]}_{l\ (q,s)}\ \hat{c}_n\wedge\hat{c}_p\wedge\hat{c}^{\ s}
\ +\ \chi^{[2]}_{l\ (p,s)}\ \hat{c}_n\wedge\hat{c}_q\wedge\hat{c}^{\ s}
\nonumber\\[0.3cm]
&&-\ \chi^{[2]}_{l\ (n,s)}\ \hat{c}_p\wedge\hat{c}_q\wedge\hat{c}^{\ s}
\ +\ \chi^{[1]\ (r)}_l\
\hat{c}_n\wedge\hat{c}_p\wedge\hat{c}_q\wedge\hat{c}_r\ .
\end{eqnarray}
$m\neq n\neq p\neq q$, the r.h.s.\ is completely antisymmetric in 
$m$, $n$, $p$ and $q$:
\begin{eqnarray}
\label{ntwofurther40d}
\hat{c}_m\hat{c}_n\hat{c}_p\hat{c}_q\hat{d}_l&=&
\chi^{[4]}_{l\ (m,n,p,q)}\ i\ \openone_4
\ +\ \chi^{[3]}_{l\ (n,p,q)}\ i\ \hat{c}_m
\ -\ \chi^{[3]}_{l\ (m,p,q)}\ i\ \hat{c}_n
\ +\ \chi^{[3]}_{l\ (m,n,q)}\ i\ \hat{c}_p\nonumber\\[0.3cm]
&&-\ \chi^{[3]}_{l\ (m,n,p)}\ i\ \hat{c}_q
\ -\ \chi^{[2]}_{l\ (p,q)}\ \hat{c}_m\wedge\hat{c}_n
\ +\ \chi^{[2]}_{l\ (n,q)}\ \hat{c}_m\wedge\hat{c}_p\nonumber\\[0.3cm]
&&-\ \chi^{[2]}_{l\ (n,p)}\ \hat{c}_m\wedge\hat{c}_q
\ -\ \chi^{[2]}_{l\ (m,q)}\ \hat{c}_n\wedge\hat{c}_p
\ +\ \chi^{[2]}_{l\ (m,p)}\ \hat{c}_n\wedge\hat{c}_q\nonumber\\[0.3cm]
&&-\ \chi^{[2]}_{l\ (m,n)}\ \hat{c}_p\wedge\hat{c}_q
\ -\ \chi^{[1]}_{l\ (q)}\ \hat{c}_m\wedge\hat{c}_n\wedge\hat{c}_p
\ +\ \chi^{[1]}_{l\ (p)}\ \hat{c}_m\wedge\hat{c}_n\wedge\hat{c}_q
\nonumber\\[0.3cm]
&&-\ \chi^{[1]}_{l\ (n)}\ \hat{c}_m\wedge\hat{c}_p\wedge\hat{c}_q
\ +\ \chi^{[1]}_{l\ (m)}\ \hat{c}_n\wedge\hat{c}_p\wedge\hat{c}_q\ .
\end{eqnarray}
In deriving the above relations we have used the
Clifford algebra relations ($\bf x$, ${\bf y}_i$ are vectors and 
$A$ is an arbitrary multivector.).
\begin{eqnarray}
\label{ntwofurther41}
{\bf x}\ A &=&{\bf x}\rfloor A\ +\ {\bf x}\wedge A
\end{eqnarray}
and ($\check{\bf y}_k$ denotes a vector to be omitted.)
\begin{eqnarray}
\label{ntwofurther42}
{\bf x}\rfloor\left({\bf y}_1\wedge\ldots\wedge {\bf y}_p\right)
&=&\sum_{k=1}^{p} (-1)^{k+1}\ ({\bf x}\cdot {\bf y}_k)
\left({\bf y}_1\wedge\ldots\wedge \check{\bf y}_k\wedge\ldots\wedge 
{\bf y}_p\right) 
\end{eqnarray}
(Cf., e.g.,  ref.\ \citep{hest1}, Chap.\ 1, p.\ 9, eq.\ (1.33). Note,
that Hestenes and Sobczyk use a somewhat different notation for the 
operation of left contraction $\rfloor$.).\\

\stepcounter{asection}
\refstepcounter{section}
\label{appcomp}
\section*{Appendix \Alph{asection}}
\setcounter{equation}{0}
\addcontentsline{toc}{section}{Appendix \Alph{asection}}

To make the present paper self-contained we repeat in this Appendix 
some definitions and mathematical results concerning compound matrices
given in \citep{scha1},
Subsec.\ II.A, p.\ 5420 and Appendix A, pp.\ 5443/5444.\\

We introduce 
${n\choose k}\times {n\choose k}$ matrices ${\sf A}^{(2k)}$ ($k = 1,\ldots,n$)
by writing (choose $l_1< l_2< \ldots <l_k$, $m_1< m_2< \ldots <m_k$)
\begin{eqnarray}
\label{M10a}
{\sf A}^{(2k)}_{LM} &=& A_{l_1 \ldots l_k,m_1 \ldots m_k}^{(2k)}\
\end{eqnarray}
(We identify the indices $L$, $M$ with the ordered strings
$l_1 \ldots l_k$, $m_1 \ldots m_k$.)
or, more generally (not requesting
$l_1< l_2< \ldots <l_k$, $m_1< m_2< \ldots <m_k$)
\begin{eqnarray}
\label{M10b}
{\sf A}^{(2k)}_{LM}\ =\ {\rm sgn}\left[\sigma_a(l_1, \ldots, l_k)\right]\
{\rm sgn}\left[\sigma_b(m_1, \ldots, m_k)\right]\
A_{l_1 \ldots l_k,m_1 \ldots m_k}^{(2k)}\ .
\end{eqnarray}
The indices $L,M$ label the equivalence classes of all permutations of the
indices $l_1, \ldots, l_k$ and $m_1, \ldots, m_k$, respectively, 
and $\sigma_a$, $\sigma_b$
are the permutations which bring the indices $l_i, m_i$ ($i = 1,\ldots,k$)
into order with respect to the $<$ relation
(i.e., $\sigma_a(l_1)< \sigma_a(l_2)< \ldots <\sigma_a(l_k)$,
$\sigma_b(m_1)< \sigma_b(m_2)< \ldots <\sigma_b(m_k)$).
The matrix elements of the matrix ${\sf A}^{(2k)}$ are arranged
according to the lexicographical order of the row and column indices
$L$, $M$
(We identify the indices $L$, $M$ with the ordered strings
$\sigma_a(l_1) \ldots \sigma_a(l_k)$, $\sigma_b(m_1) \ldots \sigma_b(m_k)$,
respectively.).\\

We also define a set of (dual)
${n\choose k}\times {n\choose k}$ matrices
${\sf A}^{(2k) \star}$ ($k = 1,\ldots,n$) by writing
\begin{eqnarray}
\label{M11}
{\sf A}^{(2k) \star} &=&{\cal E}^{(k)}  {\sf A}^{(2k) T} {\cal E}^{(k) T}
\end{eqnarray}
where the ${n\choose k}\times {n\choose k}$ matrix ${\cal E}^{(k)}$
is defined by
\begin{eqnarray}
\label{M12a}
{\cal E}^{(k)}_{LM}&=&\epsilon_{l_1 \ldots l_{n-k} m_1 \ldots m_k}\ ,
\end{eqnarray}
consequently,
\begin{eqnarray}
\label{M12b}
{\cal E}^{(k) T}&=&(-1)^{(n-k)k}\ {\cal E}^{(n-k)}
\end{eqnarray}
(Quite generally, for any ${n\choose k}\times {n\choose k}$ matrix ${\sf B}$
we define ${\sf B}^\star$ by ${\sf B}^\star =
{\cal E}^{(k)}  {\sf B}^T {\cal E}^{(k) T}$.).
It holds ($\openone_r$ is the $r\times r$ unit matrix)
\begin{eqnarray}
\label{M13a}
{\cal E}^{(k)}\ {\cal E}^{(k) T}&=&\openone_{n\choose k}\ ,\\[0.3cm]
\label{M13b}
{\cal E}^{(k) T}\ {\cal E}^{(k)}&=&\openone_{n\choose k}\ .
\end{eqnarray}

We can now give some formulas for compound matrices\footnote{In the
context of projective geometry, these matrices and their elements
often are referred
to as Pl\"ucker-Grassmann coordinates.}.
Let ${\sf B}$, ${\sf D}$ be $n\times n$ matrices.
The {\it compound matrix} $C_k\left({\sf B}\right)$,
$0\le k\le n$, is a ${n\choose k}\times {n\choose k}$
matrix of all order $k$
minors of the matrix ${\sf B}$. The indices of the compound
matrix entries are given by ordered strings of length $k$.
These strings are composed from the row and
column indices of the matrix elements of the matrix ${\sf B}$
the given minor of the matrix ${\sf B}$ is composed of.
Typically, the entries
of a compound matrix are ordered lexicographically with respect
to the compound matrix indices (We also apply this convention.).
The {\it supplementary} (or {\it adjugate})
{\it compound matrix} $C^{\; n-k}\left({\sf B}\right)$
(sometimes also referred to as the {\it matrix of the $k$th cofactors})
of the matrix ${\sf B}$ is defined by
the equation (cf.\ eq.\ (\ref{M11}))
\begin{eqnarray}
\label{A1d}
C^{\; n-k}\left({\sf B}\right)&=&C_{n-k}\left({\sf B}\right)^\star\ .
\end{eqnarray}
The components of the supplementary compound matrix
$C^{\; n-k}\left({\sf B}\right)$ can also be defined by means
of the following formula
(here, $l_1 < l_2 <\ldots < l_k$,
$m_1 < m_2 <\ldots < m_k$; \citep{muir2}, Chap.\ IV, \S 89,
p.\ 75, \citep{vein1}, Chap.\ 3, p.\ 18)\footnote{Note, that
in the eqs.\ (A.2), p.\ 5443, (31)-(33), p.\ 5421 in \citep{scha1} the 
indices $l$ and $m$ on the r.h.s.\ should be interchanged to correct
the display of these equations.}.
\begin{eqnarray}
\label{A1f}
C^{\; n-k}\left({\sf B}\right)_{LM}&=&\frac{\partial}{
\partial {\sf B}_{m_1 l_1}}\ldots
\frac{\partial}{\partial {\sf B}_{m_k l_k}}\ \det{\sf B}
\end{eqnarray}
This comparatively little known definition of (matrices of)
cofactors (supplementary
compound matrices) is essentially due to Jacobi \citep{jaco2}, \S 10,
p.\ 301, p.\ 273 of the `Gesammelte Werke', p.\ 25 of the German transl.\
(also see
the corresponding comment by Muir in \citep{muir1}, Part I, Chap.\ IX,
pp.\ 253-272, in particular pp.\ 262/263).\\

For compound matrices holds ($\openone_r$ is the $r\times r$ unit matrix,
$\alpha$ some constant)
\begin{eqnarray}
\label{A1b}
C_k\left(\alpha \openone_n\right)&=&\alpha^k\ \openone_{n\choose k}\ .
\end{eqnarray}
Important relations are given by the
{\it Binet-Cauchy formula}
\begin{eqnarray}
\label{A1}
C_k\left({\sf B}\right) C_k\left({\sf D}\right)
&=& C_k\left({\sf B D}\right)
\end{eqnarray}
from which immediately follows
\begin{eqnarray}
\label{A1c}
C_k\left({\sf B}^{-1}\right) &=& C_k\left({\sf B}\right)^{-1}\ ,
\end{eqnarray}
the
{\it Laplace expansion}
\begin{eqnarray}
\label{A2}
C_k\left({\sf B}\right) C^{\; n-k}\left({\sf B}\right)\ =\
C^{\; n-k}\left({\sf B}\right) C_k\left({\sf B}\right)&=&\nonumber\\[0.3cm]
C_k\left({\sf B}\right) C_{n-k}\left({\sf B}\right)^\star\ =\
C_{n-k}\left({\sf B}\right)^\star C_k\left({\sf B}\right)
&=&\det {\sf B}\ \openone_{n\choose k}\ ,
\end{eqnarray}
{\it Jacobi's theorem} (a consequence of the eqs.\ (\ref{A2}) and
(\ref{A1c}))
\begin{eqnarray}
\label{A3}
C_k\left({\sf B}^{-1}\right) &=&
\frac{1}{\det {\sf B}}\ C^{\; n-k}\left({\sf B}\right)\ =\
\frac{1}{\det {\sf B}}\ C_{n-k}\left({\sf B}\right)^\star\ ,
\end{eqnarray}
and the
{\it Sylvester-Franke theorem}
\begin{eqnarray}
\label{A4}
\det C_k\left({\sf B}\right)&=&\left(\det {\sf B}\right)^{n-1\choose k-1}\ .
\end{eqnarray}

Compound matrices are treated in a number of references.
A comprehensive discussion of compound matrices can be found in
\citep{wedd}, Chap.\ V, pp.\ 63-87, \citep{aitk}, Chap.\ V,
pp.\ 90-110, and, in a modern treatment, in \citep{fied}, Chap.\ 6,
pp.\ 142-155 of the English translation. 
More algebraically oriented modern treatments can
be found in \citep{marc1}, Part I, Chap.\ 2, Sect.\ 2.4, pp.\ 116-159,
Part II, Chap.\ 4, pp.\ 1-164 (very thorough),
\citep{jaco}, Chap.\ 7, Sect.\ 7.2, pp.\ 411-420, and
\citep{cohn}, Vol.\ 3, Chap.\ 2, Sect.\ 2.4, pp.\ 58-68. Concise
reviews of the properties of compound matrices are given
in \citep{bout,prel}. Also note \citep{barn}.\\

\stepcounter{asection}
\refstepcounter{section}
\label{appc}
\section*{Appendix \Alph{asection}}
\setcounter{equation}{0}
\addcontentsline{toc}{section}{Appendix \Alph{asection}}

In this Appendix, we first give a concrete example for the complex 
antisymmetric matrices introduced in Subsec.\ \ref{double}:
\begin{eqnarray}
\label{ntwofurtheradd801}
\Gamma_{(-1)}^{\hspace{-0.5cm}+\atop\ }\ =\
\Gamma_{(-1)}^{{\hspace{-0.5cm}-\atop\ }\dagger}\ =\ 
-\ \Gamma_{(-1)}^{\hspace{-0.5cm}-\atop\ }&=&\ \ \; \left(
\begin{array}{*{4}{c}}
0&
0&
0&
-1\\
0&
0&
-1&
0\\
0&
1&
0&
0\\
1&
0&
0&
0\end{array}
\right)\ ,\\[0.3cm] 
\label{ntwofurtheradd802}
\Gamma_0^{\hspace{-0.5cm}+\atop\ }\ =\ 
\ \ \Gamma_0^{{\hspace{-0.5cm}-\atop\ }\dagger}\ =\ 
\ \ \ \ \Gamma_0^{\hspace{-0.5cm}-\atop\ }\ \ &=&i\ \left(
\begin{array}{*{4}{c}}
0&
0&
0&
1\\
0&
0&
-1&
0\\
0&
1&
0&
0\\
-1&
0&
0&
0\end{array}
\right)\ ,\\[0.3cm]
\label{ntwofurtheradd803}
\Gamma_1^{\hspace{-0.5cm}+\atop\ }\ =\ 
\ \ \Gamma_1^{{\hspace{-0.5cm}-\atop\ }\dagger}\ =\ 
\ \ \ \ \Gamma_1^{\hspace{-0.5cm}-\atop\ }\ \
&=&i\ \left(
\begin{array}{*{4}{c}}
0&
1&
0&
0\\
-1&
0&
0&
0\\
0&
0&
0&
-1\\
0&
0&
1&
0\end{array}
\right)\ ,\\[0.3cm]
\label{ntwofurtheradd804}
\Gamma_2^{\hspace{-0.5cm}+\atop\ }\ =\ 
\ \ \Gamma_2^{{\hspace{-0.5cm}-\atop\ }\dagger}\ =\
\ \ \ \ \Gamma_2^{\hspace{-0.5cm}-\atop\ }\ \ &=&i\ 
\left(\begin{array}{*{4}{c}}
0&
0&
1&
0\\
0&
0&
0&
1\\
-1&
0&
0&
0\\
0&
-1&
0&
0\end{array}\right)\ ,\\[0.3cm]
\label{ntwofurtheradd805}
\Gamma_3^{\hspace{-0.5cm}+\atop\ }\ =\ 
\ \ \Gamma_3^{{\hspace{-0.5cm}-\atop\ }\dagger}\ =\
\ -\ \Gamma_3^{\hspace{-0.5cm}-\atop\ }\ &=&\ \ \; \left(
\begin{array}{*{4}{c}}
0&
-1&
0&
0\\
1&
0&
0&
0\\
0&
0&
0&
-1\\
0&
0&
1&
0\end{array}
\right)\ ,\\[0.3cm]
\label{ntwofurtheradd806}
\Gamma_4^{\hspace{-0.5cm}+\atop\ }\ =\ 
\ \ \Gamma_4^{{\hspace{-0.5cm}-\atop\ }\dagger}\ =\ 
\ -\ \Gamma_4^{\hspace{-0.5cm}-\atop\ }\ &=&\ \ \ \left(
\begin{array}{*{4}{c}}
0&
0&
-1&
0\\
0&
0&
0&
1\\
1&
0&
0&
0\\
0&
-1&
0&
0\end{array}
\right)\ .
\end{eqnarray}
They represent spin space bivectors ($\in\bigwedge^2 (\mathbb{C}_4)$).
For $k = 0,1,2$, the relation
$\Gamma_k^{\hspace{-0.5cm}+\atop\ } 
= \Gamma_k^{\hspace{-0.5cm}-\atop\ }$ applies, consequently, these 
three matrices span the eigenspace of the Hodge operator to the 
eigenvalue -1 [cf.\ eq.\ (\ref{ntwofurtheradd302})]. 
For $k = -1,3,4$, the relation
$\Gamma_k^{\hspace{-0.5cm}+\atop\ }
= - \Gamma_k^{\hspace{-0.5cm}-\atop\ }$ applies and, 
consequently, these three matrices span the eigenspace 
of the Hodge operator to the eigenvalue 1 (For a related discussion
see \citep{kozl3}.). The 
antisymmetric matrices $\Gamma_k^{\hspace{-0.5cm}+\atop\ }$
with indices $k = 0,1,2$ and $k = -1,3,4$  are 
related to a matrix representation of two commuting copies, 
${\bf I}^\prime_1$, ${\bf J}^\prime_1$, ${\bf K}^\prime_1$ 
(${\bf I}^\prime_1 {\bf J}^\prime_1 = {\bf K}^\prime_1$, etc.)
and ${\bf I}^\prime_2$, ${\bf J}^\prime_2$, ${\bf K}^\prime_2,$ 
(${\bf I}^\prime_2 {\bf J}^\prime_2 = {\bf K}^\prime_2$, etc.), of the
system of quaternions (\citep{kwal2}, p.\ 329, eq.\ (5)).
We attach here a prime to the quaternionic units in order to 
distinguish them from the different choice made in 
the eqs.\ (\ref{ntwofurtheraddz67a})-(\ref{ntwofurtheraddz67f})
in Subsec.\ \ref{structures}.
With some hindsight we set
\begin{eqnarray}
\label{ntwofurtheradd807a}
{\bf I}^\prime_1&=&\ \ i\ \Gamma_2^{\hspace{-0.5cm}+\atop\ }\ \ \;
\ = \ \ \ \ i\ \Gamma_2^{\hspace{-0.5cm}-\atop\ }\ ,\\[0.3cm]
\label{ntwofurtheradd807b}
{\bf J}^\prime_1&=&-i\ \Gamma_1^{\hspace{-0.5cm}+\atop\ }\ \ \;
\ = \ -i\ \Gamma_1^{\hspace{-0.5cm}-\atop\ }\ ,\\[0.3cm]
\label{ntwofurtheradd807c}
{\bf K}^\prime_1&=&\ \ i\ \Gamma_0^{\hspace{-0.5cm}+\atop\ }\ \ \;
\ = \ \ \ \ i\ \Gamma_0^{\hspace{-0.5cm}-\atop\ }\ ,\\[0.3cm]
\label{ntwofurtheradd807d}
{\bf I}^\prime_2&=&\ \ \ \ \Gamma_3^{\hspace{-0.5cm}+\atop\ }\ \ \
\ = \ -\ \Gamma_3^{\hspace{-0.5cm}-\atop\ }\ ,\\[0.3cm]
\label{ntwofurtheradd807e}
{\bf J}^\prime_2&=&\ \ \ \ \Gamma_4^{\hspace{-0.5cm}+\atop\ }\ \ \
\ = \ -\ \Gamma_4^{\hspace{-0.5cm}-\atop\ }\ ,\\[0.3cm]
\label{ntwofurtheradd807f}
{\bf K}^\prime_2&=&\ \ \ \ \Gamma_{(-1)}^{\hspace{-0.5cm}+\atop\ }
\ = \ -\ \Gamma_{(-1)}^{\hspace{-0.5cm}-\atop\ }\ .
\end{eqnarray}
Note, that these relations are valid only if
the matrices  $\Gamma_k^{\hspace{-0.5cm}+\atop\ }$
with the indices $k = 0,1,2$ and $k = -1,3,4$, respectively, span
the two eigenspaces of the Hodge operator to the eigenvalues $-1$ and $1$. 
This will not be the case in general.
Taking into account the above identifications we find the following 
relations of the quaternionic units to the elements of the biparavector 
space\footnote{A related approach has been used in 
\citep{have1}, Sec.\ 4.1, pp.\ 99-102,
however, without relating the quaternionic units to the bivectors of 
spin space. The approach in \citep{have1} has rather some similarities 
with the one used by Milner \citep{miln2}. Incidentally, a paper by Rau 
\citep{rau1}, p.\ 4, eq.\ (14), contains some related considerations.}.
\begin{eqnarray}
\label{ntwofurtheradd808a}
{\bf I}^\prime_1&=&{\bf J}^\prime_1 {\bf K}^\prime_1
\ =\ -\ \Gamma_0^{\hspace{-0.5cm}+\atop\ } 
\Gamma_1^{\hspace{-0.5cm}-\atop\ }\ \ \ 
\ =\ -\ \hat{\hat{\hat{c}\,}}_{01}
\ =\ \hat{c}_1 \\[0.3cm]
\label{ntwofurtheradd808b}
{\bf J}^\prime_1&=&{\bf K}^\prime_1 {\bf I}^\prime_1\;
\ =\ -\ \Gamma_0^{\hspace{-0.5cm}+\atop\ } 
\Gamma_2^{\hspace{-0.5cm}-\atop\ }\ \ \
\ =\ -\ \hat{\hat{\hat{c}\,}}_{02}
\ =\ \hat{c}_2\\[0.3cm]
\label{ntwofurtheradd808c}
{\bf K}^\prime_1&=&{\bf I}^\prime_1 {\bf J}^\prime_1\; \; 
\ =\ -\ \Gamma_1^{\hspace{-0.5cm}+\atop\ } 
\Gamma_2^{\hspace{-0.5cm}-\atop\ }\ \ \
\ =\ -\ \hat{\hat{\hat{c}\,}}_{12} 
\ =\ \hat{c}_1 \hat{c}_2\\[0.3cm]
\label{ntwofurtheradd808d}
{\bf I}^\prime_2&=&{\bf J}^\prime_2 {\bf K}^\prime_2\:
\ =\ \ \ \ \Gamma_{(-1)}^{\hspace{-0.5cm}+\atop\ } 
\Gamma_4^{\hspace{-0.5cm}-\atop\ }
\ =\ \hat{\hat{\hat{c}\,}}_{(-1)4}
\ =\ i\ \hat{c}_1 \hat{c}_2 \hat{c}_3\\[0.3cm]
\label{ntwofurtheradd808e}
{\bf J}^\prime_2&=&{\bf K}^\prime_2 {\bf I}^\prime_2\
\ =\ -\ \Gamma_{(-1)}^{\hspace{-0.5cm}+\atop\ } 
\Gamma_3^{\hspace{-0.5cm}-\atop\ }
\ =\ -\ \hat{\hat{\hat{c}\,}}_{(-1)3}
\ =\ i\ \hat{c}_1 \hat{c}_2 \hat{c}_4\\[0.3cm]
\label{ntwofurtheradd808f}
{\bf K}^\prime_2&=&{\bf I}^\prime_2 {\bf J}^\prime_2\; \
\ =\ -\ \Gamma_0^{\hspace{-0.5cm}+\atop\ } 
\Gamma_2^{\hspace{-0.5cm}-\atop\ }\ \ \
\ =\ -\ \hat{\hat{\hat{c}\,}}_{34} 
\ =\ -\ \hat{c}_3 \hat{c}_4
\end{eqnarray}
For the generators of the Clifford algebra $C(0,4)$ we 
obtain\footnote{This choice corresponds to the 
Cartan extension of the Clifford algebra
$C(0,2)$ generated by $\hat{\hat{\hat{c}\,}}_1$, $\hat{\hat{\hat{c}\,}}_2$, 
cf.\ \citep{budi1}, Sec.\ 6.3, p.\ 80, eq.\ (6.26).}
\begin{eqnarray}
\label{ntwofurtheradd809a}
\hat{\hat{\hat{c}\,}}_1&=&\hat{c}_1\ =\ {\bf I}^\prime_1\ ,\\[0.3cm]
\label{ntwofurtheradd809b}
\hat{\hat{\hat{c}\,}}_2&=&\hat{c}_2\ =\ {\bf J}^\prime_1\ ,\\[0.3cm]
\label{ntwofurtheradd809c}
\hat{\hat{\hat{c}\,}}_3&=&\hat{c}_3\ =\ 
i\ {\bf K}^\prime_1 {\bf I}^\prime_2\ ,\\[0.3cm]
\label{ntwofurtheradd809d}
\hat{\hat{\hat{c}\,}}_4&=&\hat{c}_4\ =\ 
i\ {\bf K}^\prime_1 {\bf J}^\prime_2\ ,
\end{eqnarray}
and, consequently,
\begin{eqnarray}
\label{ntwofurtheradd809e}
\hat{\hat{\hat{c}\,}}_{(-1)}&=&i\ {\bf K}^\prime_1{\bf K}^\prime_2\ .
\end{eqnarray}
The above identifications are up to minor detail the same as given by
Schouten \citep{scho3}, p.\ 106, eqs.\ (5), (6), and 
Lema\^{\i}tre \citep{lema2}, p.\ 170
(For related discussions also see 
\citep{eddi5}, Sec.\ 2.3, p.\ 23 and Sec.\ 3.8, p.\ 47,
\citep{eddi6}, Chap.\ VI, \S\ 53, p.\ 108, \citep{miln2}.).\\

\stepcounter{asection}
\refstepcounter{section}
\label{appd}
\section*{Appendix \Alph{asection}}
\setcounter{equation}{0}
\addcontentsline{toc}{section}{Appendix \Alph{asection}}

Here we calculate traces
of products of the operators $\hat{\hat{\hat{c}\,}}_k$ 
[For related results see \citep{macf1}, p.\ 136, eqs.\ (2.12)-(2.21).].
These expressions are necessary for the discussion in Subsec.\ \ref{double}
[see eqs.\ (\ref{ntwofurtheradd318})-(\ref{ntwofurtheradd319b})].
We find
\begin{eqnarray}
\label{ntwofurtheradd309}
-\ {\rm tr}\;\hat{\hat{\hat{c}\,}}_l\ =\ 
{\rm tr}\left(\Gamma_0^{\hspace{-0.5cm}+\atop\ }
\Gamma_l^{\hspace{-0.5cm}-\atop\ }\right)&=&4\ \delta_{0l}
\end{eqnarray}
which is a special case of
\begin{eqnarray}
\label{ntwofurtheradd310}
{\rm tr}\left(\hat{\hat{\hat{c}\hspace{0.3mm}}}_k^\dagger\,
\hat{\hat{\hat{c}\,}}_l\right)\ =\ 
{\rm tr}\left(\Gamma_k^{\hspace{-0.5cm}+\atop\ }
\Gamma_l^{\hspace{-0.5cm}-\atop\ }\right)&=&4\ \delta_{kl}\ .
\end{eqnarray}
Furthermore, we have
\begin{eqnarray}
\label{ntwofurtheradd311}
-\ {\rm tr}\left(\hat{\hat{\hat{c}\,}}_l\,
\hat{\hat{\hat{c}\hspace{0.3mm}}}_m^\dagger\, 
\hat{\hat{\hat{c}\,}}_n\right)\ =\ 
{\rm tr}\left(\Gamma_0^{\hspace{-0.5cm}+\atop\ }
\Gamma_l^{\hspace{-0.5cm}-\atop\ }\Gamma_m^{\hspace{-0.5cm}+\atop\ }
\Gamma_n^{\hspace{-0.5cm}-\atop\ }\right)&=&4\left(
\delta_{0l}\ \delta_{mn} - \delta_{0m}\ \delta_{ln} +
\delta_{0n}\ \delta_{lm}\right)\ \ \ \ 
\end{eqnarray}
which is a special case of
\begin{eqnarray}
\label{ntwofurtheradd312}
{\rm tr}\left(\hat{\hat{\hat{c}\hspace{0.3mm}}}_k^\dagger\, 
\hat{\hat{\hat{c}\,}}_l\,
\hat{\hat{\hat{c}\hspace{0.3mm}}}_m^\dagger\, 
\hat{\hat{\hat{c}\,}}_n\right)\ =\ 
{\rm tr}\left(\Gamma_k^{\hspace{-0.5cm}+\atop\ }
\Gamma_l^{\hspace{-0.5cm}-\atop\ }\Gamma_m^{\hspace{-0.5cm}+\atop\ }
\Gamma_n^{\hspace{-0.5cm}-\atop\ }\right)&=&4\left(
\delta_{kl}\ \delta_{mn} - \delta_{km}\ \delta_{ln} +
\delta_{kn}\ \delta_{lm}\right)\ , \ \ \ \ \ 
\end{eqnarray}
and 
\begin{eqnarray}
\label{ntwofurtheradd313}
-\ {\rm tr}\left(
\hat{\hat{\hat{c}\,}}_l\,
\hat{\hat{\hat{c}\hspace{0.3mm}}}_m^\dagger\, \hat{\hat{\hat{c}\,}}_n
\hat{\hat{\hat{c}\hspace{0.3mm}}}_p^\dagger\, \hat{\hat{\hat{c}\,}}_q\right)&=&
{\rm tr}\left(\Gamma_0^{\hspace{-0.5cm}+\atop\ }
\Gamma_l^{\hspace{-0.5cm}-\atop\ }\Gamma_m^{\hspace{-0.5cm}+\atop\ }
\Gamma_n^{\hspace{-0.5cm}-\atop\ }\Gamma_p^{\hspace{-0.5cm}+\atop\ }
\Gamma_q^{\hspace{-0.5cm}-\atop\ }\right)\nonumber\\[0.3cm]
&=&4 \left[ i\ \epsilon_{0lmnpq}\right.\nonumber\\[0.3cm]
&&+\ \delta_{0l}\ \left(
\delta_{mn}\ \delta_{pq} - \delta_{mp}\ \delta_{nq} +
\delta_{mq}\ \delta_{np}\right)\nonumber\\[0.3cm]
&&-\ \delta_{0m}\ \left(
\delta_{ln}\ \delta_{pq} - \delta_{lp}\ \delta_{nq} +
\delta_{lq}\ \delta_{np}\right)\nonumber\\[0.3cm]
&&+\ \delta_{0n}\ \left(
\delta_{lm}\ \delta_{pq} - \delta_{lp}\ \delta_{mq} +
\delta_{lq}\ \delta_{mp}\right)\nonumber\\[0.3cm]
&&-\ \delta_{0p}\ \left(
\delta_{lm}\ \delta_{nq} - \delta_{ln}\ \delta_{mq} +
\delta_{lq}\ \delta_{mn}\right)\nonumber\\[0.3cm]
&&+\ \left.\delta_{0q}\ \left(
\delta_{lm}\ \delta_{np} - \delta_{ln}\ \delta_{mp} +
\delta_{lp}\ \delta_{mn}\right)\right] 
\end{eqnarray}
which is a special case of
\begin{eqnarray}
\label{ntwofurtheradd314}
{\rm tr}\left(\hat{\hat{\hat{c}\hspace{0.3mm}}}_k^\dagger\, 
\hat{\hat{\hat{c}\,}}_l\,
\hat{\hat{\hat{c}\hspace{0.3mm}}}_m^\dagger\, 
\hat{\hat{\hat{c}\,}}_n
\hat{\hat{\hat{c}\hspace{0.3mm}}}_p^\dagger\, 
\hat{\hat{\hat{c}\,}}_q\right)&=& 
{\rm tr}\left(\Gamma_k^{\hspace{-0.5cm}+\atop\ }
\Gamma_l^{\hspace{-0.5cm}-\atop\ }\Gamma_m^{\hspace{-0.5cm}+\atop\ }
\Gamma_n^{\hspace{-0.5cm}-\atop\ }\Gamma_p^{\hspace{-0.5cm}+\atop\ }
\Gamma_q^{\hspace{-0.5cm}-\atop\ }\right)\nonumber\\[0.3cm]
&=&4 \left[ i\ \epsilon_{klmnpq}\right.\nonumber\\[0.3cm]
&&\ \ \ \ +\ \delta_{kl}\ \delta_{mn}\ \delta_{pq} - 
\delta_{kl}\ \delta_{mp}\ \delta_{nq} +
\delta_{kl}\ \delta_{mq}\ \delta_{np}\nonumber\\[0.3cm]
&&\ \ \ \ -\ \delta_{km}\ \delta_{ln}\ \delta_{pq} + 
\delta_{km}\ \delta_{lp}\ \delta_{nq} -
\delta_{km}\ \delta_{lq}\ \delta_{np}\nonumber\\[0.3cm]
&&\ \ \ \ +\ \delta_{kn}\ \delta_{lm}\ \delta_{pq} - 
\delta_{kn}\ \delta_{lp}\ \delta_{mq} +
\delta_{kn}\ \delta_{lq}\ \delta_{mp}\nonumber\\[0.3cm]
&&\ \ \ \ -\ \delta_{kp}\ \delta_{lm}\ \delta_{nq} + 
\delta_{kp}\ \delta_{ln}\ \delta_{mq} -
\delta_{kp}\ \delta_{lq}\ \delta_{mn}\nonumber\\[0.3cm]
&&\ \ \ \ +\ \left.\delta_{kq}\ \delta_{lm}\ \delta_{np} - 
\delta_{kq}\ \delta_{ln}\ \delta_{mp} +
\delta_{kq}\ \delta_{lp}\ \delta_{mn} \right]\ .\ \ \ \ \ 
\end{eqnarray}
Further traces can easily be obtained from the above results by means
of the relation [This is a special version of 
eq.\ (\ref{ntwofurtheradd303}) in disguise.]
\begin{eqnarray}
\label{ntwofurtheradd315}
\hat{\hat{\hat{c}\hspace{0.3mm}}}_k^\dagger\ +\ \hat{\hat{\hat{c}\,}}_k
&=&-2\ \delta_{k0}\ \openone_4\ .
\end{eqnarray}

\stepcounter{asection}
\refstepcounter{section}
\label{appe}
\section*{Appendix \Alph{asection}}
\setcounter{equation}{0}
\addcontentsline{toc}{section}{Appendix \Alph{asection}}

In this Appendix we give the explicit relation between the 
representation of $SU(4)$ matrices discussed by \"Ostlund and Mele
\citep{ostl1} (cf.\ Subsec.\ \ref{ostlund}) and our parametrization
(\ref{ntwofurtheradd316}). The coefficients of the two 
parametrizations transform into each other in 4 groups of 4
coefficients.\\
\noindent
Group 1 reads:
\begin{eqnarray}
\label{appost1a}
X_{(-1)(-1)}&=&
\frac{1}{4}\left(T_0 - i\ T_{(-1)0} - i\ T_{12} - i\ T_{34}\right)\ ,
\hspace{2.6cm}\ \\[0.3cm] 
\label{appost1b}
X_{00}&=&
\frac{1}{4}\left(T_0 - i\ T_{(-1)0} + i\ T_{12} + i\ T_{34}\right)\ ,
\\[0.3cm] 
\label{appost1c}
X_{11}&=&
\frac{1}{4}\left(T_0 + i\ T_{(-1)0} - i\ T_{12} + i\ T_{34}\right)\ ,
\\[0.3cm] 
\label{appost1d}
X_{22}&=&
\frac{1}{4}\left(T_0 + i\ T_{(-1)0} + i\ T_{12} - i\ T_{34}\right)\ ,
\end{eqnarray}
and the inverse relations are
\begin{eqnarray}
\label{appost1e}
T_0&=&\frac{1}{4}\left(X_{(-1)(-1)} + X_{00} + X_{11} + X_{22}\right)\ =\
\frac{1}{4}\ {\rm tr}\; X\ ,
\\[0.3cm] 
\label{appost1f}
T_{(-1)0}&=&\frac{i}{4}\left(X_{(-1)(-1)} + X_{00} - X_{11} - X_{22}\right)\ ,
\\[0.3cm] 
\label{appost1g}
T_{12}&=&\frac{i}{4}\left(X_{(-1)(-1)} - X_{00} + X_{11} - X_{22}\right)\ ,
\\[0.3cm] 
\label{appost1h}
T_{34}&=&\frac{i}{4}\left(X_{(-1)(-1)} - X_{00} - X_{11} + X_{22}\right)\ .
\end{eqnarray}
Group 2 reads:
\begin{eqnarray}
\label{appost2a}
X_{(-1)0}&=&
\frac{1}{4}\left(- T_{13} - i\ T_{14} - i\ T_{23} + T_{24}\right)\ ,
\hspace{2.5cm}\ \\[0.3cm] 
\label{appost2b}
X_{0(-1)}&=&
\frac{1}{4}\left(\ \  T_{13} - i\ T_{14} - i\ T_{23} - T_{24}\right)\ ,
\\[0.3cm] 
\label{appost2c}
X_{12}&=&
\frac{1}{4}\left(\ \  T_{13} - i\ T_{14} + i\ T_{23} + T_{24}\right)\ ,
\\[0.3cm] 
\label{appost2d}
X_{21}&=&
\frac{1}{4}\left(- T_{13} - i\ T_{14} + i\ T_{23} - T_{24}\right)\ ,
\end{eqnarray}
and the inverse relations are
\begin{eqnarray}
\label{appost2e}
T_{13}&=&
\frac{1}{4}\left(- X_{(-1)0} + X_{0(-1)} + X_{12} - X_{21}\right)\ ,
\hspace{1.2cm}\ \\[0.3cm] 
\label{appost2f}
T_{14}&=&
\frac{i}{4}\left(\ \ X_{(-1)0} + X_{0(-1)} + X_{12} + X_{21}\right)\ ,
\\[0.3cm] 
\label{appost2g}
T_{23}&=&
\frac{i}{4}\left(\ \ X_{(-1)0} + X_{0(-1)} - X_{12} - X_{21}\right)\ ,
\\[0.3cm] 
\label{appost2h}
T_{24}&=&
\frac{1}{4}\left(\ \ X_{(-1)0} - X_{0(-1)} + X_{12} - X_{21}\right)\ .
\end{eqnarray}
Group 3 reads:
\begin{eqnarray}
\label{appost3a}
X_{(-1)1}&=&
\frac{1}{4}\left(- T_{(-1)3} - i\ T_{(-1)4} - i\ T_{03} + T_{04}\right)\ ,
\hspace{1.7cm}\ \\[0.3cm] 
\label{appost3b}
X_{02}&=&
\frac{1}{4}\left(\ \ T_{(-1)3} - i\ T_{(-1)4} - i\ T_{03} - T_{04}\right)\ ,
\\[0.3cm] 
\label{appost3c}
X_{1(-1)}&=&
\frac{1}{4}\left(- T_{(-1)3} + i\ T_{(-1)4} - i\ T_{03} - T_{04}\right)\ ,
\\[0.3cm] 
\label{appost3d}
X_{20}&=&
\frac{1}{4}\left(\ \ T_{(-1)3} + i\ T_{(-1)4} - i\ T_{03} + T_{04}\right)\ ,
\end{eqnarray}
and the inverse relations are
\begin{eqnarray}
\label{appost3e}
T_{(-1)3}&=&
\frac{1}{4}\left(- X_{(-1)1} - X_{02} + X_{1(-1)} + X_{20}\right)\ ,
\hspace{1.7cm}\ \\[0.3cm] 
\label{appost3f}
T_{(-1)4}&=&
\frac{i}{4}\left(\ \ X_{(-1)1} - X_{02} + X_{1(-1)} - X_{20}\right)\ ,
\\[0.3cm] 
\label{appost3g}
T_{03}&=&
\frac{i}{4}\left(\ \ X_{(-1)1} + X_{02} + X_{1(-1)} + X_{20}\right)\ ,
\\[0.3cm] 
\label{appost3h}
T_{04}&=&
\frac{1}{4}\left(\ \ X_{(-1)1} - X_{02} - X_{1(-1)} + X_{20}\right)\ .
\end{eqnarray}
Group 4 reads:
\begin{eqnarray}
\label{appost4a}
X_{(-1)2}&=&
\frac{1}{4}\left(\ \ T_{(-1)1} + i\ T_{(-1)2} + i\ T_{01} - T_{02}\right)\ ,
\hspace{1.7cm}\ \\[0.3cm] 
\label{appost4b}
X_{01}&=&
\frac{1}{4}\left(- T_{(-1)1} + i\ T_{(-1)2} - i\ T_{01} - T_{02}\right)\ ,
\\[0.3cm] 
\label{appost4c}
X_{10}&=&
\frac{1}{4}\left(\ \ T_{(-1)1} + i\ T_{(-1)2} - i\ T_{01} + T_{02}\right)\ ,
\\[0.3cm] 
\label{appost4d}
X_{2(-1)}&=&
\frac{1}{4}\left(- T_{(-1)1} + i\ T_{(-1)2} + i\ T_{01} + T_{02}\right)\ ,
\end{eqnarray}
and the inverse relations are
\begin{eqnarray}
\label{appost4e}
T_{(-1)1}&=&
\frac{1}{4}\left(\ \ X_{(-1)2} - X_{01} + X_{10} - X_{2(-1)}\right)\ ,
\hspace{1.7cm}\ \\[0.3cm] 
\label{appost4f}
T_{(-1)2}&=&
\frac{i}{4}\left(- X_{(-1)2} - X_{01} - X_{10} - X_{2(-1)}\right)\ ,
\\[0.3cm] 
\label{appost4g}
T_{01}&=&
\frac{i}{4}\left(- X_{(-1)2} + X_{01} + X_{10} - X_{2(-1)}\right)\ ,
\\[0.3cm] 
\label{appost4h}
T_{02}&=&
\frac{i}{4}\left(- X_{(-1)2} - X_{01} + X_{10} + X_{2(-1)}\right)\ .
\end{eqnarray}
\newpage

\end{document}